\documentclass[ms]{informs3} 

\OneAndAHalfSpacedXI
\usepackage{amsmath}

\usepackage{natbib}
 \bibpunct[, ]{(}{)}{,}{a}{}{,}%
 \def\bibfont{\small}%

\usepackage{stackengine}
\usepackage{soul}
\usepackage{multirow}
\usepackage{tikz}

\usepackage{bbm}
\usepackage{cases, enumitem}
\usepackage{hyperref}
\usepackage{pgfplots}
\pgfplotsset{width=10cm,compat=1.9}

\TheoremsNumberedThrough     

\EquationsNumberedThrough    

\newcommand{\defeq}{\mathrel{\mathop:}=}

\def\po{\textsc{polyra}}
\def\bucketOneOne{B_{1,1}}
\def\bucketOneTwo{B_{1,2}}
\def\bucketTwoTwo{B_{2,2}}
\def\bucketTwoOne{B_{2,1}}

\def\bucketThree{B_{3}}

\def\bucketTwo{B_{2}}

\def\bucketMPlusOne{B_{M+1}}
\def\bucketK{B_{K}}

\def\bucketTwoOne{B_{2,1}}
\def\bucketMOne{B_{M,1}}
\def\bucketOneTwo{B_{1,2}}
\def\bucketTwoTwo{B_{2,2}}
\def\bucketMTwo{B_{M,2}}

\newcommand{\bucketij}[2]{B_{#1,#2}}
\newcommand{\bucket}[1]{B_{#1}}

\def\fbucketOneOne{b_{1,1}}
\def\fbucketOneTwo{b_{1,2}}
\def\fbucketTwoOne{b_{2,1}}
\def\fbucketTwoTwo{b_{2,2}}
\def\fbucketThree{b_{3}}

\def\fbucketTwo{b_{2}}

\def\fbucketMPlusOne{b_{M+1}}

\def\fbucketK{b_{K}}
\def\fbucketOnei{b_{1, i}}

\def\fbucketMOne{b_{M, 1}}
\def\fbucketK{b_{K}}
\def\fbucketTwoTwo{b_{2,2}}  
\def\fbucketMTwo{b_{M, 2}}

\def\stateB{\mathbf{B}}

\def\feasible{\mathcal{B}}

\def\consistent{valid}
\def\nestOne{{n}_1}
\def\nestTwo{n_2}

\def\nestK{n_K}

\newcommand{\nest}[1]{n_{#1}}
\def\feasiblenest{\mathcal{B}_{\text{nest}}}
\def\nestedPolyra{\mathcal{B}_{\text{nest}}}

\def\optupper{s}
\def\optOneOne{s_{1,1}}
\def\optOneTwo{s_{1,2}}
\def\optTwoTwo{s_{2,2}}
\def\optOnet{s_{1,t}}
\def\optTwot{s_{2,t}}
\def\optTwoOne{s_{2,1}}

\def\deltan{\Delta n}
\newcommand{\nestedvar}[2]{s_{#1}^{#2}}
\newcommand{\numopt}[1]{y_{#1}}
\newcommand{\numinput}[1]{x_{#1}}
\newcommand{\optvar}[2]{s_{#1, #2}}

\newcommand{\algserveone}[1]{z_{#1}}
\newcommand{\algservetwo}[1]{w_{#1}}
\def\maxtype{k_0}
\def\crr{\Gamma}
\usepackage{amsmath,calc}
\newlength{\LPlhbox}

\def\feasibleMOne{\mathcal{B}^{(3, 1)}}
\def\feasibleMTwo{\mathcal{B}^{(3, 2)}}
\def\up{\text{up}}
\def\lpp{\text{LP}}
\def\alg{\mathcal A}

\usepackage{amsmath,xparse,zref-savepos}
\makeatletter
\@ifundefined{zsaveposx}{\let\zsaveposx\zsavepos}{}
\newcounter{hposcnt}
\renewcommand*{\thehposcnt}{hpos\number\value{hposcnt}}
\NewDocumentCommand{\lplabel}{o m}{%
	\stepcounter{hposcnt}%
	\zsaveposx{\thehposcnt l}%
	\zref@refused{\thehposcnt l}%
	\zref@refused{hpos0l}%
	\makebox[0pt][r]{\makebox[\dimexpr\zposx{\thehposcnt l}sp-\zposx{hpos0l}sp][l]{#2}}%
	\IfNoValueF{#1}
	{\def\@currentlabel{#2}\ltx@label{#1}}
}
\makeatother
\AtBeginDocument{\zsaveposx{hpos0l}}

\usepackage{enumitem}

\usepackage{caption}
\usepackage{subcaption}

\begin{document}




\TITLE{Upfront Commitment in Online Resource Allocation with Patient Customers}
%
%
%

\ARTICLEAUTHORS{
	\AUTHOR{Negin Golrezaei}
	\AFF{Sloan School of Management, Massachusetts Institute of Technology, \EMAIL{golrezae@mit.edu}, \URL{}}
	\AUTHOR{Evan Yao}
	\AFF{Operations Research Center, Massachusetts Institute of Technology, \EMAIL{evanyao@mit.edu}, \URL{}}
}

\ABSTRACT{%
{\color{black}In many  on-demand online platforms such as ride-sharing, grocery delivery, or shipping, 
 some arriving demand points (agents) may be patient and willing to wait a short amount of time for the resource or service as long as there is an
upfront guarantee that service will be ultimately provided within a certain delay. Motivated by this, we study the online resource allocation problem in the presence of such flexible agents that we refer to \emph{partially patient}. We present a setting with partially patient and impatient agents who seek a resource or service that replenishes periodically. Impatient agents demand the resource immediately upon arrival while partially patient agents are willing to wait a short period conditioned on an upfront commitment to receive the resource. We study this setting under adversarial arrival models and  using a relaxed notion of competitive ratio, which is known as resource/feature augmented competitive ratio. Under this notion, introduced  by \cite{kalyanasundaram2000speed},  online algorithms have access to additional resources/features (in our case agents' flexibility) than the optimal offline algorithm to which it is compared, allowing us to measure the value of having such flexible agents.}

We present a class of POLYtope-based Resource Allocation (\po) algorithms that achieve optimal or near-optimal competitive ratios. Such \po{} algorithms work by consulting a particular polytope and only making decisions that guarantee the algorithm's state remains feasible in this polytope. When the number of agent types is either two or three, \po{} algorithms can obtain the optimal competitive ratio.  We design these polytopes by constructing an upper bound on the competitive ratio of any algorithm, which is characterized via a linear program (LP) that considers a collection of overlapping worst-case input sequences. Our designed \po{} algorithms then mimic the optimal solution of this upper bound LP via its polytope's definition, obtaining the optimal competitive ratio. When there are more than three types, our overlapping worst-case input sequences do not necessarily result in an attainable competitive ratio, adding an additional challenge to the problem. Considering this, we present a near-optimal \emph{nested} \po{} algorithm which achieves at least $80\%$ of the optimal competitive ratio while having a simple and interpretable nested structure. {\color{black} {\color{black}We complement our theoretical studies with numerical analysis which shows the efficiency of our algorithms beyond adversarial arrivals and feature-augmented competitive ratio metric.} }}


\maketitle

\section{Introduction}
{\color{black}Online resource allocation is a fundamental problem that arises in almost every {\color{black}industry} from government and healthcare (e.g., during a pandemic crisis), to e-commerce and online services. In such problems, the decision-maker strives to best allocate a set of scarce resources to agents with varying levels of value. While even an offline resource allocation problem can be difficult, the online setting brings extra challenges because we must make irrevocable decisions without having full information regarding the future. The critical property of the online setting is that arriving agents must be allocated the resource immediately or otherwise leave the system forever. Such a property creates the following trade-off: allocating a resource now to a low-value agent has a potential opportunity cost in the future, while saving resources for the future can hurt if no higher-valued agents arrive.}

{
\color{black}
In practical applications, one way of mitigating the challenge of the online setting is to exploit agents' patience. The assumption that all agents demand the resource immediately can be relaxed if some agents are patient and willing to wait a short period  for the resource. If a decision-maker were able to delay the acceptance/rejection decision for patient agents, then they would be certainly able to make better decisions as they can observe more of the demand. However, in many applications, it is unreasonable to assume that  agents are willing to wait for a resource and risk being ultimately rejected. This is especially true in the allocation of scarce services in on-demand online platforms such as ride-sharing, grocery delivery, or shipping. An agent may agree to receive their ride, grocery, or package within a longer time window, but only if there is an upfront guarantee that service will be ultimately provided within this window. We refer to such agents who are patient only if provided with an upfront guarantee as \textit{partially patient} or flexible. 

The existence of partially patient agents is challenging because we face a new dilemma: accepting a partially patient agent results in greater short-term flexibility in our decision-making process, but it comes at the cost of a future resource commitment that  cannot be allocated to a future higher-reward agent. We present a novel model for online resource allocation with the existence of such partially patient agents and  provide simple algorithms which achieve optimal or near-optimal performance under an adversarial arrival process.
}

\subsection{Our Contributions}
We describe the contributions of our work as follows.
 
{\color{black}\textbf{A novel extension of a classic revenue management model to include partially patient agents who require commitment (Section \ref{sec:model}).} A famous model for online resource allocation is the single-leg revenue management model under adversarial arrivals by \cite{BallandQueyrenne2009}. In this problem, a decision-maker has  $C$ units of a single resource and must allocate them to arriving agents with heterogeneous rewards $r_1 < \ldots < r_K$ in an online fashion. The goal is to design algorithms which maximize the competitive ratio, the worst-case ratio of an online algorithm's cumulative reward to that of a clairvoyant optimal solution that knows the arrival sequence in advance. We present an extension of this model with the following main modifications, which we define formally in Section \ref{sec:model}.}

\begin{itemize} [leftmargin=*]
    \item {\color{black}\textbf{Multiple periods with partially patient agents. } We generalize the model to a multi-period setting with $T \geq 1$ periods, during each of which agents arrive in an online fashion. Each time period $t \in [T]$ is divided into an arrival phase and service phase. During the arrival phase, agents arrive in an online fashion and we still must make an irrevocable decision about whether  to accept or reject each one. During the service phase, we must choose how to allocate the $C$ capacity we have in each period to the accepted agents.  When an agent is accepted, we provide an upfront \textit{commitment} to providing service to that agent in the near future. Agents in our new model differ not only in terms of reward but also in their level of patience. Out of these $K$ types of agents, the first $M$ are partially patient, while the remaining $K-M$ are impatient. For impatient agents accepted in period $t \in [T]$, they must be serviced in period $t$'s service phase.  For any accepted partially patient agents in time period  $t \in [T]$,  we are committed to serving  them either in  period $t$ or $t+1$'s service phase. See Section \ref{sec:application} for more details.}

    \item {\color{black}\textbf{Competitive ratio (CR) definition.} While we continue to study adversarial arrival sequences, we employ a relaxed CR analysis called \textit{resource augmented CR}. Introduced by \cite{kalyanasundaram2000speed}, this relaxed notion of CR (that {\color{black}we refer to} as feature-augmented CR) bounds the worst-case performance of an online algorithm augmented with an additional feature versus an optimal offline solution that cannot utilize the feature. In our case, the feature is the time-flexibility of the first $M$ types of agents. Such a relaxed  notion of CR is standard/common in the literature (see e.g, \cite{resource_augmentation_1, resource_augmentation_2, resource_augmentation_3} or \cite{roughgarden} for a survey) and is often necessary to characterize the benefit of an additional feature because features often disproportionately benefit the offline clairvoyant benchmark, resulting in a decreased {\color{black}CR} after introducing the feature (see Section \ref{sec:resource_augmented_cr} for details). As we will show in this work, considering this relaxed notion of CR results in  practical and easy-to-implement algorithms with great performance in realistic scenarios under both feature-augmented and traditional notions of CR  (see our numerical studies in Section \ref{sec:numerical_studies}).}  
\end{itemize}

\textbf{Polytope-based resource allocation (POLYRA) algorithm (Section \ref{sec:polyra}).} We present a general class of resource allocation algorithms that consist of two components: a state $\stateB$ and a feasible region $\feasible$. At any moment in the arrival sequence, the state $\stateB$ encapsulates all the decisions the algorithm has made so far. For our specific model, $\stateB$ keeps track of the number of agents of each type that the algorithm has accepted but not yet served, including how many of them require resource in the current time period. A \po{} algorithm makes decisions in a greedy fashion while maintaining the invariant that $\stateB \in \feasible$; see Section \ref{sec:polyra} for more details. By cleverly defining the state $\stateB$ and constructing the feasible polytope $\feasible$, we design optimal or near-optimal algorithms for our model. (We will provide more details later.) {\color{black}Some of our designed algorithms---called nested \po{} algorithms---have a simple \emph{nested} structure and can be viewed as an extension of the nested booking policies from \cite{BallandQueyrenne2009} that are designed for a setting without flexibility.
The nested booking policies  reserve a capacity for higher reward agents in a nested fashion. We generalize this notion of nested booking limits to our multi-period setting with partially patient agents through the definition of a nested polytope (Definition \ref{def:nested}). This generalization also leads to the following benchmark algorithm.}

{\color{black}\textbf{Benchmark algorithm.}
The key challenge in designing a nested booking policy is choosing the size of the booking limits, or nests as we will refer to them. The  benchmark algorithm that we refer to as the \emph{BQ nested \po{} algorithm} uses the nest sizes from the nested booking policies of \cite{BallandQueyrenne2009}, which are designed to achieve good performance in a setting without partially patient agents. We will compare this benchmark algorithm with our nested \po{} algorithms whose nest sizes are carefully optimized through solving a linear program (LP). Theoretically, we show that optimizing these nest sizes results in a strictly better theoretical performance in terms of feature augmented CR. (See Theorem \ref{theorem:our_alg_beats_bq}.) {\color{black} Numerically, we show that our optimized nested \po{} algorithms outperform the BQ nested \po{} algorithm in terms of both the our feature-augmented CR as well as the more traditional CR definition (Section \ref{sec:numerical_studies}). While obtaining a higher feature-augmented CR was expected given our theoretical results, it is quite interesting that our optimized nested \po{} algorithms yield higher traditional CR than the BQ nested \po{} algorithm.}}

\textbf{Optimality of POLYRA algorithms for two types (Section \ref{section:two_type}).} When there are two types of agents with rewards $r_1 < r_2$ and type 1 partially patient and type 2 impatient, we characterize a simple nested \po{} algorithm that achieves the optimal {\color{black}feature-augmented}  
CR of $\frac{2}{3 - r_1/r_2}$ (Theorem \ref{thm:opt_2}). 
In contrast, our benchmark, the BQ nested \po{} algorithm obtains a strictly smaller CR of $\frac{1}{2 - r_1/r_2}$. {\color{black}We present a concrete example (Example 
\ref{example:two_types}) demonstrating why optimizing the nest sizes is crucial to achieve the higher CR, compared with the BQ nested \po{} algorithm.}

\textbf{Optimal POLYRA algorithms with three types (Section \ref{section:three_type}).} When there are three types of agents with one  or two partially patient types, we present two {\color{black}(feature-augmented CR)} optimal \po{} algorithms with feasible polytopes $\feasible^{(K, M)}$ for $K =3$ and $M\in \{1, 2\}$.  See Theorem \ref{thm:opt_alg_M=1} for  $\feasible^{(3, 1)}$ and  Theorem \ref{theorem:three_type_M=2} for $\feasible^{(3, 2)}$. To design both these feasible polytopes, we first present an upper bound on the feature-augmented CR of any non-anticipating resource allocation algorithm (Theorem \ref{thm:3_type_upper_bound}). This upper bound, which is characterized by an LP, is constructed by considering four overlapping arrival sequences (across two time periods) that are truncated versions of a particular worst-case arrival sequence. Our \po{} algorithms then mimic the optimal solution to this LP in constructing their feasible polytopes, yielding the optimal CR. Interestingly, the feasible polytope $\feasible^{(3, 1)}$  is not  of the simple nested structure (as per Definition \ref{def:nested}), highlighting the challenges of exploiting agents' temporal flexibility optimally beyond two types of agents. We provide intuitions for why this is the case in Section \ref{section:three_type}.


\textbf{Near optimality of nested POLYRA algorithms for more than three types (Section \ref{section:beyond_three}).} When there are at least four types of agents, a generalization of our upper bound LP from three types is no longer tight as it is difficult to characterize the worst-case input sequence. Considering this challenge, we turn our attention back to the class of nested polytopes and show that finding the nested polytope with the highest CR  of $\crr_{\text{nest}}^*$ can be done by solving a simple LP (Theorem \ref{theorem:polyra_nested_performance}). We also show that while characterizing a tight upper bound is challenging, we can easily characterize a looser upper bound $\crr_{\text{up}}$ on the CR of any non-anticipating algorithm for any number of types (Theorem \ref{thm:loose_upper_bound}). Finally, we show that $\crr_{\text{nest}} / \crr_{\text{up}} \geq 0.8$, which proves that the class of nested polytopes is near-optimal (Theorem \ref{thm:optimality_gap}) in the sense that it obtains at least $80\%$ of the optimal CR.

{\color{black} \textbf{Numerical Studies (Section \ref{sec:numerical_studies}).} 
We've undertaken extensive numerical studies to supplement our theoretical findings. Specifically, when contrasting our optimized nested \po{} algorithm with the BQ nested \po{} algorithm, both under traditional and feature-augmented CRs, the results are illuminating. Our algorithm clearly outpaces the BQ nested \po{} algorithm across both metrics, with improvements ranging from 1\% to 19\% in average (traditional) CR and up to 45\% in worst-case (traditional) CR, highlighting its superior capability in harnessing flexibility.

To ensure a holistic understanding, we evaluated our algorithm's performance against the BQ nested \po{} algorithm in terms of the distribution of traditional CRs, and notably, in its worst-case value. The results are consistent: our algorithm stochastically surpasses the BQ nested \po{} algorithm in the distribution of traditional CRs, a trend that also holds for the feature-augmented CR.

\subsection{Managerial Insights}
 Together, our theoretical results and numerical studies lead to the following managerial insights.

\textbf{Trade-off between commitment and flexibility.} The presence of partially patient customers, who require an upfront commitment, presents a unique trade-off between \textit{flexibility in the current period} and \textit{burden in future periods}. An effective algorithm must delicately balance this trade-off: while accepting many partially flexible agents (with low rewards) provides robustness against uncertainty in the current period's arrival sequence, the commitment to serve such agents places a burden on future resources. Despite the inherent challenges in understanding this trade-off, our study offers optimal or near-optimal algorithms that achieve this delicate balance.

\textbf{Recognizing Flexibility in Algorithm Design and its Value.} The enhanced performance of our optimized nested \po{} algorithm, compared to the BQ nested \po{} algorithm, accentuates the significance of tailoring algorithms with flexible agents in consideration. While flexibility is undeniably valuable, it necessitates thoughtful exploitation. Our numerical analyses affirm this: our approach consistently surpasses the BQ nested \po{} algorithm in both feature-augmented and traditional CR metrics. Moreover, our findings underscore the intrinsic value of flexibility. Introducing flexibility has been observed to boost the traditional CR by margins ranging from 0.8\% to 12\%, contingent on the number of flexible agents $M$ and fluctuations in the arrival sequence. Notably, the gain amplifies with an increase in both $M$ and sequence volatility.

\textbf{Merits of increased flexible types.} Our findings challenge the widely accepted idea that ``a little bit of flexibility goes a long way." In contrast, as we introduce more flexible agent types, the growth in both the traditional and feature-augmented CRs remains linear, not showing signs of diminishing returns. This suggests that businesses stand to gain by encouraging a broader base of customers to adopt flexibility.

\textbf{Design of simple nested policies.} Nested booking limit policies are a proven approach for resource allocation in traditional settings without flexible agents. Our research indicates that such a paradigm can be extended to settings with partially flexible agents. Although the introduction of partially flexible agents adds complexity to the optimization of the nested booking limits, we demonstrate that this optimization is tractable, resulting in algorithms that achieve strong performance both theoretically and empirically.

\textbf{Adversarial design enhancing real-world performance.} Designing algorithms with adversarial arrivals as a foundation often proves beneficial, even in more typical stochastic environments. Essentially, these algorithms, primed for worst-case scenarios, tend to outshine expectations in standard conditions, a trend our numerical studies echo. Delving into the intricacies of arrival sequence volatility in our numerical studies, we observed performance pressures on both our algorithm and the BQ nested \po{} algorithm as volatility surges. However, as volatility escalates, the performance disparity between our algorithm and the BQ nested \po{} algorithm grows more pronounced. This emphasizes the pivotal role flexibility plays in managing unpredictable agent inflows. Moreover, our algorithm often surpasses its theoretical benchmarks, especially in the context of the feature-augmented CR.
}




\section{Related Works}\label{sec:related_work}
Our work fits in with existing literature in the following research areas:

\textbf{Single-leg revenue management.} Our work is an extension of the single-leg revenue management (RM) problem to a setting where agents not only have different rewards, but also different temporal flexibility levels. The single-leg RM problem, which is a fundamental RM problem, involves allocating multiple units of a single resource to arriving agents of various types (i.e. demand), each with some fixed reward. The main challenge is deciding in an online fashion how many low-reward agents to accept while making sure there is enough capacity for potential higher-reward agents. A classic result from \cite{littlewood1972} shows that the optimal number of lower type agents to accept can be given by a simple threshold-based policy. Extensive further work has been done on this model, with numerous variations such as adding a network resource structure,  customers' choice models, and sample information and machined-learned advice about the arrival process (see e.g., \cite{ Talluri1998Network, Talluri2004choicemodel, Perakis2010Robust, golrezaei2014real,Gallego2015OnlineCustomerChoice, Ke2019networkRMpricing, Ma2020Network, Sun2020QBNRM, golrezaei2022online, balseiro2022single, golrezaei2023online}, and \cite{Strauss2018ChoiceSurvey} for a survey). 

\textbf{Adversarial demand modeling.} Most previous work in the single-leg RM problem and its variations model demand using a stochastic process (e.g. \cite{asadpour2020limitedflexibility, Talluri1998Network, Talluri2004choicemodel}) or robust uncertainty sets (e.g. \cite{Perakis2010Robust,birbil2006robust}). In both cases, we must make certain assumptions about how demand behaves. However, in an adversarial framework with CR analysis, no assumptions about demand are made as the performance of any algorithm is given by its worst-case performance over any possible demand arrival. One of the first lines of work in adversarial demand modeling was \cite{BallandQueyrenne2009}, whose work is a special case of our model. The authors show that the optimal policy initially accepts all arriving agents, but gradually only accepts higher reward agents once certain capacity thresholds are reached. Their solution structure is similar to that of Littlewood in that both reserve a certain fraction of the capacity exclusively for higher type reward agents, but differ in that the thresholds are set based on knowing the demand distribution, {\color{black}not only based on the ratios between agent rewards}. Other works in this area have explored variations such as setting posted prices instead of making binary accept/reject decisions in \cite{eren2010pricing}, relaxing the adversarial arrival order by assuming that {\color{black} part of the arrival sequence} is stochastic in \cite{Hwang2018partiallypredictable} or by removing the assumption of independent demand classes (the idea that higher type agents never choose the lower-priced option) in \cite{ma2021singleleg}. Our work contributes to the literature on adversarial single-leg RM problems by studying how to optimally exploit  demand time-flexibility in  adversarial single-leg RM problems.  

Our work is also related to the literature on flexibility. Papers in this area can be divided into three streams: supply side, demand side, and time-based flexibility. 

\textbf{Supply-side flexibility. } Supply side flexibility refers to the ability of a firm to satisfy demand for a product with more than a single combination of raw goods or underlying resources. This kind of production-side flexibility has been well studied in the literature. In particular, it has been shown that a limited amount of flexibility (e.g. the long chain design) is almost as good as a fully flexible system (see e.g. \cite{jordan1995principles,simchi2012understanding, simchi2015worst, desir2016sparse, chen2019optimal, devalve2021understanding}). (See \cite{petrick2012survey} for a survey).\footnote{See also \cite{mak2009stochastic,lu2015reliable, shen2019reliable} for works that study supply-side flexibility in the presence of disruptions to  demand and supply and \cite{armony2004customer,castro2020matching} for works that study flexibility in queuing systems.} While most work understanding production-side flexibility has been in an offline setting, the adage ``a little flexibility goes a long way'' also seems to be true in the online setting from \cite{asadpour2020limitedflexibility}. 

\textbf{Demand-side flexibility. }On the other hand, demand-side flexibility refers to customers' willingness to accept one of many products that are usually considered substitutes (e.g. two flights that depart within an hour). The main insights from consumer flexibility is that when customers reveal their indifference to a firm (sometimes for a discount or incentive), it greatly improves capacity utilization by helping the firm overcome uncertainty as seen in \cite{Gallego2004FlexibleProducts}. In addition, the potential discount offered also induces more demand as a larger segment of the population may now be able to afford the good or service. One concrete example of demand-side flexibility is the idea of opaque products in which customers specify a bundle of specific goods that they are indifferent between (such as the color of a notebook) and the firm chooses the exact good for them \cite{Elmachtoub2015Opaque1,Elmachtoub2019Opaque2}. 

\textbf{Patient (timing-flexible) customers.} Patient customers have been studied in many applications healthcare (e.g. scheduling non-urgent elective surgeries, see \cite{gerchak1996emergency, huh2013electiveemergent} or organ transplantation (e.g.,\cite{bertsimas2013fairness}) and ride-sharing (e.g. pooled rides, see \cite{Dogan2019rideshare,Jacob2017rideshare}). Time-flexible (patient) customers have also studied in the context of pricing problems (see, e.g.,  \cite{katta2005pricing,chen2018robust, golrezaei2020dynamic,lobel2020dynamic,abhishek2021strategic}). In dynamic pricing literature with time-flexible customers, it is assumed that customer's valuation for an item decays with their waiting time. These customers, however, time their purchasing decision strategically and the goal is to identify an optimal sequence of prices over time. This line of work investigates the impact of having time-flexible customers in pricing decisions, while in our work we investigate the benefit of time-flexible customers in resource allocation problems. (See also \cite{chawla2017stability} for work that studies pricing problems in the presence of customers with deadlines.)


{\color{black} \textbf{Upfront commitment.} A related body of work with a  notion of upfront commitment is lead time quotation for online job scheduling (see, e.g., \cite{lead_time_1, lead_time_2} and \cite{slotnick2011order} for a survey). In one version of lead-time quotation, a quote for when the job finishes must be provided immediately upon accepting an order while in an easier version of the model,  that quote can be provided any time before the order expires. Not surprisingly, \cite{lead_time_1} shows that the former setting is significantly more challenging than the latter as the best algorithms proposed for the former exhibit larger optimality gaps. Our setting is  closer to the former stringent online model where the acceptance/rejection and the lead time quotation decisions   must be made immediately. Recall that in our setting,  the acceptance/rejection needs to be made immediately and once we accept a partially patient agent, we are committed  to serving them by the next time period. Thus, at a high-level, acceptance and scheduling decisions in the aforementioned works bear some resemblance to our acceptance and serving decisions. However, the differences between our model  do not allow us to use the methodologies  developed in this line of work. In our setting, we can serve agents in parallel contrasting to serving agents sequentially as it is done in job scheduling problems. In addition, unlike job scheduling problems, the notion of processing time for an agent is not relevant in our setting.\footnote{See also \cite{azar2015truthful} for online job scheduling with deadlines.}} 

{\color{black}It is worth noting a recent follow-up study \cite{xie2022benefits} that investigates online resource allocation when decisions are made immediately, with a commitment to fulfilling promises within the next few periods. This research delves into the benefits of delaying real-time decisions under a stochastic framework. Notably, it finds that a slight delay yields the most advantages, and the benefit of delay diminishes as delays grow longer. A primary distinction between this study and ours lies in the modeling of demand: while our work considers adversarial demand, their setting assumes stochastic demand. Under the stochastic model, it is possible, as shown in \cite{xie2022benefits}, to design an algorithm with zero regret, which is not possible in adversarial models. Furthermore, in our work, we handle a mixture of flexible and inflexible agents, whereas in \cite{xie2022benefits}, all requests are flexible. As another difference, while in our work, flexibility allows us to delay serving the flexible agents by one period, in \cite{xie2022benefits}, agents' services can be delayed by $k$ periods. The authors study the impact of $k$ (their level of flexibility) on the obtained regret, showing that a little bit of flexibility goes a long way. In our case, the level of flexibility is determined by the number of flexible agents, and as observed in numerical studies, the performance of our algorithms increases linearly with the level of flexibility. } 


{\color{black}\textbf{Timing-flexibility and matching.} 
Timing-flexibility has been extensively studied in matching problems within stochastic settings, as opposed to our focus on adversarial settings. Notable studies in this domain include  \cite{ashlagi2017min},  \cite{aouad2020dynamic},  \cite{akbarpour2020thickness}, and \cite{hu2021dynamic}, where customers arrive and depart based on stochastic patterns. To design effective algorithms, these studies optimize matching timing while balancing market thickness and customer abandonment risk. One key distinction from our work is that we require immediate acceptance/rejection decisions upon agent arrival, whereas other studies allow decisions at any time before agent departure. This upfront commitment to service is unique to our problem setup.

{\color{black}In a related work by \cite{feng2021batching}, similar trade-offs are explored under adversarial arrivals. They introduce algorithms for multi-stage bipartite allocations with batch arrivals, but unlike our scenario, they allow delays in acceptance decisions until the entire batch arrives, without upfront commitments.}



\section{Problem Setup}
\label{sec:model}
In this work, we study a model where a service-provider (decision-maker) must allocate a scarce service (or resource)  to agents arriving in an online fashion. The agents differ in both their rewards when served as well as their timing flexibility (i.e. patience). We detail the model as follows.

\subsection{Model Definition}
\label{sec:model:model_definitions}
Our model consists of $T$ discrete time periods and $K$ types of agents with rewards $0 < r_1 < \ldots < r_K$ in which the first $M \in [K-1]$ types are \textit{partially patient} (flexible), while the remaining $K - M$ types are \textit{impatient} (inflexible).\footnote{This assumption that the reward for any partially patient agent is less than that of any impatient agent is a reasonable assumption as agents who more urgently need a resource also value it more. We also assume that $M < K$ meaning that there is always at least 1 impatient type as in most applications there is always a type of agent who urgently needs the resource} (We define this precisely later.) Note that for any integer $m$, $[m]= \{1, 2, \ldots, m\}$.  Each time period $t \in [T]$ consists of an \textit{arrival phase} and a \textit{service phase}. When we accept a partially patient agent, we make a \textit{commitment} to provide them service and we must adhere to this commitment.

\textbf{Arrival phase.} For each $t \in [T]$, let $I_t = \{z_{t, 1}, \ldots, z_{t, |I_t|}\}$ be the sequence of agent arrivals in the arrival phase, where $z_{t,n} \in [K]$ is the type of the $n$th arrival in time period $t$. Upon the arrival of the $n$-th agent, we observe its type $z_{t,n}$ and must immediately {\color{black}decide}   whether to accept or reject the agent. Accepted agents must be provided service in an upcoming service phase, while rejected agents leave the system forever.

\textbf{Service Phase. } After all agents in period $t \in [T]$ have arrived, the service-provider can provide service to up to $C$ accepted agents, and collect a reward $r_i$ for each depending on their type $i \in [K]$. When we choose which $C$ agents to serve, we must follow each agents' time-flexibility. All impatient agents (i.e. types $M+1, \ldots, K$) accepted in period $t \in [T]$'s arrival phase must be serviced in period $t$'s service phase, while partially patient agents (i.e. types $1, 2, \ldots, M$) can be serviced either in period $t$'s service phase or carried over to period $t+1$ and serviced during its service phase. 

Our model can be interpreted as an extension of \cite{BallandQueyrenne2009}, incorporating the elements of partially patient agents within a multi-period setting. \cite{BallandQueyrenne2009} delves into a specific case of our model where \( M = 0 \) (indicating all agents are impatient) and \( T = 1 \) (a single time period). Mirroring the approach of \cite{BallandQueyrenne2009}, our goal is to formulate algorithms ensuring reward guarantees in the face of adversarial arrival sequences. However, our model's distinctive features, including the flexibility of certain agents and an upfront commitment, necessitate that all agents receive service timely. {\color{black} We assume that the number of periods $T$ is not known a priori to an algorithm and can be chosen adversarially. In practice, if $T$ is known, then an algorithm may be able to make better choices knowing the ``end is near". However, since we consider adversarial arrivals, knowledge of $T$ does not improve the worst-case instances as an adversary can simply choose an inflated $T$ and stop sending agents after a certain point to simulate a smaller $T$.}

We'd now like to shed light on two  assumptions related to our model:
\begin{enumerate}[leftmargin=*]
    \item In order to allow algorithms to take advantage of flexibility in the last period $T$, we add a dummy period $T+1$ where $I_{T+1} = \emptyset$, but we still have $C$ capacity available to serve partially patient agents from period $T$. This period is necessary for otherwise partially patient agents that arrive in the last period would not be any different from impatient agents and so an adversary could choose $T = 1$, reducing us to a problem where all agents are impatient.
    \item This work will focus on the continuous version of the resource allocation problem, where we are allowed to partially accept and serve agents. Although the continuous problem is less realistic than the discrete one, its analysis is much simpler and we can still capture all the core ideas regarding time-flexibility. For the discrete model, all of the algorithms from this paper can still be applied and are asymptotically correct as $C \to \infty$. In addition, to deal with small values of $C$, one can apply a technique used in \cite{BallandQueyrenne2009} to randomly accept or reject agents, obtaining the same performance guarantees in expectation.
\end{enumerate}
Before we discuss our objective in detail, we provide some application examples to make our model more concrete.

\subsection{Application Examples } 
\label{sec:application}
{ \color{black} 
The principle of upfront commitment is especially relevant in scenarios that involve \emph{opaque service timings}. Here, while customers receive an assurance of service delivery, the precise timing remains undisclosed until later. Two real-world examples that exemplify this concept are Instacart and UPS.


Instacart is a grocery delivery platform that has recently introduced an opaque service option. Typically, the platform provides delivery within the next hour, but recently has broadened their offerings to include a 2-hour delivery window for more patient customers. Those who opt for the 2-hour delivery are analogous to our partially patient customers, while those demanding immediate delivery fit the impatient category. We can interpret each our $T$ time periods as an hourly slot with $C$ couriers available. If we accept an inflexible customer, we must serve them using one of the $C$ couriers in the current period, while for flexible agents we can also utilize couriers in the next period. Should the situation call for it, Instacart can exercise its right to limit access by deactivating specific delivery windows, thereby rejecting certain customers.

Shipping services like UPS also operate on this opaque principle. The notion of multi-day shipping, whether for end consumers via UPS or businesses through platforms like Amazon, inherently possesses this opaque quality. For instance, a UPS store offering next-day and 2-day shipping services dispatches packages every evening via a truck with a capacity of $C$ packages. While next-day shipping orders must be loaded onto this truck, the 2-day orders have some flexibility: they can either join the same truck if there's room or wait for the next day's truck. Similar to Instacart, UPS can regulate service by marking certain shipping options unavailable when required.

Beyond Instacart and UPS, there are many other arenas where services with opaque timing can leverage agent patience. For instance, an airline or travel agency might offer opaque flights aimed at customers who are indifferent to the precise timing of their departure. Those planning vacations might have the option to secure a booking for the upcoming week or select a more flexible two-week window. Similarly, for ticketed events like concerts, a model can be implemented for fans indifferent between two  consecutive show-times. In service scheduling, such as plumbing or doctor's appointments, there's potential to boost efficiency by catering to customers or patients who are flexible regarding their time slots.

In these scenarios, instead of immediately committing a customer to a specific flight, concert time, or appointment, there's significant value in offering a flexible choice. Here, agents receive a firm assurance of service, with the particulars – be it the exact flight, concert timing, or appointment – disclosed a bit later. Collectively, these examples underscore the significance of adeptly managing opaque services in contemporary business frameworks and the nuanced balancing act essential for their optimization.

}

With the model definition and some motivating applications in mind, we now proceed to describe how we measure the performance of an online resource allocation algorithm under this model.

{\color{black}\subsection{Objective and Benchmark}
\label{sec:resource_augmented_cr}
We measure the performance of our algorithms using worst-case analysis under a \textit{resource augmentation} framework (see e.g., \cite{resource_augmentation_1, resource_augmentation_2, resource_augmentation_3} or \cite{roughgarden} for a survey). To avoid confusion with the term resource, which we already use to describe our capacity, we will use the term ``feature-augmented" instead. A feature augmentation guarantee bounds the worst-case ratio between the reward of an online algorithm $\mathcal{A}$ that has access to additional feature versus the optimal offline algorithm $\textsc{opt}$ which cannot access the additional feature. 
The feature-augmented CR allows us to characterize the value of a feature for the algorithm designer. ({\color{black}Refer to Section \ref{sec:compare} for an exploration of the distinctions between the traditional  and feature-augmented CRs.})

{\color{black}In our particular setting, the feature in question is the flexibility of types $1$ through $M$. Our feature augmentation bound compares the performance of an online algorithm $\mathcal{A}$ which can utilize the flexibility of these $M$ partially patient types with the offline optimal $\textsc{opt}$ which treats all $K$ types as impatient. For an algorithm $\mathcal A$, arrival sequence  $\{I_t\}_{t\in [T]}$, and $\tau \in [T]$, let $\text{Rew}_{\mathcal {A}, \tau}\left(\{I_t\}_{t\in [T]}\right)$ be the reward of algorithm $\mathcal A$ in time period $\tau$. Here, $\mathcal{A}$ is able to utilize the flexibility of types $1, 2, \ldots, M$ and potentially delay their service by one time period. Let $\textsc{opt}(I_t)$ be the optimal reward that can be obtained from $I_t$ when we treat all types as impatient. We define the \textit{per-period feature-augmented CR} of an algorithm $\mathcal{A}$ as the following:
\begin{align}
  \text{FA-CR}_{\mathcal A} \defeq \inf_{T, \{I_{t}\}_{t\in [T]}, \tau\in [T]}\left\{\frac{\text{Rew}_{\mathcal {A}, \tau}\left(\{I_t\}_{t\in [T]}\right)}{\textsc{opt}(I_\tau)}\right\}\,.
  \label{definition:racr}
\end{align}}




The per-period feature-augmented CR bounds the ratio of the reward \textit{in each period} $\tau \in [T]$ over all input sequences $\{I_t\}_{t\in [T]}$ of arbitrary length $T$. {\color{black} \color{black} Choosing to consider per-period reward (instead of total reward) is mostly for analytical convenience as the resulting algorithms are much easier to describe and interpret. Additionally, considering this benchmark results in an algorithm that has an equally good performance across all the time periods. We note that if $\mathcal{A}$ achieves a per-period CR  of $\crr$ by equation \eqref{definition:racr}, then it is also the case that:
$ \inf_{T, \{I_{t}\}_{t\in [T]}} \frac{\sum_{\tau \in [T]} \textsc{rew}_{\mathcal{A}, \tau}( \{I_t\}_{t \in [T]}) }{\sum_{\tau \in [T]} \textsc{opt}(I_\tau)} \geq \crr \,, $
implying that the cumulative reward of $\alg$ is also at least $\crr$ times that of the optimal. Finally, ensuring  high per-period rewards can be helpful for small businesses that need steady income to keep going. Nonetheless, in our numerical studies presented in Section \ref{sec:numerical_studies}, we show that our algorithms that are designed based on the per-period feature-augmented CR notion manage to obtain high total reward.}

Observe that  that when there are no partially patient agents, that is $M = 0$, then our definition in equation \eqref{definition:racr} is exactly the traditional CR definition applied to each time period. When $M = 0$, there is no interaction between periods, so we can assume without loss of generality that $T = 1$, reducing our model to the model from \cite{BallandQueyrenne2009}, in which the optimal CR is \begin{equation}
    \label{equation:bq_cr}
    \crr^{BQ} \defeq \left( K - \sum_{i=1}^{K-1} \frac{r_i}{r_{i+1}} \right)^{-1}\,.
\end{equation} 

{\color{black}\subsection{Feature-Augmented CR versus the Traditional CR}
\label{sec:compare}

In this section, we demonstrate that when using the traditional CR definition, even in scenarios where almost every type is partially patient (i.e., $M = K-1$), no algorithm can outperform $\crr^{BQ}$ in CR. Intuitively, since the traditional CR benchmark compares an online algorithm with an offline algorithm that can also utilize flexibility, as our algorithms improve with increasing $M$, so does the benchmark. On the other hand, with the feature-augmented CR, we precisely quantify how leveraging the flexibility of these partially patient agents leads to enhanced online performance.

To be more precise, let
\begin{equation} 
\label{equation:conventional_benchmark}
\text{TR-CR}_{\mathcal{A}} = \inf_{\{I_t\}_{t \in [T]}} \left\{ 
\frac{\sum_{\tau = 1}^T \textsc{rew}_{\mathcal{A}, \tau}(\{I_t\}_{t \in [T]})  }{ \textsc{opt-total} (\{ I_t \}_{t\in[T]}) }
\right\}\,
\end{equation}
represent the traditional CR of an algorithm $\mathcal{A}$. Here, the numerator calculates the total reward over $T$ periods of algorithm $\mathcal{A}$, while the denominator is the total reward of the optimal clairvoyant solution that knows $\{I_t\}_{t \in [T]}$ in advance and can utilize the flexibility of types $1$ through $M$.

Theorem \ref{flex_is_bad} establishes that in situations where nearly every agent type is patient ($M = K - 1$), no non-anticipating algorithm can achieve a CR greater than $\crr^{BQ}$.

\begin{theorem}
\label{flex_is_bad}
For any $K \geq 2$ and $M = K - 1$, no non-anticipating algorithm $\mathcal{A}$ can achieve a traditional CR (as given by  equation \eqref{equation:conventional_benchmark}) greater than $\crr^{BQ}$, i.e., $\text{TR-CR}_{\mathcal{A}} \leq \crr^{BQ}$, where $\crr^{BQ} = \frac{1}{K - \sum_{i=1}^{K-1} \frac{r_{i}}{r_{i+1}}}$.
\end{theorem}

This theorem implies that under the traditional CR, we cannot observe the value of flexibility in scenarios where nearly every agent type is patient. By endowing the offline benchmark with the ability to leverage patient agents, it becomes such a strong benchmark that it negates any additional competitive advantage an algorithm might seek from agent flexibility.

Moreover, we would like to highlight that while our inflexible offline benchmark, used in the definition of the feature-augmented CR, has a closed-form solution making its analysis more straightforward, the flexible offline benchmark does not admit a closed-form solution. Characterizing the flexible offline benchmark $\textsc{opt-total} (\{ I_t \}_{t\in[T]})$ requires the formulation of dynamic programming or a mathematical  program to elucidate the behavior of this optimal flexible clairvoyant solution.

Although the majority of our paper focuses on theoretical results for the feature-augmented CR, we provide an extensive numerical study in Section \ref{sec:numerical_studies} where we demonstrate that the algorithms we design to optimize for the feature-augmented CR also performs well under the traditional CR definition. In the rest of this paper, for the sake of conciseness, we will simply use the term CR to refer to the feature-augmented CR from equation \eqref{definition:racr}, even though it is a slight misnomer. In Section \ref{sec:numerical_studies} when we discuss both CR definitions, we will be very clear which one we're referring to.}



\section{Polytope-based Resource Allocation (POLYRA) Algorithms}
\label{sec:polyra}
We present an intuitive class of algorithms called polytope-based resource allocation (\po) algorithms. Such algorithms keep track of a state $\stateB \in \mathbb{R}_+^{K+M}$ which tracks the number of agents of each type that our algorithm has accepted so far, as well as their deadlines. Each \po{} algorithm instance is parameterized by a feasible polytope $\feasible \subset \mathbb{R}_+^{K+M}$ which intuitively describe the set of all states that our algorithm is allowed to be in. As we will explain in more detail below, a \po{} algorithm makes greedy decisions while maintaining the invariant that $\stateB \in \feasible$.

\begin{figure}
  \centering
  \begin{equation*}
    \begin{array}{r|ccccccc}
      \text{Type} \qquad  & 1 \quad               & 2 \quad               & \ldots \quad & M \quad               & M+1 \quad          & \ldots \quad & K          \\ \hline
      \text{Row 1} \qquad & \bucketij{1}{1} \quad & \bucketij{2}{1} \quad & \ldots \quad & \bucketij{M}{1} \quad & \bucket{M+1} \quad & \ldots \quad & \bucket{K} \\
      \text{Row 2} \qquad & \bucketij{1}{2} \quad & \bucketij{2}{2} \quad & \ldots \quad & \bucketij{M}{2} \quad &                    &              &
    \end{array}
  \end{equation*}
  \caption{The $K + M$ buckets in our state $\stateB$. Buckets in the first row contain agents that have been allocated a resource from the current time period, while buckets in the second row contain only partially patient agents  who have accepted that we tentatively {plan on carrying over} to period $t+1$ and servicing in that period's service phase.}
  \label{fig:buckets}
\end{figure}

\textbf{State.} At any point in time, our algorithm keeps track of a state $\stateB$ which consists of the following $K+M$ dimensional vector of non-negative numbers:
\begin{equation} 
\label{bucketdefinition}
\stateB=(\bucketOneOne, \bucketTwoOne, \ldots, \bucketMOne, \bucketMPlusOne, \ldots, \bucketK; \bucketOneTwo, \bucketTwoTwo, \ldots, \bucketMTwo)\,.
\end{equation}
We will refer to each index of the state as a \textit{bucket}. All accepted agents are assigned to a bucket and the bucket keeps track of the number of such agents assigned to it. These buckets are illustrated more visually in Figure \ref{fig:buckets}. We use $\bucketij{k}{j}$, $k\in [M]$ and $j\in [2]$, or $\bucket{k}$, $k\in \{M+1, \ldots, K\}$ to refer to both the bucket itself or as well as the number of agents assigned to it.
\begin{itemize}[leftmargin=*]
  \item \textbf{First row.}
        In any given time period $t$, the first row of buckets from Figure \ref{fig:buckets} keeps track of the number of agents of each type that have been accepted and \textit{we plan on serving} in period $t$'s service phase. More precisely, for $i \in [M]$, bucket $B_{i,1}$ and for $i \in \{M+1,\ldots, K\}$ bucket $\bucket{i}$ store agents of type $i$ that we definitely going to serve using the $C$ capacity available at the end of the current period $t$.

  \item \textbf{Second row. } 
        Each bucket in the second row $\bucketij{k}{2}$,
        $k\in [M]$ captures the number of type $k$ agents we have accepted that we tentatively \textit{plan on carrying over} to period $t+1$ and servicing in that period's service phase. 
\end{itemize}

\noindent In other words, at any moment in time, the state $\stateB$ describes how we are planning on servicing the agents we have accepted and are still waiting for service. We now describe how we update the states during the arrival phase and service phase.

{\color{black}\textbf{Arrival phase: acceptance rule.} Upon the arrival of an agent of type $i$, we will accept them if there is a way to update the state while maintaining feasibility. More precisely, we split into two cases based on whether type $i$ is impatient or partially patient.

\begin{itemize}[leftmargin= *]
  \item \textbf{Impatient  types. }If $i \in \{M+1, \ldots, K\}$, accepting such an agent would imply that we must service them using period $t$'s capacity, which means that $\bucket{i}$ would need to be incremented in $\stateB$. If this new state is feasible, then we accept this type $i$ agent, otherwise we reject them. \footnote{More precisely, under our continuous model, we find the largest value of $\epsilon \leq 1$ such that: $\stateB + \epsilon \cdot e_{i} \in \feasible$, where $e_i$ is a   vector with $K+M$ length  and all $0$'s except a $1$ at the $i$-th index.} 

  \item \textbf{Partially patient types. }If type $i \in [M]$, then we consider two possibilities, which corresponding to when we plan on servicing this agent. We first check to see if bucket $\bucketij{i}{2}$ can be incremented while maintaining feasibility. If so, we accept this agent and assign them to bucket $\bucketij{i}{2}$. Otherwise, we try the same with $\bucketij{i}{1}$ and if not, we reject the agent. \footnote{More precisely, we first find the largest $\epsilon_2 \leq 1$ for which $\stateB' = \stateB + \epsilon_2 \cdot e_{K+i}$ is feasible, where $e_{K+i}$ is a vector of all $0$'s and a 1 corresponding to bucket $\bucketij{i}{2}$ (recall equation \ref{bucketdefinition}). If $\epsilon_2 = 1$, then we are done. Otherwise, we find the largest $\epsilon_1 \leq 1 - \epsilon_2$ such that the state $\stateB' = \stateB + \epsilon_2 \cdot e_{K+i} + \epsilon_1 \cdot e_{i}$ is feasible, which corresponds to assigning a fractional $\epsilon_1$ of an agent to bucket $\bucketij{i}{1}$.}
\end{itemize}

\textbf{Service phase: updating bucket state $\stateB$}. During the service phase, we will choose to serve all the agents assigned to the first row of Figure \ref{fig:buckets}, as per the definition of the first row of buckets. Using the remaining  $\ell \defeq C - \left( \bucketij{1}{1} + \ldots + \bucketij{M}{1} + \bucket{M+1} +\ldots+ \bucket{K} \right)$, we will greedily serve the agents assigned to buckets $B_{1,2}, \ldots, B_{M, 2}$ (i.e. the second row) by prioritizing agents with higher rewards. This greedy action  can be described  using  the optimal solution to the following LP: 
\begin{align}
    (s_i^*)_{i=1}^M = \text{argmax}_{(s_i)_{i=1}^M} \quad & \sum_{i=1}^M r_i \cdot s_i \label{serving_lp}
   \qquad  \text{s.t.} \quad  \sum_{k=1}^M s_i \leq \ell \quad \text{ and } \quad s_i \in [0, \bucketij{i}{2}], \quad  i\in [M] \,. 
\end{align}
Here, $s_i^*$ is the number of agents of type $i$ that upon acceptance,  were assigned to bucket $B_{i,2}$ in time period $t$, and are served in the same time period.  

Our total reward from servicing would be $\sum_{i=1}^M r_i \cdot (s_i^* + \bucketij{i}{1}) + \sum_{i=M+1}^K r_i \cdot \bucket{i}$. We would update our state as follows:
\begin{equation}
  \widehat{\stateB} = (\bucketij{1}{2} - s_1^*, \ldots, \bucketij{M}{2} - s_M^*, \underbrace{0, \ldots, 0}_{K-M};  \underbrace{0, \ldots, 0}_{M})\,. \label{serving_rule_update_state}
\end{equation}
Here, the first $M$ coordinates represent the remaining partially patient agents that we did not service in period $t$, so we must service in period $t+1$. All other coordinates are empty because they contain agents that were served (the state $\stateB$ only keeps track of agents accepted but not yet served). }

In order for the algorithm above to be well defined, we present the following definition for any feasible polytope $\feasible$ we consider.

\begin{definition} [Valid Polytopes] \label{def:valid}
  A polytope $\feasible$ is \textit{valid} if:
  \begin{enumerate}
    \item [(i)] \label{def:valid:never_overaccept} For all $\stateB \in \feasible$, we must have the sum of the first row of $\stateB$ be at most $C$, i.e. $\sum_{i=1}^M \bucketij{i}{1} + \sum_{i=M+1}^K \bucket{i} \leq C$. Since the first row contains agents who must be served in the current time period, we certainly cannot promise this to more than $C$ agents. 
    \item [(ii)] \label{def:valid:never_overaccept_row_two} For any time period $t\in [T]$, and (feasible) buckets state $\stateB$, the updated buckets state after applying our serving rule (i.e., $\widehat \stateB$ presented in equation \eqref{serving_rule_update_state} and finding the optimal solution $s_i^*$ to equation \eqref{serving_lp}) remains feasible (i.e., $\widehat \stateB \in \feasible$). Recall that \po{} algorithms maintain the invariant that $\stateB \in \feasible$ is always true, so certainly our updating rule in equation \eqref{serving_rule_update_state} must maintain this invariant.
  \end{enumerate}
\end{definition}

Any polytope we present in this work will be valid as per Definition \ref{def:valid}. 

 


\section{Nested POLYRA Algorithm} 
To show the versatility of our \po{} algorithms, we now present a special class of \po{} algorithms that we refer to  as \emph{nested \po{}} algorithms. These \po{} algorithms use a special class of  
 polytopes called \textit{nested polytopes} that we will define shortly.  As it will become more clear later, the nested \po{} algorithms can be viewed as a  generalization of the well-known nested booking limits policy from \cite{BallandQueyrenne2009}.

\begin{definition}[Nested Polytopes] \label{def:nested}
  When there are $K$ types with the first $M$ partially patient, for any $(n_1, \ldots, n_K)$ with $0 \leq n_1 \leq \ldots \leq n_K = C$, we define the nested polytope with nest sizes $(n_1, \ldots, n_K)$ to be $\feasiblenest(\nestOne, \nestTwo, \ldots,  \nestK)$ given by the vector of $$(\fbucketOneOne, \fbucketTwoOne, \ldots, \fbucketMOne, \fbucketMPlusOne, \ldots, \fbucketK; \fbucketOneTwo, \fbucketTwoTwo, \ldots, \fbucketMTwo)\subset \mathbb{R}_+^{K+M}$$ that satisfies the following two sets of constraints:
  \begin{align}
    \sum_{i=1}^k b_{i, j}                        & \leq n_k & k \in [M], j \in [2] \notag \\
    \sum_{i=1}^M b_{i, 1} + \sum_{i=M+1}^k b_{i} & \leq n_k & k \in \{M+1, \ldots, K\}
    \label{eq:nested:polytope}\tag{$\feasiblenest(\nestOne, \nestTwo, \ldots,  \nestK)$}\,.
  \end{align}
\end{definition}


A nested polytope $\feasiblenest(\nestOne, \nestTwo, \ldots,  \nestK)$ imposes \emph{booking limits} on the $C$ resources we have in each time period, restricting type $i$ agents to only $n_i$ out of the $C$ resources. The limits $n_i$ are nested in a way that higher reward agents have access to more resources (note that $r_1 < \ldots < r_K$ and $n_1 \leq \ldots \leq n_K)$. The two sets of constraints above enforce these limits: the first set of constraints enforces the booking limit for partially patient agents in \textit{both rows} of our state $\stateB$ (since the same booking limits apply to both agents we plan on serving in period $t$ as well as $t+1$). The second set of constraints enforces booking limits across the entire first row for each impatient type. It is also easy to see that nested polytopes are \consistent{} as per Definition \ref{def:valid}. Property (i) is explicitly enforced as $n_K = C$ in Definition \ref{def:nested} and property (ii) is satisfied because the first set of constraints above are duplicated for $j =2$.

Observe that  when $M = 0$, our nested polytope only consider the current period's resources (since no agents are flexible) and we recover exactly their nested booking limit policy from \cite{BallandQueyrenne2009}. 
We now provide a concrete example of a nested polytope and its interpretation as booking limits. 

\begin{figure}
    \centering
    \begin{subfigure}[c]{0.45\textwidth}
        \centering
        \includegraphics[width=\textwidth]{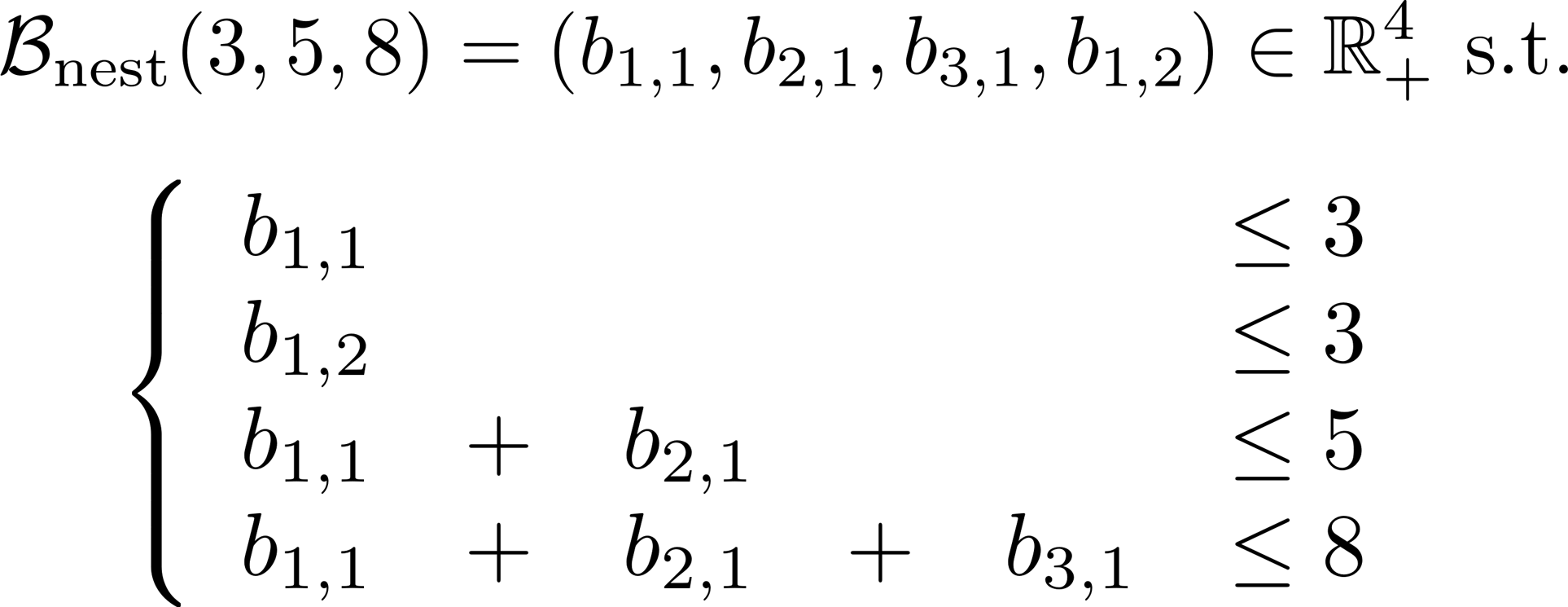}
    \end{subfigure}
    \begin{subfigure}[c]{0.45\textwidth}
        \centering
        \includegraphics[width=\textwidth]{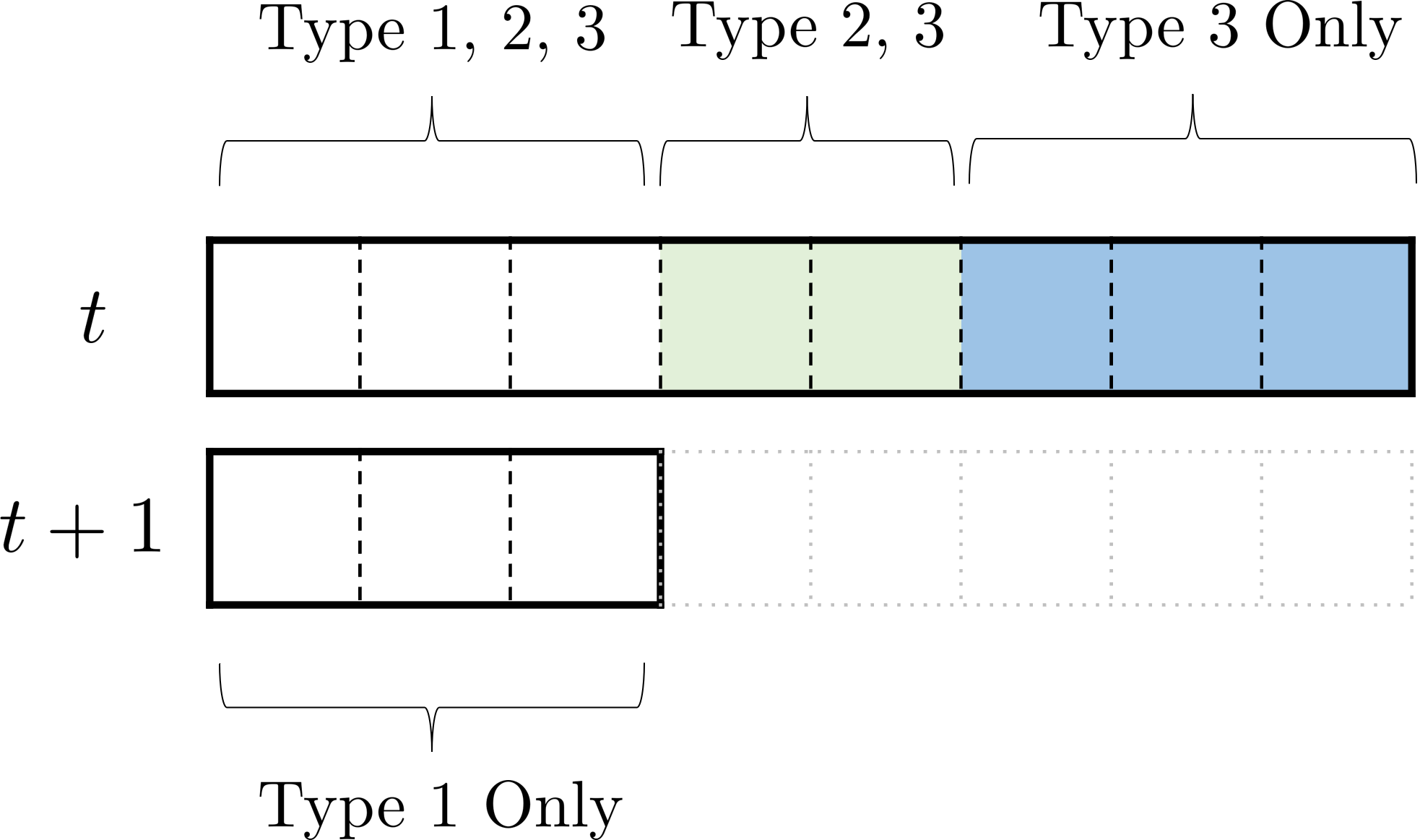}
    \end{subfigure}
    \vspace{0.5em}
    \caption{Example of the nested polytope $\nestedPolyra(3, 5, 8)$ with $K = 3, M = 1, C = 8$. }
    \label{fig:nested_polytope_example}
\end{figure}

{
\color{black}
\begin{example}
    \label{example:nested_example}

    Let's consider an example with $K = 3$ agent types having rewards $r_1 < r_2 < r_3$. Here, type $1$ is flexible, meaning $M = 1$. Given a capacity $C = 8$, we can explore the nested polytope $\nestedPolyra(3, 5, 8)$. The feasible region of this polytope is illustrated on the left of Figure \ref{fig:nested_polytope_example}. The accompanying diagram on the right provides a visual representation of this polytope. In this diagram, the top row indicates the $C = 8$ units of capacity available in the current period, while the bottom row displays the $C = 8$ resources set for the next period.

The core notion of a \textit{booking limit} revolves around directing how these $C = 8$ resources are designated among type $1, 2$, and $3$ agents, ensuring specific capacities are reserved for higher reward agents. In the current period (represented by the top row in Figure \ref{fig:nested_polytope_example}), the first 3 units can cater to any agent type. The subsequent 2 units (highlighted in light green) are reserved for types 2 and 3 agents, while the remaining 3 units (in blue) are exclusively for type 3 agents. The constraints $b_{1,1} \leq 3$, $b_{1,1} + b_{2,1} \leq 5$, and $b_{1,1} + b_{2,1} + b_{3,1} \leq 8$ uphold these booking limits. As an illustration, the constraint $b_{1,1} + b_{2,1} \leq 5$ means our state $(b_{1,1}, b_{2,1}, b_{3,1}, b_{1,2})$ is feasible in $\nestedPolyra(3, 5, 8)$ only if a minimum of $8-3=5$ units of the current period's capacity is retained for type $3$ agents. For the next period's resource, we see a comparable booking limit, planning to serve up to 3 type 1 agents. Hence, the constraint $b_{1,2} \leq 3$ is in place. 

It's important to note that these booking limits primarily apply during the arrival phase when making decisions about agent acceptance. In the service phase, after attending to agents allocated to the top row, any remaining resources are used for agents in the bottom row, irrespective of previous booking limits.

\end{example}
}

\subsection{An LP Characterization of the CR of any Nested \textsc{POLYRA} Algorithms} \label{sec:CR_nested}
Here, we present a general way of computing the CR of any given instance of the nested \po{} algorithm.  
That is, for any given values of $n_1, \ldots, n_K$ such that $0 \le n_1 \le n_2\le \ldots \le n_K$, we show how to compute the CR achieved by a \po{} algorithm with feasible polytope $\nestedPolyra(n_1, \ldots, n_K)$.

\begin{theorem}[CR of Nested \textsc{POLYRA} Algorithms]
  \label{theorem:polyra_nested_performance}
  For any given values of $n_1, \ldots, n_K$ such that $0 \le n_1 \le n_2\le \ldots \le n_K = C$,  the CR of the nested \po{} algorithm with  polytope $\nestedPolyra(n_1, \ldots, n_K)$ is given by $\crr_{\text{nest}}(\{\nest{i}\}_{i\in [K]})$,  which is  defined by the following LP
  \begin{align}
    \crr_{\text{nest}}(\{\nest{i}\}_{i\in [K]}) \defeq \max_{\crr, \{s_i^j\}_{j \in [K], i \in [j]}} \quad & \crr \nonumber \\
    \text{s.t} \quad                              & \crr \cdot C \cdot r_j \leq \sum_{i=1}^j \nestedvar{i}{j} \cdot r_i \label{nested_lp_revenue_target} & j \in [K]                                     \\
    \lplabel[lp:nested_polytope]{\textsc{(nest)}} & \sum_{i=1}^j \nestedvar{i}{j} \leq C                                                                 & j \in [K] \label{nested_lp_capacity}          \\
      & \deltan_i \leq \nestedvar{i}{j} \leq 2 \cdot \deltan_i                                           & j \leq M, i \in [j] \label{nested_lp_j_le_M}  \\ \label{nested_lp_j_ge_M}
      & \nestedvar{i}{j} = \deltan_i                                                                      & j > M, i \in [j] \\
      & \Gamma\ge 0, s_i^j\ge 0 & j\in [K], i\in [j]\,, \label{eq:positive}
  \end{align}
  Here, $\deltan_i = \nest{i}-\nest{i-1}$ and $\nest{0}$ is defined to be zero. When $\{n_i\}_{i \in [K]}$ is understood from the context, we omit it and just write $\crr_{\text{nest}}$.
\end{theorem}

\begin{figure}
    \centering
    \includegraphics[width=0.9\textwidth]{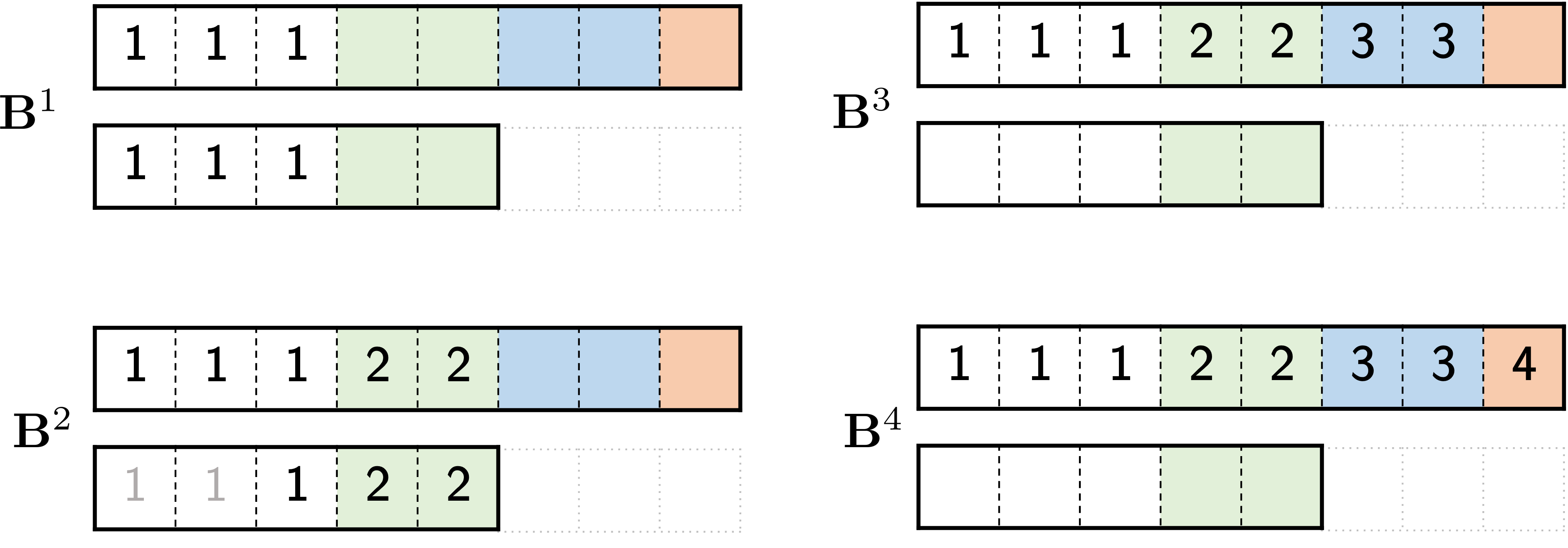}
    \caption{Worst case states $\stateB^1, \stateB^2, \stateB^3, \stateB^4$ for the nested polytope $\feasiblenest(3, 5, 7, 8)$ with $K = 4$ and $M = 2$. The shading represents the capacity that is reserved for each type. The numbers in each block correspond to the agent type designated to that block. Black numbers represent agents who were accepted and subsequently served by the end of the current period, while gray numbers signify those that were accepted but could not be served in the current period.  }
    \label{fig:nested_polytope_worst_cases}
\end{figure}

\textbf{Intuition for Theorem \ref{theorem:polyra_nested_performance} and its proof sketch}. 
We can characterize any \po{} algorithm's CR with the nested polytope $\nestedPolyra(n_1, \ldots, n_K)$ by considering $K$ worst-case scenarios. If our \po{} algorithm performs well in each of these $K$ worst-case scenarios (achieving a CR of $\crr$), each associated with a state $\stateB^j$, $j\in [K]$, then it will perform well for any arbitrary arrival sequence (achieving a CR of $\crr$).

These states $\stateB^j$ are defined as follows:
\begin{itemize}
\item If $j \in [M]$, then $\stateB^j_{i,1} = \stateB^j_{i,2} = \deltan_i$ for $i \in [j]$, and all other buckets are empty.
\item If $j \in \{M+1, \ldots, K\}$, then $\stateB^j_{i,1} = \deltan_i$ for all $i \in [M]$, $\stateB^j_{i} = \deltan_i$ for all $i \in {M+1, \ldots, j}$, and all other buckets are empty.
\end{itemize}
Each scenario $\stateB^j$ represents a state where no more type $j$ agents can be accommodated, and among such states, $\stateB^j$ minimizes the total reward of agents by maximizing the number of type $1, 2, \ldots, j-1$ agents. Let $s^j_i$ be the total number of type $i$ agents serviced from $\stateB^j$ (assuming this is the state at the end of the arrival phase). The reward obtained during this period is $\sum_{i=1}^j r_i \cdot s^j_i$. To ensure that our choice of nest sizes $n_1, \ldots, n_K$ yields a $\crr$-competitive algorithm, we establish the condition $\sum_{i=1}^j r_i \cdot s^j_i \geq \crr \cdot C\cdot r_j$ (constraint \eqref{nested_lp_revenue_target}). Additionally, we must have $\sum_{i=1}^j s_{i}^j \leq C$ to allocate no more than $C$ resources in scenario $j$ (constraint \eqref{nested_lp_capacity}). Constraints \eqref{nested_lp_j_le_M} and \eqref{nested_lp_j_ge_M} limit the range of $s_{i}^j$, ensuring we serve agents in the first row and no more agents than the total between the first and second rows for scenarios $j \leq M$.

{
\color{black}
\begin{example}
We illustrate the proof of Theorem \ref{theorem:polyra_nested_performance} with a concrete example shown in Figure \ref{fig:nested_polytope_worst_cases}. Similar to Example \ref{example:nested_example}, the shading represents the resources that are ``reserved" for higher reward agents (e.g., the light green region can only contain types $2, 3$, and $4$). The numbers indicate the type of agent we plan on serving with that specific resource. This example matches the definition of $\stateB^j$ for $j \in [4]$ defined above. For each $\stateB^j$, we discuss the agents serviced if $\stateB^j$ is the final state at the end of the arrival phase.

\begin{itemize}
    \item $\stateB^1$: We will service all 6 type 1 agents. No other agents have been accepted. Thus: $s_{1}^1 = 6, s_{2}^{1} = s_3^1 = s_4^1 = 0$. The total reward is $6 \cdot r_1$. 
    \item $\stateB^2$: We service all agents on the top row and use the remaining $8-5=3$ capacity to greedily service the agents in the second row. This results in $s_1^2 = 4, s_2^2 = 4, s_3^2 = s_4^2 = 0$. The total reward is $4 \cdot r_1 + 4 \cdot r_2$. 
    \item $\stateB^3$: We service all agents, thus $s_1^3 = 3, s_2^3 = 2, s_3^3 = 2$ for a total reward of $3 \cdot r_1 + 2 \cdot r_2 + 2 \cdot r_3$. 
    \item $\stateB^4$: Similarly, we service all agents and thus $s_1^4 = 3, s_2^4 = 2, s_3^4 = 2, s_4^4 = 1$ for a total reward of $3 \cdot r_1 + 2 \cdot r_2 + 2 \cdot r_3 + r_4$. 
\end{itemize}

Plugging the above values of $s_{i}^j$ into the LP in Theorem \ref{theorem:polyra_nested_performance}, we get that the optimal $\crr$ and thus the CR of $\nestedPolyra(3, 5, 7, 8)$ is: 
$$ \crr_{\text{nest}}(3, 5, 7, 8) = \min \left( \frac{6 \cdot r_1}{8 \cdot r_1}, \frac{4 \cdot r_1 + 4 \cdot r_2}{8 \cdot r_2}, \frac{3 \cdot r_1 + 2 \cdot r_2 + 2 \cdot r_3}{8 \cdot r_3}, \frac{3 \cdot r_1 + 2 \cdot r_2 + 2 \cdot r_3 + 1 \cdot r_4}{8 \cdot r_4}  \right)\,.  $$
For any rewards $r_1 < r_2 < r_3 < r_4$, Theorem \ref{theorem:polyra_nested_performance} claims that the above minimum of 4 expressions is the CR of this nested \po{} algorithm. 
\end{example}
}

{\color{black}
\subsection{Benchmark Algorithm: BQ Nested POLYRA Algorithm}
Having defined nested \po{} algorithms, we now present a natural benchmark algorithm for our model that we refer to as \emph{BQ Nested \po{} Algorithm}. This benchmark is inspired by the algorithm designed in  \cite{BallandQueyrenne2009} for the setting without partially patient agents. 
Recall that when $M = 0$, \cite{BallandQueyrenne2009} showed that the optimal algorithm uses a nested polytope and gives an explicit formula for these nest sizes. In general, for the case when $M > 0$, one can design a nested polytope using these \textit{same} nest sizes that are optimized for the $M = 0$ case. We define this naive nested polytope formally below.

\begin{definition}[BQ Nested Polytope] 
\label{def:bq_nest}
For any number of agents $K$ with reward $r_1 < \ldots < r_K$ where the first $M \in [K-1]$ are partially patient, we define the BQ nested polytope as follows: 
$$
\feasiblenest^{BQ} \defeq \feasiblenest(n^{BQ}_1, \ldots, n^{BQ}_K)\,,
$$ 
where $(n^{BQ}_1, \ldots, n^{BQ}_K)$ are the  nest sizes from \cite{BallandQueyrenne2009}:
\begin{align} \label{eq;bQ:nest size}
 {\color{black}\qquad n_i^{BQ} \defeq \left(i - \sum_{j=1}^{i-1} \frac{r_j}{r_{j+1}} \right) \cdot \crr^{BQ} \cdot C \qquad i\in[K]\,.}
\end{align}
Here, as shown by \cite{BallandQueyrenne2009}, $\crr^{BQ}$ is 
the CR of the corresponding booking limit policy when there is no partially patient agent (i.e., $M=0$), where we recall 
$
    \crr^{BQ} = \left(K - \sum_{i=1}^{K-1} \frac{r_i}{r_{i+1}}\right)^{-1}
$ is defined in equation \eqref{equation:bq_cr}.

\end{definition}

A natural question to ask would be how well does the BQ nested \po{}  algorithm perform when $M \geq 1$? From \cite{BallandQueyrenne2009}, we only know that this algorithm achieves a CR of $\crr^{BQ}$ when $M=0$. Note that when $M = 0$, the feature-augmented CR is the same as the traditional CR. The following theorem states that this nested \po{} algorithm achieves a (feature-augmented) CR of $\crr^{BQ}$   when $M \geq 1$. 

\begin{theorem}[CR of the BQ Nested POLYRA Algorithm]
\label{theorem:bq_polytope_same_cr}
  For any number of agents $K$ with the first $M \geq 1$ partially patient, a \po{} algorithm with the BQ nested polytope (defined in Definition \ref{def:bq_nest}) achieves a (feature-augmented) CR of $\crr^{BQ}$, where $\crr^{BQ}$ is defined in equation \eqref{equation:bq_cr}.
\end{theorem}

 The proof of Theorem \ref{theorem:bq_polytope_same_cr} can be found in Appendix \ref{appendix:bq_polytope_same_cr_proof}. Theorem \ref{theorem:bq_polytope_same_cr} is a special case of Theorem \ref{theorem:polyra_nested_performance}, which characterizes the CR for \textit{any} nested algorithm given its nest sizes. Theorem \ref{theorem:bq_polytope_same_cr} implies that the BQ nested \po{} algorithm achieves the same CR of $\crr^{BQ}$  for any value of $M$. As we will see later in Section \ref{section:beyond_three}, where we focus on nested polytopes, in general for $M \geq 1$, we can achieve a CR of $\crr^* > \crr^{BQ}$ by simply optimizing the nests according to an LP that we wil present later. 

We now turn our attention to characterizing \textit{optimal} polytopes for the special case of $K= 2, 3$. As we will see, some of them are nested while others are not.
}

\section{Optimal POLYRA Algorithm with Two or Three Types}

In this section, we show that a \po{} algorithm is able to achieve the optimal CR for two and three types. For two types, the polytope follows the nested structure described in Section \ref{sec:polyra}, but unfortunately, this is not necessarily the case for three types.

\subsection{Optimal POLYRA Algorithm with Two Types}
\label{section:two_type}
Consider a special case of our model for $K = 2$ types with rewards $r_1 < r_2$, where type $1$ is partially patient and type $2$ is impatient. It turns out that nested \po{} algorithms  are optimal here. We show in the following theorem that choosing the right nest sizes achieves the optimal CR. We then provide some intuition for why choosing the BQ nested polytope (per Definition \ref{def:bq_nest}) provides a sub-optimal algorithm.

\begin{theorem}[Optimal Algorithm with Two Types] 
\label{thm:opt_2}
    When the number of types $K = 2$ and the number of partially patient types $M = 1$ with rewards $r_1 < r_2$ and per-period capacity $C$. Then, the nested \po{} algorithm with the nested   polytope $\feasible^{(2,1)}$ achieves a CR of $\frac{2}{3 - r_1 / r_2}$, where
    \begin{equation} 
    \label{eq:feasible_two_type}
    \feasible^{(2, 1)} = \nestedPolyra \left( \frac{C}{3 - r_1 / r_2}, C \right).
    \end{equation}
    Furthermore, no {\color{black} deterministic} online algorithm can achieve a higher CR than $\frac{2}{3 - r_1 / r_2}$ hence proving that our \po{} algorithm with polytope $\feasible^{(2,1)}$ is optimal.
\end{theorem}

To get an intuition for this polytope $\feasible^{(2,1)}$, it is useful to compare it to $\nestedPolyra^{BQ}$, which is defined as follows.  
\begin{align} \label{eq:two_type_bq}
\nestedPolyra^{BQ} = \nestedPolyra \left( \frac{C}{2 - r_1 / r_2}, C \right).
\end{align}
Based on Theorem \ref{theorem:bq_polytope_same_cr}, the above polytope achieves a CR of $\frac{1}{2 - r_1/r_2}$, which is strictly smaller than $\frac{2}{3 - r_1 / r_2}$ from Theorem \ref{thm:opt_2}. Notice that the only difference between $\feasible^{(2,1)}$ and $\nestedPolyra^{BQ}$ is the choice of the first nest size $n_1$: $\frac{C}{3 - r_1/r_2}$ versus $\frac{C}{2 - r_1/r_2}$.  We now present an example to show how two polytopes' performance differ, as well as, a proof sketch for why our polytope $\feasible^{(2,1)}$ is optimal.

\begin{figure}
    \centering
    \includegraphics[scale=0.5]{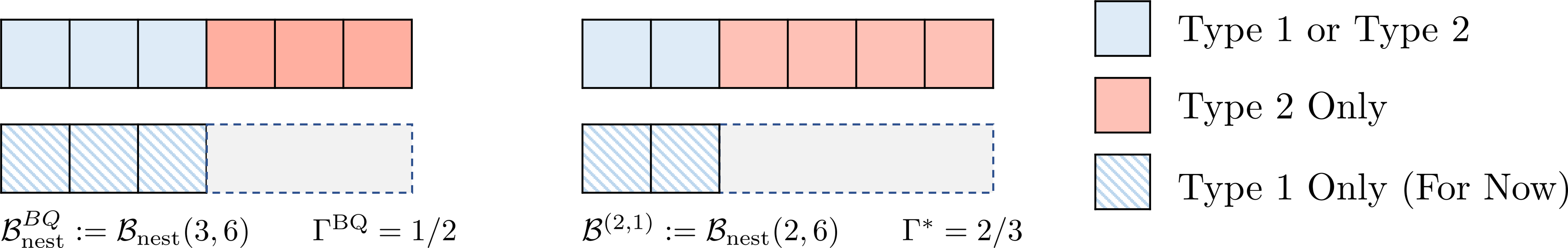} 
    \vspace{0.5em}
    \caption{Comparing $\nestedPolyra^{BQ}$ (left) and $\feasible^{(2,1)}$ (right) in the special case of $(K, M) = (2, 1)$, $r_1 = 1, r_2 = 1000$ and $C = 6$. In this case, $\nestedPolyra^{BQ} = \nestedPolyra(3, 6)$ and $ \feasible^{(2,1)} = \nestedPolyra(2, 6)$.
    The first row represents the 6 capacity in the current time period, while the second row is for the next period. The 6 capacity in the top row is divided into light blue (can be assigned to either type 1 or type 2) and red (can be assigned to type 2 only). Type 1 agents are also allowed to occupy the striped blue region in the second row (but type 2 agents are not at the moment since type 2 agents are impatient). 
    }
    \label{fig:nested_comparison}
\end{figure}

\begin{example} 
\label{example:two_types}
Consider a concrete example with $r_1 = 1, r_2 = 1000$, $T = 1$, and $C = 6$. Here, $r_2$ is significantly larger than $r_1$, approaching infinity in the limit. In this scenario, our nested polytope $\feasible^{(2,1)}$ (from equation \eqref{eq:feasible_two_type}) and the BQ nested polytope (from equation \eqref{eq:two_type_bq}) become, asymptotically as $r_2 / r_1 \to \infty$, $\nestedPolyra(2, 6)$ and $\nestedPolyra(3, 6)$ respectively (see Figure \ref{fig:nested_comparison}). Our optimal polytope reserves $C-n_1 = 6-2 = 4$ units of capacity for higher reward agents in each period, while the BQ polytope only reserves $C- n_1 = 6 - 3 = 3$ units. Now, let's examine how the \po{} algorithms with feasible polytope $\feasible^{(2,1)}$ and $\nestedPolyra^{BQ}$ react to a simple arrival sequence $I_1$, which includes a dummy second period with no agent arrivals: $I_1 = \{1, 1, 1, 1, 1, 1, 2, 2, 2, 2, 2, 2\}, I_2 = \{\}$.

Our \po{} algorithm with polytope $\feasible^{(2,1)} = \nestedPolyra(2, 6)$ accepts $2\cdot n_1 = 4$ out of $6$ type $1$ agents from $I_1$ and then accepts $C- n_1 = 4$ out of 6 type $2$ agents. It then provides service to 2 type $1$ agents and 4 type $2$ agents for a total reward of approximately $4000$, achieving a $2/3$ approximation to the optimal reward of $6000$ in $I_1$.

The BQ nested \po{} algorithm with polytope $\nestedPolyra^{BQ} = \nestedPolyra(3, 6)$ accepts $2\cdot n_1 = 6$ type $1$ agents and 3 type $2$ agents from $I_1$, and provides service to 3 type $1$ agents and $C- n_1 = 3$ type $2$ agents for a total reward of approximately $3000$, achieving a $1/2$ approximation to the optimal reward in $I_1$.

The BQ nested \po{} algorithm services only 3 type $2$ agents, while our algorithm services 4. This difference arises because the BQ nested \po{} algorithm accepts too many type $1$ agents (6 instead of 4 in our algorithm). It is essential for any good algorithm to accept some type $1$ agents because a non-anticipating algorithm cannot predict if type $2$ agents will arrive after observing the first 6 type $1$ agents. However, the BQ nested \po{} algorithm is overly pessimistic about the arrival of type $2$ agents, accepting too many type $1$ agents. This leads to poor performance if type $2$ agents do arrive. In contrast, our nested \po{} algorithm achieves good performance in both cases, resulting in better worst-case performance.

{\color{black}We remark that while achieving an optimal CR does not guarantee our algorithm outperforms the BQ algorithm in \textit{every} instance, our numerical studies in Section \ref{sec:numerical_studies} demonstrate that our optimized nested algorithm surpasses the BQ benchmark algorithm in more realistic input sequences that are not stylized under both feature-augmented and traditional notions of CRs.}
\end{example}

\subsection{Optimal POLYRA Algorithm with Three Types}

\label{section:three_type}
In this section, we consider a special case with $K = 3$ and $M\in \{1, 2\}$. Our main result in this section is an explicit characterization of two polytopes $\feasible^{(3, 1)}$ and $\feasible^{(3, 2)}$ (defined in equations \eqref{eq:feasibleMOne} and \eqref{feasibleMTwo}) which give optimal \po{} algorithms for the case when $M = 1$ and $M = 2$, respectively. We present an innovative technique where we first construct an upper bound on the CR of any online algorithm through an LP, and then use its optimal solution values to build our polytopes $\feasible^{(3, 1)}$ and $\feasible^{(3, 2)}$.

\subsubsection{An Upper Bound for Three Types}

In order to prove the optimality of our algorithms for $K = 3$, we first need to establish an upper bound. To do so, for each of $M = 1$ and $M = 2$, we construct  \textit{worst-case arrival sequences}, which best exemplify the main challenge that an online algorithm must face: the trade-off between a moderate reward now versus waiting for a potentially higher future reward. We then argue that to obtain a CR of $\crr$ on these worst-case arrival sequences, the behavior of non-anticipating algorithm needs to satisfy certain conditions. These conditions  lead to  an LP construction, where  the objective of the LP maximizes the CR (i.e., $\crr $ in the LP) and the constraints enforce the aforementioned conditions.

\begin{theorem}[Upper Bound on the CR with Three Types]
  \label{thm:3_type_upper_bound}
  Suppose that there are three types with $M\in \{1, 2\}$ and rewards $r_1 < r_2 < r_3$. Then, the CR of any non-anticipating algorithm $\mathcal {A}$, i.e.,  $\text{CR}_{\mathcal A}$, is at most the optimal value of the following LP, denoted by $\crr_{up}$. 
  \begin{align}
   \crr_{up} \defeq \max_{\crr, \{\optupper_{i,t}\}_{i\in[3], t\in[2]}} \quad && \crr & \nonumber \\
    \text{s.t.} \quad &   & \optOnet + \optTwot + \optupper_{3,t} & \leq C                                                                         & t \in [2] \label{align:Ia_capacity} \\
                      &   & \crr \cdot r_3 \cdot C                & \leq \optOnet \cdot r_1 + \optTwot \cdot r_2 + \optupper_{3,t} \cdot r_3       & t \in [2] \label{align:Ia_revenue}  \\
    \lplabel[3_class_ub]{\textsc{(upper3)}}  && \crr \cdot r_1 \cdot C &\leq(\optOneOne + \optOneTwo) \cdot r_1 \label{align:Id_revenue} \\
    && \crr \cdot r_2 \cdot C &\leq \left(C - \sum_{t=1}^M \optTwot \right) \cdot r_1 + \left( \sum_{t=1}^M \optTwot \right) \cdot r_2 \label{align:Ic_capacity}\\
    && \crr \cdot r_2 \cdot C &\leq \left(\optOneOne + \optOneTwo \right) \cdot r_1 + \left( \sum_{t=1}^M \optTwot \right) \cdot r_2 \label{align:Ic_revenue}\\
                      &   & \crr \cdot r_2 \cdot C                & \leq \optOneTwo \cdot r_1 + \optupper_{2,2} \cdot r_2 \label{align:Ib_revenue} & \text{if $M = 1$}\\
                      & & {\color{black} s_{i, t}} & {\color{black}\ge 0\, ,\quad  \crr \ge 0}\, &i\in [3], t\in [2]
  \end{align}
\end{theorem}
Next, we present a proof sketch of Theorem \ref{thm:3_type_upper_bound}. The complete proof of this theorem and the proof of all other statements in this section are presented in Section \ref{secproofthree}.

\begin{proof}{Proof Sketch of Theorem \ref{thm:3_type_upper_bound}.} 
  A fundamental limitation of a non-anticipating (online) algorithm is that it cannot differentiate between two arrival sequences $I$ and $I'$ until their first divergence. Leveraging this, we structure our upper bounds using overlapping multi-period arrival sequences $\mathcal{I}$, and stipulate conditions for a $\crr$-competitive algorithm $\alg$ concerning each sequence $\{I_t\}_{t \in [T]} \in \mathcal{I}$. These conditions, when expressed as an LP, allow us to maximize $\crr$ to derive our upper bound.

  Illustratively, for $K = 3$, $M = 1$, and $C = 3$, the arrival sequence is:
    \[
        \begin{array}{ll}
            I_1 = \{ 1, 1, 1, \: 2, 2, 2, \:3, 3, 3 \} & I_2 = \{ 2, 2, 2,\: 3, 3, 3  \}
        \end{array}
    \]
    Such an arrival sequence is difficult because lower reward agents arrive before higher reward ones, forcing an algorithm to commit to lower reward agents before knowing whether higher reward agents will arrive. We can define the following \textit{truncations} of $\{I_1, I_2\}$ which consist of abruptly ending the entire input sequence after a certain type of agent has arrived. For example, $\{I_1^2, I_2^2\}$ represents the truncation of $\{I_1, I_2\}$ where the entire 2-period input suddenly ends after type $2$ agents arrive in the first time period.
    \[\arraycolsep=1.4pt\def\arraystretch{1.5}
        \begin{array}{ll}
            I_1^1 = \{ 1, 1, 1\} \quad                                    & I_2^1 = \{  \}         \\
            I_1^2 = \{ 1, 1, 1, \: 2, 2, 2\} \quad                        & I_2^2 = \{  \}         \\
            I_1^3 = \{ 1, 1, 1, \: 2, 2, 2, \:3, 3, 3\} \quad             & I_2^3 = \{  \}         \\
            I_1^4 = \{ 1, 1, 1, \: 2, 2, 2, \:3, 3, 3 \} \quad & I_2^4 = \{2, 2, 2  \}  \\
        \end{array}
    \]

 We now give a sketch of how the constraints in \ref{3_class_ub} come about. Suppose we are given an algorithm $\alg$ which is $\crr$ competitive. Let $s_{i,t}$, $i \in [3], t \in [2]$, be the number of type $i$ agents that are serviced in period $t$ under $\alg$. Since types $2$ and $3$ are impatient, $s_{i,t}$ also represents the number of type $2$ and $3$ agents that $\alg$ accepts from $I_t$ for $t\in [2]$, while for type $1$, $s_{1,1} + s_{1,2}$ is the number of type $1$ agents that $\alg$ accepts from $I_1$. Since $\textsc{opt}_t(\{I_1, I_2\}) = r_3 \cdot C$, to be $\crr$-competitive, we must have  $\textsc{rew}_{\alg, t}(\{I_1, I_2\}) = r_1 \cdot s_{1,t} + r_2 \cdot s_{2,t} + r_3 \cdot s_{3,t} \geq \crr \cdot r_3 \cdot C$, which gives rise to constraint \eqref{align:Ia_revenue}. In addition, we certainly must have $s_{1,t} + s_{2,t} + s_{3,t} \leq C$, as given in \eqref{align:Ia_capacity}. We apply a similar reasoning to arrival sequences $\{I_1^1, I_2^1\}, \{I_1^2, I_2^2\}$ and $\{I_1^4, I_2^4\}$, to get the $\crr \cdot r_i \cdot C$ (for $i= 1$ or $2$) reward lower bounds in constraints \eqref{align:Id_revenue}, \eqref{align:Ic_revenue}, \eqref{align:Ic_capacity}, and \eqref{align:Ib_revenue}.\footnote{Note that $\{I^3_1, I^3_2\}$ is actually not useful as it would imply the constraint $r_1 \cdot s_{1,1} + r_2 \cdot s_{2,1} + r_3 \cdot s_{3,1} \geq \crr \cdot r_3 \cdot C$, which was already captured by $\{I_1, I_2\}$.} 
  \Halmos
\end{proof}

We will now state two \po{} algorithms which achieve the $\crr_{up}$ CR from the upper bound, one for the case where $M =1$ and another for $M = 2$. Both of these optimal algorithms require us to first compute the upper bound presented above, get an optimal solution $(\crr_{up}, \{\optupper_{i,t}^*\}_{i\in[3], t\in [2]})$ and use some of these values as constants in the feasible polytopes $\feasibleMOne$ for $M = 1$ and $\feasibleMTwo$ for $M = 2$. We will require some properties of this optimal upper bound solution, which are described in Lemmas \ref{lemma:3_class_tight_properties} and \ref{lemma:3_type_additional_properties} in Appendix \ref{sec:lemma:3_class_tight_properties} and \ref{sec:lemma:3_type_additional_properties}. We highlight that the upper bound only characterizes an optimal algorithm's behavior against our constructed worst-case arrival sequences, not any arbitrary arrival sequences over $T$ periods. Nonetheless, we show that how to use the upper bound solution, which represents a particular way of dealing with the worst-case arrival sequence, to construct an optimal algorithm for any arrival sequence.

\subsubsection{Optimal Three Type Algorithm for \texorpdfstring{$M = 1$}{M=1}}
With $M=1$, type $1$ agents are partially patient while types $2$ and $3$ are impatient. 
Let  $\Delta_{i,j} \defeq r_i - r_j$ and $(\crr_{up}, \{\optupper_{i,t}^*\}_{i\in[3], t\in[2]})$ be the optimal solution to the upper bound LP \ref{3_class_ub}. We define our feasible polytope $\feasibleMOne$ as follows
\begin{align} \label{eq:feasibleMOne}  \feasible^{(3,1)} = \left\{ (\fbucketOneOne, \fbucketOneTwo, \fbucketTwo, \fbucketThree) \subset \mathbb{R}_+^4: \quad  \begin{array}{cll}
    \fbucketOneOne + \fbucketTwo + \fbucketThree                       & \leq s_{1,1}^* + s_{2,1}^* + s_{3,1}^*                           &           \\
    \Delta_{3,1} \cdot \fbucketOneOne + \Delta_{3,2} \cdot \fbucketTwo & \leq \Delta_{3,1} \cdot s_{1,1}^* + \Delta_{3,2} \cdot s_{2,1}^* &           \\
    \fbucketOnei                                                       & \leq \optupper_{1,i}^*                                           & i \in [2]\end{array} \right\}\,.
\end{align}

\begin{theorem}[Optimality of \textsc{POLYRA} Algorithm with  Feasible  Polytope $\feasibleMOne$]
  \label{thm:opt_alg_M=1} When the number of types $K = 3$ and the number of partially patient types $M = 1$, with $r_1< r_2< r_3$, and  per-period capacity $C$,  the \po{} algorithm with feasible polytope $\feasibleMOne$ achieves a CR of $\crr_{up}$, and hence, is optimal. Here, $\crr_{up}$ 
  is the optimal objective value of the LP \ref{3_class_ub}.
\end{theorem}

\textbf{Discussion on Feasible Polytope $\feasibleMOne$.} 
We provide insight into the constraints of $\feasibleMOne$ and its achievement of a CR of $\crr_{up}$. The first constraint ensures no over-allocation in a time period, governed by $C$ as dictated by \eqref{align:Ia_capacity} in the LP \ref{3_class_ub}. It turns out that $\optupper_{1,1}^* + \optupper_{2,1}^* + \optupper_{3,1}^* = C$ always olds (see Lemma \ref{lemma:3_class_tight_properties}). The second constraint emphasizes accommodating type $3$ agents in $\bucketThree$, by limiting a linear combination of agents in buckets $\bucketOneOne$ and $\bucketTwo$. The last constraint ensures adequate space for types $2$ and $3$ by regulating the number of type $1$ agents.Interestingly, $\feasibleMOne$ isn't a nested polytope per Definition \ref{def:nested}. For achieving the optimal CR of $\crr_{up}$, the algorithm must consider a linear combination of agents in $\bucketOneOne$ and $\bucketTwo$, rather than their mere sum.

By employing our \po{} algorithm with polytope $\feasibleMOne$ on the critical arrival sequence from $\{I_1, I_2\}$ (as outlined in the proof sketch of Theorem \ref{thm:3_type_upper_bound}), the algorithm's performance aligns with the optimal solution $\{s_{i,t}^*\}$. Essentially, our \po{} algorithm replicates the upper bound for worst-case scenarios (achieving exactly the CR), and we show it does better for all other arrivals (Theorem \ref{thm:opt_alg_M=1}). Next, we delve into the polytope construction for three types when $M = 2$.

\subsubsection{Optimal Three Type Algorithm for \texorpdfstring{$M = 2$}{M=2}}
With $M=2$, types $1,2$ agents are partially patient while type $3$ agents are impatient. 
To construct the optimal algorithm for three types with two partially patient types, we  first compute the optimal solution $\left(\crr_{up}, \{s_{i,t}^*\}_{i \in [3], t \in [2]} \right)$ from the upper bound LP \ref{3_class_ub} in Theorem \ref{thm:3_type_upper_bound} and use it to construct our polytope $\feasibleMTwo$, which is defined in the following theorem.

\begin{theorem}[Optimal Algorithm for $K=3$ and $M=2$]
  \label{theorem:three_type_M=2}
 When the number of types $K = 3$ and the number of partially patient types $M = 2$, with $r_1< r_2< r_3$, and  per-period capacity $C$,  the nested \po{} algorithm with the {nested} polytope $\feasibleMTwo$, defined below, achieves a CR of $\crr_{up}$, and hence, is optimal:
  \begin{equation}\label{feasibleMTwo}
    \feasible^{(3,2)} \defeq \nestedPolyra\left(\frac{1}{2} \cdot (\optupper_{1,1}^* + \optupper_{1,2}^* ), \frac{1}{2} \cdot \left( \optupper_{1,1}^* + \optupper_{1,2}^* + \optupper_{2,1}^* +\optupper_{2,2}^*  \right)  , C \right) \, .
  \end{equation}
  Here, $(\crr_{up}, \{\optupper_{i,t}^*\}_{i\in[3], t\in[2]})$ is the optimal solution to the upper bound LP \ref{3_class_ub} and nested polytopes $\nestedPolyra$ are defined in Definition \ref{def:nested}.
\end{theorem}

\textbf{Discussion on the nested polytope $\feasibleMTwo$ and the optimal algorithm.} 
For two partially patient types ($M=2$), the optimal strategy is a straightforward nested \po{} algorithm, as indicated by Definition \ref{def:nested}. One can show that if we apply our \po{} algorithm with feasible polytope $\feasibleMTwo$ to the corresponding worst-case arrival sequence for $M = 2$, our \po{} algorithm will exactly serve $s_{i,t}^*$ agents of type $i\in[3]$ in time period $t\in [2]$.  This approach effectively mirrors the upper bound, ensuring our algorithm's $\crr_{up}$-competitiveness on worst-case arrival sequences.  To verify the nested polytope's CR of $\crr_{up}$, we use Theorem \ref{theorem:polyra_nested_performance} that utilizes an LP to characterize  the CR of any nested polytope.

While $\feasibleMOne$ is not a nested polytope, $\feasibleMTwo$ is. Recall from the earlier discussion that $\feasibleMOne$ is not nested because it has a constraint of the form $c_1 \cdot b_{1,1} + c_2 \cdot b_2 \leq c_3$ where $c_1 \neq c_2$. On the other hand, all of  $\feasibleMTwo$'s constraints involve a linear combination of variables with coefficients all equal to $1$. This can be explained by the following intuition: Both polytopes $\feasibleMOne$ and $\feasibleMTwo$ have constraints which make sure that we do not accept too many type $1$ or $2$ agents, thus making sure enough space is reserved for type $3$ agents. When type $1$ is partially patient and type $2$ is not (i.e. $M = 1$), this constraint is of the form $\Delta_{3,1} \cdot b_{1,1} + \Delta_{3,2} \cdot b_{2}$ where the coefficients $\Delta_{3,1}\neq \Delta_{3,2}$. This is because type $1$ and $2$ differ in their flexibility so our linear combination must weigh the two differently. On the other hand, when both type $1$ and type $2$ are partially patient (i.e $M = 2$), restricting the simple sum of $b_{1,1}$ and $b_{2,1}$ is sufficient.

\section{A Near-optimal Nested POLYRA  Algorithm{} for \texorpdfstring{$K > 3$}{K > 3}}
\label{section:beyond_three}
We proceed with the case where $K > 3$. For $K = 2$ and $K = 3$, we were able to find \po{} algorithms with optimal polytopes through presenting a \textit{tight} upper bound on the CR of any non-anticipating algorithm and constructing a \po{} algorithm whose CR matches that upper bound. Recall that to construct our upper bounds for $K = 2$ and $K = 3$, we presented  \textit{worst-case} arrival sequences over $T = 2$ time periods 
that present a significant challenge for any non-anticipating algorithm  that has  decide how many lower reward agents to accept without knowing if higher reward agents will actually arrive. 
While our worst-case arrival sequences   can certainly be extended to $K > 3$, unfortunately it may not lead to a tight upper bound (i.e. one that is actually attainable by an algorithm). See Appendix \ref{appendix:not_tight_example} for more details. There, we show that slight modifications of our worst-case arrival sequences can lead to a tighter upper bound, implying that our worst-case construction does not lead to an attainable upper bound. 


Given the challenge of constructing tight upper bounds, for $K > 3$, we turn our attention back to nested \po{} algorithms. We show that it is easy to optimize the size of the nests of a nested \po{} algorithm in order to achieve a CR of $\crr^*$ which is the highest CR out of all nested algorithms (Definition \ref{def:optimal_nested_algorithm}). We show theoretically that this optimal nested algorithm has two nice properties: 
\begin{enumerate}
    \item For $M \geq 1$, $\crr^{*} > \crr^{BQ}$, meaning that optimizing our nest sizes results in a better theoretical performance than using our benchmark, the BQ nested polytope (Theorem \ref{theorem:our_alg_beats_bq}). 
    \item $\crr^*$ is at least 80\% of the optimal CR (Theorem \ref{thm:optimality_gap}). We show this by exhibiting an upper bound $\overline{\crr}_{up}$ and showing that $\crr^* \geq 0.8 \cdot \overline{\crr}_{up}$.
\end{enumerate}

\subsection{ Optimal Nested POLYRA Algorithms and Their Performance }
While the LP from Theorem \ref{theorem:polyra_nested_performance} can be used to compute the CR of any nested \po{} algorithm, we notice that this LP is also linear in the variables $\deltan_i$ (and hence also in $n_i$). Therefore, we can solve the following LP to find the nested \po{} that results in the best CR across all nested \po{} algorithms. 

\begin{definition}[Optimized Nested POLYRA Algorithms] \label{def:optimal_nest}
Consider the following LP
    \label{def:optimal_nested_algorithm}
    \begin{align*}
        \crr^* \defeq \max_{\Gamma, \{s_i^j\}_{j \in [K], i \in [j]}, \{\deltan_i\}_{i \in [K]}} \quad & \crr \\
        \text{s.t.} \quad &  \eqref{nested_lp_revenue_target}, \eqref{nested_lp_capacity}, \eqref{nested_lp_j_le_M}, \eqref{nested_lp_j_ge_M}, \eqref{eq:positive}\,.
    \end{align*}
We define $\crr^*$ to be the optimal value of the above LP and $n_k^* \defeq \sum_{i=1}^k \deltan_i^*$ where $(\deltan_i^*)_{i=1}^K$ are the optimal solution for the variables $(\deltan_i)_{i=1}^K$. We define the nested polytope with nests $(n_k^*)_{k=1}^K$ to be the optimal nested polytope $\nestedPolyra^*$:
\begin{equation}
    \label{eq:nested_polyra_definition}
    \nestedPolyra^* \defeq \nestedPolyra(n_1^*, \ldots, n_K^*)\,.
\end{equation}
The \po{} algorithm using $\nestedPolyra^*$ is called the \textit{optimized nested \po{} algorithm} and by our construction, achieves a CR of $\crr^*$.
\end{definition} 

{\color{black}The following theorem shows that the CR of our optimized nested \po{} algorithm  (i.e., $\crr^*$) is strictly greater than the CR of the BQ nested \po{} algorithm, denoted by $\crr^{BQ}$, justifying the need to optimize the nest sizes. See also our numerical studies for a comparison of  our optimized nested \po{} algorithm and the BQ nested \po{} algorithm.}  
{
\color{black}
\begin{theorem}[Optimized Nested POLYRA Alg. vs. the BQ Nested POLYRA Alg.]
    \label{theorem:our_alg_beats_bq}
    For any $K$ types of agents with rewards $r_1 < \ldots < r_K$ with $M \geq 1$ partially patient types, we have that $\crr^* > \crr^{BQ}$, where $\crr^* $, defined in Definition \ref{def:optimal_nested_algorithm}, is the CR of our optimized nested \po{} algorithm and $\crr^{BQ}$, defined in equation \eqref{equation:bq_cr}, is the CR of the BQ nested \po{} algorithm. 
\end{theorem}
The proof of this statement can be found in Appendix \ref{appendix:nested_vs_bq}. In the proof, we show that 
when $M = 0$ (i.e., there are no partially patient agents), $\crr^*$ obtained by solving the LP in Definition \ref{def:optimal_nested_algorithm} is the same as $\crr^{BQ}$. However, when $M \geq 1$, the LP we solve in Definition \ref{def:optimal_nested_algorithm}  accounts for the $M$ partially patient types,  leading to  a strictly  higher objective value (CR). This demonstrates the fact that the BQ nested \po{} algorithm is optimized for the $M = 0$ case and our nested polytope can do strictly better than for $M \geq 1$ (i.e. when there are partially patient agents). 
}

In addition to being strictly better than the CR of the BQ nested \po{} algorithm, the CR of our optimized nested \po{} algorithm is also guaranteed to be close to optimal, i.e., the best possible CR.

\begin{theorem}[CR of our Optimized POLYRA Alg.]
\label{theorem:main_optimality_gap}
For any rewards $r_1 < \ldots < r_K$ and flexible types $M \in [K-1]$, let $\crr^*$ be the CR of the optimized nested \po{} algorithm (per Definition \ref{def:optimal_nested_algorithm}). Then, for any non-anticipating deterministic algorithm with a CR  of $\crr$, it is the case that $\crr^* \geq 0.8 \cdot \crr\,$.
\end{theorem}

{\color{black}
\noindent In Appendix \ref{sec:theorem:main_optimality_gap}, we confirm the validity of this statement by following these steps:

\begin{enumerate}
    \item 
In Lemma \ref{thm:loose_upper_bound}, we introduce an approximate closed-form upper bound, denoted as $\overline{\crr}_{up}$, for the CR of any deterministic non-anticipating algorithm.
    \item Subsequently, we provide a specific example of a nested polytope that has nested sizes represented by $(\overline{\deltan_i})_{i=1}^K$. While this polytope may not necessarily be the optimal nested configuration, it does achieve a CR denoted as $\crr_{nest}$. It's important to note that $\crr_{nest}$ satisfies the condition $\crr_{nest} \leq \crr^*$ as defined in Definition \ref{lem:bar_gamma}.
    \item Following that, we demonstrate that $\crr_{nest}$ has a lower bound of at least 80\% of $\overline{\crr}_{up}$ (as shown in Lemma \ref{thm:optimality_gap}). This finding also implies that $\crr^*$ (the CR of the optimized nested \po{} algorithm) cannot be less than 80\% of $\overline{\crr}_{up}$, thereby providing proof for Theorem \ref{theorem:main_optimality_gap}.
\end{enumerate}
}

{\color{black}
\section{Numerical Studies}
\label{sec:numerical_studies} 
In this section, we complement our theoretical results with numerical studies. We evaluate our optimized nested \po{} algorithms (as defined in Theorem \ref{theorem:polyra_nested_performance}) in real-world inspired settings where the arrival sequence is no longer adversarial and compare its performance with the BQ nested \po{} algorithm (Definition \ref{def:bq_nest}). We evaluate the performance of the two algorithms not only using our feature-augmented CR definition, but also under a more traditional CR definition. 

\subsection{Setup.} 
We fix the number of types at $K = 5$, the total number of resources in each time period $C=100$, the number of time periods is $T= 10$, and the number of partially patient types $M\in \{0, 1, 2, 3, 4\}$. 

\textbf{Agent Rewards.} We fix $r_1 = 1$. For each $i \in \{2, 3, 4, 5\}$, we draw $r_i \mid r_{i-1} \sim \left(r_{i-1} \cdot \text{Unif}(1, 3)\right)$. In other words, the reward of each agent type ranges uniformly from 1 to 3 times that of the previous agent type, in increasing order. 

\textbf{Arrival Sequence Generation.} 
First, we draw a \( K \)-dimensional probability distribution \( p \) from a Dirichlet distribution with uniform parameters, i.e., \( p \sim \text{Dirichlet}(1, \ldots, 1) \). This distribution represents the likelihood of the arrival of each of the \( K \) types.

For every period \( t \) ranging from 1 to \( T \), we determine the total number of arrivals, denoted by \( |I_t| \). This number is drawn from a distribution given by \(\max(0, \lfloor \text{Norm}(1.1 \cdot C, \sigma \cdot C)\rfloor) \), where ``Norm" refers to the Gaussian distribution. The mean of this distribution suggests that the demand (total number of arriving agents) is approximately 10\% higher than the available capacity \( C \) and the parameter \( \sigma \) controls the variability of demand in each period.

Each individual agent that arrives in period \( t \) is represented in sequence as \( z_{t, 1}, \ldots, z_{t, |I_t|} \). The type of every arriving agent is determined independently based on the categorical distribution derived from \( p \). To be precise, the probability that an agent \( z_{t,n} \) is of type \( k \) is given by \( P(z_{t,n} = k) = p_k \), where we recall that \( p \sim \text{Dirichlet}(1, \ldots, 1) \).

 In our experiments, we explore the effects of varying both \( M \) and \( \sigma \). A single instance of our simulation study involves drawing the agent rewards $(r_1, \ldots, r_5)$ and an arrival sequence $\{I_t\}_{t=1}^T$. We run our optimized nested algorithm (notated by $\alg^*$) and the BQ nested \po{} algorithm (denoted by $\alg^{BQ}$) on this arrival sequence and compute the following metrics.
\begin{itemize}[leftmargin=*]
    \item \textbf{Theoretical FA-CR Lower Bound}. Let $\crr^*$ and $\crr^{BQ}$ be the theoretical lower bounds on the feature-augmented CR of $\alg^*$ and $\alg^{BQ}$ respectively. Note that this metric is calculated purely based on $r_1, \ldots, r_5$ and the number of flexible types $M\in \{0, 1, 2, 3, 4\}$ irrespective of the actual arrival sequence. 
    \item \textbf{Empirical FA-CR}. For each arrival sequence, we can compute the empirical feature-augmented CR of that instance by measuring the worst-case ratio between an algorithm's reward and the optimal reward (without flexibility) over each of the $T$ periods. For each of the two algorithms $\alg \in \{\alg^*, \alg^{BQ}\}$, we define:
    \begin{align*}
        \textsc{fa-cr}_{\alg}(\{I_t\}_{t=1}^T) \defeq \min_{\tau \in [T]} \frac{\text{Rew}_{\alg, \tau}(\{I_t\}_{t=1}^T)}{\textsc{opt}(I_t)}\,.
    \end{align*}
    Recall that $\textsc{opt}(I_t)$ is the optimal offline reward from the agents arriving in period $t$ without access to flexibility for types $1, 2, \ldots, M$. 

    \item \textbf{Empirical TR-CR}. We introduce a more conventional CR notion termed the empirical total reward or traditional (TR) CR. It's defined as the ratio between the total reward our algorithm achieves over the $T$ periods and the optimal total reward of an algorithm that knows the arrival sequence in advance and \textbf{has the ability to utilize flexibility}. For any arrival sequence ${I_t}$, let $\textsc{opt-total}(\{I_t\}_{t=1}^T)$ represent the optimal total reward in hindsight achievable from $\{I_t\}_{t=1}^T$, where flexibility of types $1, 2, \ldots, M$ is available. There are several methods to determine $\textsc{opt-total}(\{I_t\}_{t=1}^T)$; one of the most straightforward is via the following mathematical  program.

     Given an arrival sequence $\{I_t\}_{t=1}^T$, let $n_{t, k}$ denote the count of type $k$ arrivals in $I_t$. We employ decision variables: $a_{t,k}$ is the count of type $k$ agents accepted from $I_t$, and $s_{t,k}$ represents the subset of $a_{t,k}$ designated for service in period $t$. The goal is to maximize the reward from all accepted agents. The formulation is detailed below.
    \begin{align}
        \textsc{opt-total}(\{I_t\}_{t=1}^T) &= \max_{\{s_{t,k}, a_{t,k}\}_{t \in [T], i \in [K]}} \sum_{t=1}^T \sum_{k=1}^K r_k \cdot a_{t,k} \nonumber \\
        \text{s.t.} \quad & \sum_{k=1}^K s_{t, k} + \sum_{k=1}^M (a_{t-1, k} - s_{t-1, k}) \leq C \label{opttotal:capacity} &t \in [T] \\
        &  a_{t,k} \leq n_{t,k} & t \in [T], k \in [K] \label{opttotal:accept} \\
        &  s_{t,k} = a_{t,k} & t \in [T], k \in \{M+1, \ldots, K\} \label{opttotal:serve} \\ 
        & s_{t, k}, a_{t,k} \in \mathbb{Z} \nonumber
    \end{align}
    For notational convenience, we define $a_{-1, k} = s_{-1, k} = 0$ for all $k \in [K]$. Constraint \eqref{opttotal:capacity} is the capacity constraint, where the first summation are the agents from $I_t$ we serve, while the second summation represent the unserved flexible agents from period $t-1$. Constraints \eqref{opttotal:accept} simply ensures that we do not accept more agents than arrive. Constraint \eqref{opttotal:serve} ensures that all inflexible agents are served in the current period. Note that although some of the variables are redundant, we believe this notation is the easiest to understand. For each of the two algorithms $\alg \in \{\alg^*, \alg^{BQ}\}$, we define the Total Reward CR (\textsc{tr-cr}):
    \begin{align*}
        \textsc{tr-cr}_{\alg}(\{I_t\}_{t=1}^T) \defeq  \frac{\sum_{\tau \in [T]} \text{Rew}_{\alg, \tau}(\{I_t\}_{t=1}^T)}{\textsc{opt-total}(\{I_t\}_{t=1}^T)}\,.
    \end{align*}
\end{itemize}

\captionsetup[sub]{font=footnotesize,justification=centering}
\captionsetup[subfigure]{font=scriptsize,labelfont=scriptsize, justification=centering}

\begin{figure}[hbt!]
    \centering
    \begin{subfigure}[c]{0.4\textwidth}
        \centering 
        \includegraphics[width=\textwidth]{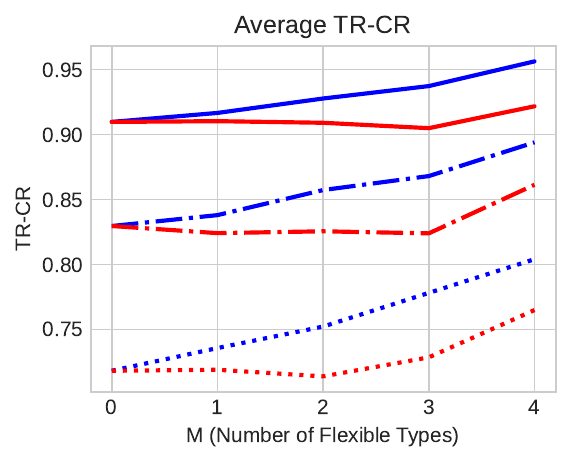}
        \subcaption{Average \textsc{tr-cr} by $M$}
        \label{sim:by_M_left}
    \end{subfigure}
    \begin{subfigure}[c]{0.4\textwidth}
        \centering 
        \includegraphics[width=\textwidth]{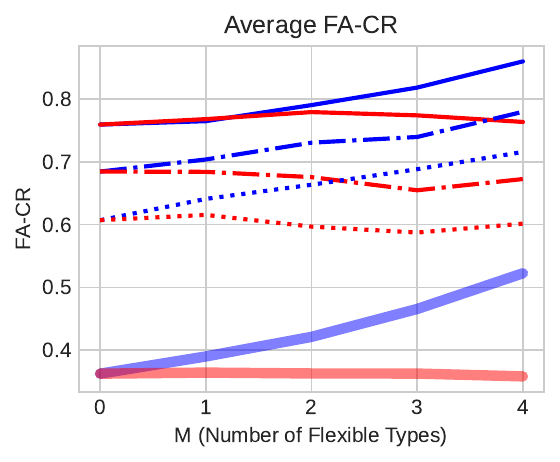}
        \subcaption{Average \textsc{fa-cr} by $M$}
        \label{sim:by_M_right}
    \end{subfigure}
    \begin{subfigure}[c]{0.15\textwidth}
        \centering 
        \includegraphics[width=\textwidth]{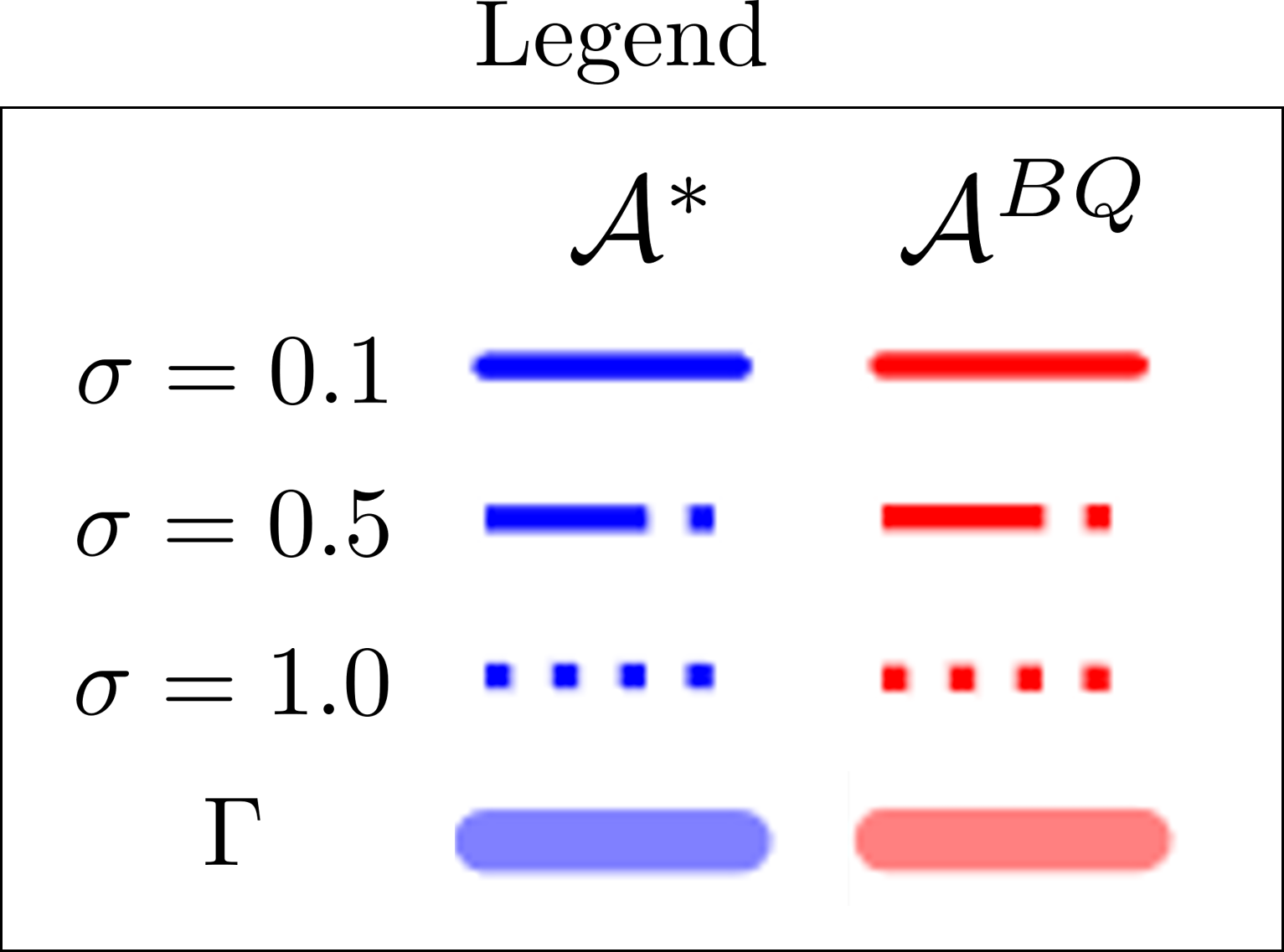}
    \end{subfigure}
    \vspace{0.5em}
    \caption{Average 
    TR-CR and FA-CR across 250 problem instances  with  $M\in\{0, 1, \ldots, 4\}$ and  $\sigma \in \{0.1, 0.5, 1.0\}$. For FA-CR, the theoretical lower bounds $\crr^*$ and $\crr^{BQ}$ are also shown.}
    \label{sim:by_M}
\end{figure} 

\begin{figure}[hbt!]
    \begin{subfigure}[c]{0.45\textwidth}
    \footnotesize
    \centering
   \begin{tabular}{l|ccc}
      $M$ & $\sigma = 0.1$ & $\sigma=0.5$ & $\sigma=1.0$  \\ \hline
        $0$ & (1.00, 1.00)   & (1.00, 1.00) & (1.00, 1.00) \\
        $1$ & (1.01, 0.98)   & (1.02, 1.01) & (1.02, 1.08) \\
        $2$ & (1.02, 1.02)   & (1.03, 1.08) & (1.05, 1.09) \\
        $3$ & (1.04, 1.04)   & (1.05, 1.09) & (1.06, 1.13) \\
        $4$ & (1.04, 1.08)   & (1.04, 1.08) & (1.05, 1.19)
    \end{tabular}
    \subcaption{\textsc{tr-cr}}
    \label{sim:improvement_left}
     \end{subfigure}
    \begin{subfigure}[c]{0.45\textwidth}
    \centering
    \footnotesize
   \begin{tabular}{l|ccc}
      $M$ & $\sigma = 0.1$ & $\sigma=0.5$ & $\sigma=1.0$  \\ \hline
        $0$ & (1.00, 1.00)   & (1.00, 1.00) & (1.00, 1.00) \\
        $1$ & (0.99, 0.88)   & (1.03, 0.97) & (1.04, 1.09) \\
        $2$ & (1.01, 0.97)   & (1.08, 1.11) & (1.11, 1.21) \\
        $3$ & (1.05, 1.24)   & (1.13, 1.22) & (1.17, 1.29) \\
        $4$ & (1.13, 1.37)   & (1.16, 1.36) & (1.19, 1.45)
    \end{tabular}
    \vspace{0.5em}
    \subcaption{\textsc{fa-cr}}
    \label{sim:improvement_right}

     \end{subfigure}
     \caption{Ratio of average and minimum (worst) empirical CR between $\alg^*$ and $\alg^{BQ}$. For each algorithm $\alg^*$ and $\alg^{BQ}$ we compute its average empirical CR and minimum empirical CR across the 250 instances. In each tuple above, the first value is the ratio of these two average CRs and the second value is the ratio of these two minimum (i.e. worst) CRs. Both TR-CR and FA-CR are shown as labeled. }
    \label{sim:improvement}
\end{figure}

\begin{figure}[hbt!]
    \centering
    \begin{subfigure}[c]{0.47\textwidth}
        \centering 
        \includegraphics[width=\textwidth]{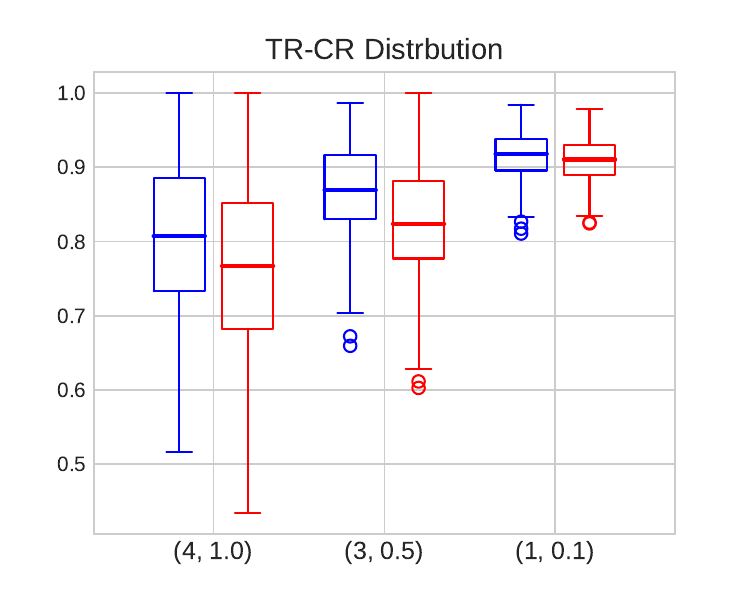}
        \subcaption{Distribution of \textsc{tr-cr}}
    \end{subfigure}
    \begin{subfigure}[c]{0.47\textwidth}
        \centering 
        \includegraphics[width=\textwidth]{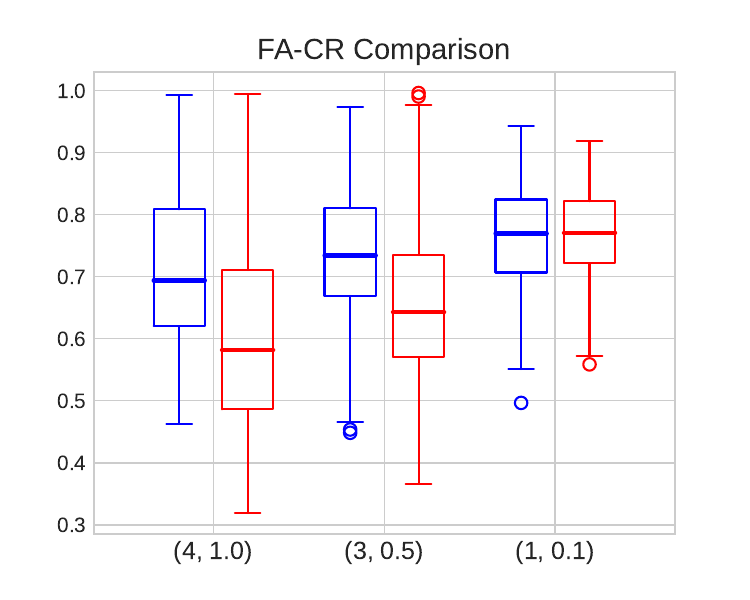}
        \subcaption{Distribution of \textsc{fa-cr}}
    \end{subfigure}
    \vspace{0.5em}
    \caption{Boxplot of the CRs obtained by $\alg^*$ in {\color{black} blue} and $\alg^{BQ}$ in {\color{red}red} across the 250 problem instances. The plot shows three values of $(M, \sigma)$, namely $(4, 1.0), (3, 0.5)$ and $(1, 0.1)$.}
    \label{sim:cdf}
\end{figure} 
 
\subsection{Results}
For each combination of 
$M \in \{0, 1, 2, 3, 4\}$ and $\sigma \in \{0.1, 0.5, 1.0\}$, we generated 250 arrival sequences. We then calculated three key metrics for both $\alg^{*}$ and $\alg^{BQ}$: the theoretical FA-CR, empirical FA-CR, and empirical TR-CR. For all figures in this section, the left-side corresponds to TR-CR while the right-side is FA-CR. The average values for these metrics are depicted in Figure \ref{sim:by_M}. Figure \ref{sim:improvement} 
provides a direct performance comparison between $\alg^*$ and $\alg^{BQ}$.  
For each $M, \sigma$ pair, the first value in the tuple is the ratio of the average CR (either FA-CR or TR-CR) of $\alg^*$ versus the average CR of $\alg^{BQ}$ over the 250 instances, which we refer to as the \emph{ratio of average CRs}. The second value denotes the ratio of the minimum (i.e. worst-performing) CR of $\alg^*$ to that of $\alg^{BQ}$ across the same instances, termed the \emph{ratio of worst CRs}. Finally, Figure \ref{sim:cdf} shows a box-plot of the CRs from $\alg^*$ and $\alg^{BQ}$ for a select number of $(M, \sigma)$ combinations. We highlight some of the insights from these figures below. 


\textbf{The relation between CRs and the number of flexible types.} Figures \ref{sim:by_M_left} and \ref{sim:by_M_right} demonstrate that as the number of flexible agent types $M$ rises, both TR-CR and FA-CR increase linearly. Contrary to the notion that ``a little bit of flexibility goes a long way", our findings suggest performance improvements are proportional to increased flexibility. 
Notably, Figure \ref{sim:by_M_left} highlights the substantial gains from introducing flexibility. For instance, in Figure \ref{sim:by_M_left}, compared to $M = 0$ (without flexibility), the TR-CR under $\alg^*$ shows improvements between 2.5\% to 12\% for $\sigma=1$ and $0.8\%$ to $5.1\%$ for $\sigma=0.1$ as $M$ increases from $1$ to $4$.

Conversely, the TR-CR and FA-CR of $\alg^{BQ}$—an algorithm with nest sizes that remain indifferent to $M$—plateau as $M$ grows. This underscores the notion that merely having flexible agents is insufficient; we must also design algorithms that adeptly harness this flexibility.

\textbf{\( \alg^* \) excels in TR-CR, even though it was primarily designed for FA-CR.}
Although \( \alg^{*} \) was designed to optimize nest sizes for the FA-CR (Definition \ref{def:optimal_nest}), it demonstrates impressive performance in both average and worst-case scenarios for TR-CR. This observation is evident in Figure \ref{sim:improvement_left}, where the ratio of averages and the worst-case ratio for TR-CRs are almost always greater than 1. By optimizing for the FA-CR, \( \alg^* \) determines nest sizes that more effectively exploit flexibility, irrespective of the specific performance metric, be it FA-CR or TR-CR.


\textbf{The performance disparity between \( \alg^* \) and \( \alg^{BQ} \) widens with greater volatility in the arrival sequence.}
As depicted in Figure \ref{sim:improvement}, the performance advantage of \( \alg^* \) over \( \alg^{BQ} \) becomes more pronounced as the volatility parameter \( \sigma \) increases for both the TR-CR and FA-CR metrics. Specifically, for any fixed \( M > 0 \), the advantage of \( \alg^* \) over \( \alg^{BQ} \) increases with \( \sigma \), by up to 19\% in average CR. Even more striking, the ratio of worst CRs surges by up to 45\% under the same conditions. This trend underscores that as the unpredictability of the arrival sequences heightens, the capacity to adapt flexibly becomes increasingly vital. Our theoretical findings bolster this observation, emphasizing that flexibility in algorithms offers enhanced resilience against uncertainties in agent arrivals. Furthermore, while exploiting flexibility improves the average case by a significant margin, it benefits the worst case by an even greater degree.

\textbf{The CR distributions for \( \alg^* \) appear to stochastically dominate those of \( \alg^{BQ} \).}
To provide a holistic view, Figure \ref{sim:cdf} presents boxplots for both FA-CR and TR-CR across three chosen configurations: \( (M=4, \sigma=1) \), \( (M=3, \sigma=0.5) \), and \( (M=1, \sigma=0.1) \). The distribution statistics, including the 1st quartile, median, and 3rd quartile, consistently indicate superior performance by \( \alg^* \) over \( \alg^{BQ} \). This trend is further substantiated in Figure \ref{sim:cdf} within Appendix \ref{appendix:K=10}, where the empirical cumulative distribution function (CDF) is plotted for the aforementioned \( (M, \sigma) \) combinations across 250 instances.

}

\section{Conclusion}
Making upfront commitments to gain flexibility has been shown to be effective in many real-world settings such as grocery delivery and shipping. Such flexibility presents a challenge to an algorithm designers who has to balance the trade-off between added flexibility in the current period and increased resource burden in the following period. To study this this trade-off, we present an online resource allocation model with a single replenishing resource and heterogeneous agent types, some of which require a resource immediately upon arrival while others are willing to wait for a short time. Our problem setting is unique in that for partially patient agents, we must make an upfront \textit{commitment} as to whether they will receive the resource, but we can delay the decision of which exact resource to allocate to them. 

We explored an adversarial arrival model and proposed a class of \po{} algorithms that perform optimally or near-optimally under a CR analysis. A \po{} algorithm keeps track of a state, which is guaranteed to be in some given feasible polytope. The linear constraints of this feasible polytope enforce that we do not accept too many lower reward agents, leaving enough space for higher reward agents. While a \po{} algorithm’s feasible polytope can be defined by arbitrarily complicated linear constraints,  a particular family of polytopes we call nested are simple to characterize and also perform optimally or near-optimally. {\color{black}Our numerical studies showed that such nested \po{} algorithms also have strong performance under stochastic arrival processes compared to the BQ nested polytope in terms of both FA-CR and TR-CR.}

Our work also opens up a number of avenues for future investigation. While partially patient customers in our model are willing to wait only a single time period, one can imagine extending our model so that certain partially patient customers are willing to wait two or more time periods. As another natural future direction, one can consider a setting in which agents are not inherently partially patient or impatient, but make a choice between the two depending on the price of each service. While the exact details of the algorithm for these new directions may be {context-specific}, the general insights provided in this work could be valuable in these settings as well.

\bibliographystyle{informs2014}
\bibliography{bibliography.bib}

\begin{APPENDICES}
\newpage
\section{Proof of Theorem \ref{flex_is_bad}} \label{section:flex_is_bad}
\begin{proof}{Proof of Theorem \ref{flex_is_bad}.}
{\color{black} We show this upper bound by constructing a collection of arrival sequences, each of which occurs over $T = 2$ periods. }

For $j \in [K]$, let $I_j$ consist of simply $2 \cdot C$ copies of all the types from $1$ through $j$ inclusive in ascending order. For example, $I_2$ would consist of $2 \cdot C$ type 1 agents followed by $2 \cdot C$ type $2$ agents.

We will use the following $K$ worst-case input sequences $I^j = \{I^j_1, I^j_2\}$. 
\begin{align*}
    I_1^1 = I_1 &\quad  I_2^1 = \emptyset \\
    I_1^2 = I_2 &\quad  I_2^2 = \emptyset \\
    \vdots & \quad \vdots \\
    I_1^{K-1} = I_{K-1} & \quad I_2^{K-1} = \emptyset \\
    I_1^K = I_K & \quad I_2^K = I_K
\end{align*}
We can see that for each $j \in [K]$, we have that: 
$$ \textsc{opt-total} \left( I^j \right) = 2 \cdot r_j \cdot C \,.$$
This is because for types $j \in [K-1]$, the optimal solution is simply to accept the last $2 \cdot C$ type $j$ agents, serve them on both days and achieve a reward of $2 \cdot C \cdot r_j$. For type $K$ agents, the optimal reward is obtained by accepting and serving $C$ of them in each of period $1$ and $2$ for a total reward of $2 \cdot C \cdot r_K$. 

Given a non-anticipating algorithm $\mathcal{A}$ which achieves a CR of $\crr$, let's define  $\{a_i\}_{i \in [K]}$ to be the total number of type $i$ agents that $\mathcal{A}$ accepts when faced with the input sequence $I^K$. For $j < K$, since $I^j$ and $I^K$ are the same in the same $2 \cdot j \cdot C$ agents in the first period, this implies that $a_1, \ldots, a_j$ is also the number of type $1$ through $j$ agents accepted in the first period when $\mathcal{A}$ faces input $I^j$. Given that $\mathcal{A}$ is $\crr$ competitive against all of $I^1, \ldots, I^K$, we must have:
$$ \sum_{i=1}^k a_i \cdot r_i \geq \crr \cdot \textsc{opt-total}(I^k) = 2 \cdot \crr \cdot r_k \cdot C $$
for all $k \in [K]$. In addition, the total number of agents we accept out of $I^K$ cannot be larger than the total capacity of $2 \cdot C$ and so we have that $\sum_{i=1}^k a_i \leq 2 \cdot C$.

Therefore, an upper bound on the performance of any online algorithm $\mathcal{A}$ is given by the following LP: 
\begin{align}
    \max_{\crr, \{a_k\}_{k \in [K]}} \quad & \crr \nonumber \\
    \text{s.t.} \quad & \sum_{i=1}^k a_i \cdot r_i \geq 2 \cdot \crr \cdot r_k \cdot C & k \in [K] \label{negative_result_revenue} \\
    & \sum_{i=1}^K a_i \leq 2 \cdot C \label{negative_result_capacity} \\
    & \crr, a_k \geq 0 & k \in [K] \nonumber
\end{align}

{\color{black}

The solution to the above LP does not depend on $C$ (as we can write all $a_k$'s as $a_k \cdot C$) and the $C$'s would cancel out in all constraints). Therefore, we just choose $C = 1$ and verify that the above LP has an optimal solution of $\crr^{BQ}$. To do so, we will write out an explicit primal and dual feasible solution which achieve the same objective value of $\crr^{BQ}$.

{ \scriptsize 
\begin{align}
\text{max} \quad  & \crr & \text{maximize} \quad & 2 \cdot t & \nonumber \\
\text{s.t.} \quad
&  2 \cdot r_k \cdot \crr \leq  \sum_{i=1}^{k} r_i \cdot a_i && \lambda_i \geq 0 \label{eq:primal_simple_1} & k \in \{1, 2, \ldots, K\} \\
&  \sum_{i=1}^K a_i \leq 2   && t \geq 0 & \label{eq:primal_simple_2} \\
& a_k \geq 0 &&t \geq r_k \cdot \left( \sum_{i=k}^K \lambda_i \right) & k \in \{1, \ldots, K\} \label{eq:dual_simple_1} \\
& \crr \geq 0 &&2 \cdot \sum_{i=1}^K r_i \cdot \lambda_i \geq 1 & \label{eq:dual_simple_2} 
\end{align}}

The optimal primal and dual solutions are:

{ \scriptsize
\begin{align*}
    \crr^* &= \crr^{BQ}  & t^* &= \frac{1}{2}\cdot \crr^{BQ} \\
    a_k^* &= 2 \cdot \crr^{*} \cdot \left(1 - \frac{r_{k-1}}{r_k} \right) 
    & \lambda^*_k &= t^* \cdot \left(\frac{1}{r_k} - \frac{1}{r_{k+1}}\right) & k \in \{1, 2, \ldots, K\}
\end{align*}
}
We define $r_0 = 0$ and $r_{K+1} = \infty$ (and so $\frac{1}{r_{K+1}} = 0$).
 It is clear that the proposed solutions above achieve the same objective value, so we just need to verify that they are each feasible. 

\textbf{Primal LP Feasibility.}  
\begin{itemize}
    \item Constraint \ref{eq:primal_simple_1}: For any $k\in \{1, 2,\ldots, K\}$, we have 
    \begin{align*}
        \sum_{j=1}^k r_j \cdot a_j^* &=  2 \cdot \crr^* \cdot \sum_{i=1}^k r_i \cdot \left(1 - \frac{r_{i-1}}{r_i} \right) = \crr^* \cdot r_k \,.
    \end{align*}
    
    \item Constraint \eqref{eq:primal_simple_2}:
    \begin{align*}
        \sum_{j=1}^K a_j^* = 2 \cdot \crr^* \cdot \left[ \sum_{i=1}^m \left(1 - \frac{r_{i-1}}{r_i} \right) \right] = 2\,.
    \end{align*}
    The term inside the square brackets above is exactly $1/\crr^{BQ}$.

    \item \textbf{Positivity}. All of our solution values $s_i^*, \crr^*$ are non-negative because $r_i / r_j < 1$ for any $i < j$.

\end{itemize}

\textbf{Dual LP Feasibility. } We examine each of the constraints separately. 
\begin{itemize}
    \item Constraint \ref{eq:dual_simple_1}: We expand the inside of the parentheses as
    \begin{align*}
        \sum_{i=k}^K \lambda_i^* = t^* \cdot \sum_{i=k}^K \left( \frac{1}{r_k} - \frac{1}{r_{k+1}} \right) = t^* \cdot \frac{1}{r_k}.
    \end{align*}

    \item Constraint \eqref{eq:dual_simple_2}: 
    \begin{align*}
        2 \cdot \sum_{i=1}^K r_j \cdot \lambda_j^* = 2 \cdot t^* \sum_{i=1}^K \left( 1 - \frac{r_{i}}{r_{i+1}}\right) = 2 \cdot t^* \sum_{i=1}^K \left( 1 - \frac{r_{i-1}}{r_{i}}\right) = 1\,.
    \end{align*}
    Note that the re-indexing in the second-to-last equality is valid because $r_0 = 0$ and $\frac{1}{r_{K+1}} = 0$. The last equation holds because  $t^* =\crr^{BQ}/2 $ and $ \crr^{BQ} = \left( K - \sum_{i=1}^{K-1} \frac{r_i}{r_{i+1}} \right)^{-1}$. 
    \item \textbf{Positivity}. All of the variables are non-negative due to the fact that $r_i < r_j$ for $i < j$ (and hence $\frac{1}{r_i} - \frac{1}{r_j} > 0$).
    
\end{itemize}
 This completes the verification of optimality. Hence $\crr^{BQ}$ is an upper bound to the performance of any online algorithm under a traditional competitive ratio benchmark using total reward (as defined in equation  \eqref{equation:conventional_benchmark}).
}
\end{proof}



\section{Proof of Statements from Section \ref{sec:polyra}}
\label{appendix:bq_polytope_same_cr_proof}
\subsection{Proof of Theorem \ref{theorem:polyra_nested_performance}}

\begin{proof}{Proof of Theorem \ref{theorem:polyra_nested_performance}}
Fix an optimal solution to the LP in Theorem \ref{theorem:polyra_nested_performance}, denoted by $(\crr_{\text{nest}}, \{\nestedvar{i}{j}\}_{j \in [K], i \in [j]})$. Throughout this proof, when we refer to $\nestedvar{i}{j}$, it will be this particular optimal solution (we avoid giving it a $*$ due to the existing superscript). In the proof, we will greatly leverage the fact that this solution $(\crr_{\text{nest}}, \{\nestedvar{i}{j}\}_{j \in [K], i \in [j]})$ satisfies all the constraints in the LP in Theorem \ref{theorem:polyra_nested_performance}.

Fix a particular time step $\tau \in [T]$ and consider the final state $\stateB$ that our \po{} algorithm with nested polytope $\nestedPolyra(n_1, \ldots, n_K)$ has at the end of the input sequence. For type $i \in [K]$, define $\numinput{i}$ to be the number of copies of type $i$ that appeared in $I_\tau$, while let $\numopt{i}$ be the number of copes of type $i$ that the optimal clairvoyant algorithm serves from $I_\tau$. Given that our optimal offline algorithm cannot utilize agents' patience, we have that $\numopt{i} \leq \numinput{i}$ for all $i \in [K]$. For notational convenience, for each $i \in [K]$, let $\algserveone{i}$ be the number of type $i$ agents that are from the first row of buckets in this final state $\stateB$ and for $i \in [M]$, let $\algservetwo{i}$ be the number of type $i$ agents that are served in $\bucketij{i}{2}$ where we define $\algservetwo{i} = 0$ if $i > M$. From our serving rule, it is clear that $\algserveone{i} = \bucketij{i}{1}$ for $i \leq M$ and $\algserveone{i} = \bucket{i}$ otherwise, and that $\algservetwo{i} \leq \bucketij{i}{2}$ for all $i \in [M]$. 

Let ${\maxtype}$ be the maximum index such that our algorithm serves as many type $k$ agents as have arrived in the input sequence.  Mathematically, we have 
\begin{equation}
    \label{definition_of_k} 
    {\maxtype} \defeq \max \{ i \in [K] \mid \algserveone{i} + \algservetwo{i} < \numinput{i} \}\,,
\end{equation}
where we define ${\maxtype} = 0$ if the set we take the $\max$ over is empty. (Note that ${\maxtype} = 0$ is not very interesting: it means that for every type $i$, we are serving as many copies of that type as have arrived in the input sequence, meaning that algorithm will certainly be $1$ competitive. Thus, we can simply assume that $\maxtype\neq 0$.)
By this definition of ${\maxtype}$, we know that $(\algserveone{i} + \algservetwo{i}) \geq \numinput{i} \geq \numopt{i}$ for all $i > {\maxtype}$.\footnote{To make the definition of $\maxtype$ more clear, consider an example with $\maxtype =3$ and the number of type $K=5$. Then, we know that no agents of type $4$ and $5$ are rejected while some of the agents of type $3$ are rejected. Further, we cannot make such statement for agents of type $k< \maxtype =3$, i.e., types $1$ and $2$.}  Our goal will be to show a lower bound on $\text{Rew}_{\nestedPolyra, \tau} \left( \{I_t\}_{t \in [T]} \right) $ and an upper bound on $\textsc{opt}_{\tau} \left( \{I_t\}_{t \in [T]} \right)$, which will allow us to lower bound their ratio. We start with presenting an upper bound on  $\textsc{opt}_{\tau} \left( \{I_t\}_{t \in [T]} \right)$: 
\begin{align}
    \textsc{opt}_{\tau}\left( \{I_t\}_{t \in [T]} \right) &= \sum_{i=1}^{{{\maxtype}}} r_i \cdot \numopt{i} + \sum_{i={\maxtype}+1}^K r_i \cdot \numopt{i} \nonumber = \sum_{i=1}^K r_i \cdot \numopt{i} + r_{\maxtype} \cdot \left(\sum_{i={\maxtype}+1}^K \numopt{i} \right) + \sum_{i={\maxtype}+1}^K \Delta_{i,{\maxtype}} \cdot \numopt{i} \nonumber \\
    &\leq C \cdot r_{\maxtype} + \sum_{i={\maxtype}+1}^K \Delta_{i,{\maxtype}} \cdot \numopt{i}  \leq C \cdot r_{\maxtype} + \sum_{i={\maxtype}+1}^K \Delta_{i,{\maxtype}} \cdot \numinput{i} \label{nested_alg_opt_upper} \,.    \intertext{Next, we will show the following lower bound on $\text{Rew}_{\nestedPolyra, \tau}\left( \{I_t\}_{t \in [T]} \right)$:}
    \text{Rew}_{\nestedPolyra, \tau}\left( \{I_t\}_{t \in [T]} \right) &\geq \crr_{\text{nest}} \cdot r_{\maxtype} \cdot C + \sum_{i={\maxtype}+1}^K \Delta_{i,{\maxtype}} \cdot \numinput{i} \, . \label{nested_alg_reward_target} 
\end{align}
Inequalities \eqref{nested_alg_opt_upper} and \eqref{nested_alg_reward_target} together would give us the desired result as the first terms of are off by a factor of $\crr_{\text{nest}}$ while the second term (i.e., $ \sum_{i={\maxtype}+1}^K \Delta_{i,{\maxtype}} \cdot \numinput{i}$) is the same. 

In the rest of this proof, we show the lower bound on $\text{Rew}_{\nestedPolyra, \tau}\left( \{I_t\}_{t \in [T]} \right)$ in equation \eqref{nested_alg_reward_target}. 
To do so, we focus on  type ${\maxtype}$. There are two reasons why  our algorithm does not serve all the type ${\maxtype}$ agents that have arrived in $I_\tau$:
\begin{enumerate}
    \item[] \textit{Case 1.} In this case,  at some point in $I_\tau$, our algorithm \textit{rejected} an agent of type ${\maxtype}$. 
    \item[] \textit{Case 2.} Our algorithm accepted all type ${\maxtype}$ agents from $I_\tau$, but it is not able to serve all of them because some of them are in bucket $\bucketij{{\maxtype}}{2}$ and we do not have enough capacity remaining after serving all the agents in buckets $\bucketij{1}{1}, \ldots, \bucketij{M}{1}, \bucket{M+1}, \ldots, \bucket{K}$. 
\end{enumerate}

We consider these two reasons separately as case 1 and case 2. 

\textbf{Case 1.} Since all nested polytopes are downward closed as per Definition \ref{def:downward_closed}, we know that if our algorithm rejected a type ${\maxtype}$ agent at some point in $I_\tau$, then the final state $\stateB$ also has the property that it would reject an incoming type ${\maxtype}$ agent. This implies that some constraint involving each of the $1$ or $2$ bucket(s) where type ${\maxtype}$ agents are allowed to go must be tight. We consider two sub-cases based on whether ${\maxtype} \leq M$ or ${\maxtype} > M$.
\begin{itemize}
    \item \textbf{Case 1a: ${\maxtype} \leq M$}. In this case, for each of $\bucketij{{\maxtype}}{1}$ and $\bucketij{{\maxtype}}{2}$, there must be at least one constraint involving them that is tight in $\feasible$. We see that all constraints involving $\bucketij{{\maxtype}}{1}$ are of the form
    \begin{align*}
        \sum_{i=1}^{\ell_1} \bucketij{i}{1} &\leq \nest{\ell_1}, \quad  \ell_1\ge \maxtype, \quad \text{or} \quad
        \sum_{i=1}^{M} \bucketij{i}{1} + \sum_{i=M}^{\ell_1} \bucket{i} \leq \nest{\ell_1}, \quad  \ell_1 \ge M+1
        \end{align*}
        while for $\bucketij{{\maxtype}}{2}$, the only constraints involving it are of the form 
        \begin{align*}
        \sum_{i=1}^{\ell_2} \bucketij{i}{2} \leq \nest{\ell_2}  \quad \ell_2 \ge \maxtype  \,.   
    \end{align*}
    Let $\ell_1, \ell_2 \in \{{\maxtype}, {\maxtype}+1, \ldots , K\}$ be indices for which the above constraints involving $\bucketij{{\maxtype}}{1}$ and $\bucketij{{\maxtype}}{2}$ respectively are tight, meaning we have
    \begin{equation}
        \label{our_tight_constraints}
        \sum_{i=1}^{\ell_1} \algserveone{i} = n_{\ell_1}, \quad \text{and}  \quad \sum_{i=1}^{\ell_2} \bucketij{i}{2} = n_{\ell_2} \,,
    \end{equation}
    where the first equality is true because by our serving rule $\algserveone{i}$ is equal to $\bucket{i}$ or $\bucketij{i}{1}$. \\
    
    For notational convenience, define $Z, Z', W,$ and  $W'$ to be the following: 
    \begin{equation}
        Z = \sum_{i=1}^{\maxtype} \algserveone{i}, \quad Z' = \sum_{i={\maxtype} + 1}^K \algserveone{i}, \quad W = \sum_{i=1}^{\maxtype} \algservetwo{i}, \quad W' = \sum_{i={\maxtype}+1}^M \algservetwo{i} \,.
    \end{equation}
(Recall that we are in the case of $\maxtype\le M$ and hence $W'$ is well-defined. Furthermore, since $\algservetwo{i}=0$ for $i> M$,    $W' = \sum_{i={\maxtype}+1}^K\algservetwo{i}$.)
Furthermore define the following quantities for the rewards from serving such agents.
    \begin{equation}
        R_Z = \sum_{i=1}^{\maxtype} r_i \cdot \algserveone{i}, \quad R_{Z'} = \sum_{i={\maxtype} + 1}^K r_i \cdot \algserveone{i}, \quad R_W = \sum_{i=1}^{\maxtype} r_i \cdot \algservetwo{i}, \quad R_{W'} = \sum_{i={\maxtype}+1}^M r_i \cdot \algservetwo{i} \,.
    \end{equation}
    It is clear that $\text{Rew}_{\nestedPolyra, \tau}\left( \{I_t\}_{t \in [T]} \right) = R_{Z} + R_{Z'} + R_{W} + R_{W'}$. Our goal is to show inequality \eqref{nested_alg_reward_target}. To do so, we will decompose $R_{Z'}$ and $R_{W'}$ into terms which have $\Delta_{i,k}$ as in \eqref{nested_alg_reward_target}. We see that: 
    \begin{align*}
        R_{Z'} + R_{W'} &= \sum_{{\maxtype} + 1}^{K} r_i \cdot \algserveone{i} + \sum_{{\maxtype}+ 1}^{M} r_i \cdot \algservetwo{i} = r_{\maxtype} \cdot (Z' + W') + \sum_{{\maxtype} + 1}^{K} \Delta_{i,{\maxtype}} \cdot \algserveone{i} + \sum_{{\maxtype}+ 1}^{M} \Delta_{i,{\maxtype}} \cdot \algservetwo{i} \\
        &\geq r_{\maxtype} \cdot (Z' + W') + \sum_{{\maxtype} + 1}^{K} \Delta_{i,{\maxtype}} \cdot x_i\,,
    \end{align*}
    where the last inequality is true because by definition of $\maxtype$, we have that that for types $i \geq \maxtype$, $\algservetwo{i} + \algserveone{i} \geq x_i$. This gives us 
     \begin{align}
        \text{Rew}_{\nestedPolyra, \tau}\left( \{I_t\}_{t \in [T]} \right) &\ge R_Z + R_{W} + r_{\maxtype} \cdot (Z' + W') + \sum_{{\maxtype} + 1}^{K} \Delta_{i,{\maxtype}} \cdot x_i\,. \label{our_reward_ZW}
    \end{align}   
    We provide a lower bound on $R_Z + r_{\maxtype} \cdot Z'$ as follows: 
    \begin{align}
        R_Z + r_{\maxtype} \cdot Z' &= \sum_{i=1}^{\maxtype} r_i \cdot \algserveone{i} + r_{\maxtype} \cdot \left( \sum_{i=\maxtype+1}^{\ell_1} \algserveone{i} + \sum_{i=\ell_1+1}^{K} \algserveone{i} \right) \nonumber \\ \nonumber
        &\geq \sum_{i=1}^{\maxtype} r_i \cdot \deltan_i + r_{\maxtype} \cdot \left(n_{\ell_1} - n_{\maxtype} + \sum_{i=\ell_1+1}^{K} \algserveone{i}\right) \nonumber \\
        &= \sum_{i=1}^{\maxtype} r_i \cdot \deltan_i + r_{\maxtype} \cdot \left(Z + Z' - n_{\maxtype} \right)\,, \label{revenue_lower_bound_Z}
    \end{align}
    where the inequality holds because of equation \eqref{our_tight_constraints} that $\sum_{i=1}^{\ell_1} \algserveone{i} = n_{\ell_1}$. To see why, observe that by the structure of the feasibility polytope ($\feasiblenest(\nest{1}, \ldots, \nest{K})$), we have $\sum_{i=1}^{k}\algserveone{i}\le \nest{i}$, $i\in [K]$. This and the fact that  $\sum_{i=1}^{\ell_1} \algserveone{i} = n_{\ell_1}$ and $r_1< r_2< \ldots< r_K$ imply that $\sum_{i=1}^{\maxtype} r_i \cdot \algserveone{i} + r_{\maxtype} \cdot \sum_{i=\maxtype+1}^{\ell_1} \algserveone{i} \ge  \sum_{i=1}^{\maxtype} r_i \cdot \deltan_i + r_{\maxtype} \cdot \left(n_{\ell_1} - n_{\maxtype} \right)$. 
    Put differently, since the sum (i.e., $\sum_{i=1}^{\ell_1} \algserveone{i} = n_{\ell_1}$) is fixed and $r_1 < r_2 < \ldots < r_{\maxtype}$, we get a lower bound by iteratively  choosing $z_1$ to be as large as possible, followed by $z_2$, etc. 
    Doing so sets all the $\algserveone{i}$'s to be equal to $\deltan_i$. 
    \footnote{
    More formally, consider the optimization problem:
  $ \min_{z_1, \ldots, z_{\ell_1}}  \left( \sum_{i=1}^{k_0} r_i \cdot z_i + r_{k_0} \cdot \sum_{i=k_0+1}^{\ell_1} z_i \right)$  subject to  $ \sum_{j=1}^i z_j \leq n_i \quad  i \in \{1, 2, \ldots, \ell_1\} $ and 
        $\sum_{i=1}^{\ell_1} z_i = n_{\ell_1}$.
    The optimal solution is clear using a water-filling argument: we start with $z_1$, which has the smallest coefficient in the objective, and raise that to the maximum. Then, we proceed with $z_2, z_3,$...etc. Doing this iteratively results in $z_i = \deltan_i$ for $i \in [\ell_1]$, which results in an objective value of 
    $ \sum_{i=1}^{k_0} r_i \cdot \deltan_i + r_{k_0} \cdot (n_{\ell_1} - n_{k_0}) $, as we used in the proof. To see that consider  the dual problem associated with the aforementioned minimization problem:  $\max_{\mu, (\lambda_i)_{i\in [\ell_1]} \ge \mathbf{0}} -\sum_{i=1}^{\ell_1}n_i\cdot  \lambda_i+\mu\cdot  n_{\ell_1}$
    subject to 
    $\mu-\sum_{j=i}^{\ell_1} \lambda_j\le r_i$ for any $i\le k_0$ and $
    \mu-\sum_{j=i}^{\ell_1} \lambda_j \le r_{k_0} $ for any  $i\in \{k_0, k_0+1, \ldots, \ell_1\}$.  Now, consider the following dual feasible solution: $\mu= \lambda_{\ell_1}+r_{k_0}$, $\lambda_i =0$ for $i\in \{k_0, k_0+1, \ldots, \ell_1-1\}$ and $\lambda_i = r_{i+1} -r_{i}$ for any $i\in \{1, 2, \ldots, k_0-1\}$. The dual objective value at this solution is $\sum_{i=1}^{k_0} r_i \deltan_i +r_{k_0}(n_{\ell}-n_{k_0})$, which is equal to the primal objective value at the suggested primal solution, verifying its optimality.  \label{footnote:optimality}

    }
    
    We take the result of inequality \eqref{revenue_lower_bound_Z} and plug it back into equation \eqref{our_reward_ZW} to get the following lower bound
     \begin{align}
        \text{Rew}_{\nestedPolyra, \tau}\left( \{I_t\}_{t \in [T]} \right) &\geq\sum_{i=1}^{\maxtype} r_i \cdot \deltan_i + {R_W + r_{\maxtype} \cdot \left(W' + Z + Z' - n_{\maxtype} \right)} + \sum_{{\maxtype} + 1}^{K} \Delta_{i,{\maxtype}} \cdot x_i \label{our_reward_ZW_2}\\
        & \ge \sum_{i=1}^{\maxtype} r_i \cdot \deltan_i+\sum_{i=1}^{\maxtype} r_i \cdot (\nestedvar{i}{{\maxtype}} - \deltan_i) +  \sum_{{\maxtype} + 1}^{K} \Delta_{i,{\maxtype}} \cdot x_i \label{first_inequality} \\
        &\geq \crr_\text{nest} \cdot r_k \cdot C  +  \sum_{{\maxtype} + 1}^{K} \Delta_{i,{\maxtype}} \cdot x_i \label{second_inequality}
    \end{align}
    where inequality \eqref{first_inequality} holds because of Lemma \ref{lem:star}, stated below, which we can apply because the total number of served agents in our  algorithm  $Z + Z' \geq n_{\ell_1} \geq n_{\maxtype}$ by equation \eqref{our_tight_constraints} . By this lemma, we know $R_W + r_{\maxtype} \cdot \left(W' + Z + Z' - n_{\maxtype} \right)\ge \sum_{i=1}^{\maxtype} r_i \cdot (\nestedvar{i}{{\maxtype}} - \deltan_i)$. Inequality \eqref{second_inequality} holds because of constraint \eqref{nested_lp_revenue_target} in the LP \ref{lp:nested_polytope}. By this constraint, we have $\sum_{i=1}^k r_i \cdot \deltan_i + \sum_{i=1}^k r_i \cdot (\nestedvar{i}{k} - \deltan_i) \geq \crr_\text{nest} \cdot r_k \cdot C$ for any $k\le M$.  Observe that the last equation is the desired result, and completes the proof for this case.

\begin{lemma} \label{lem:star} Recall that ${\maxtype} \defeq \max \{ i \in [K] \mid \algserveone{i} + \algservetwo{i} < \numinput{i} \}$,  
$R_Z = \sum_{i=1}^{\maxtype} r_i \cdot \algserveone{i}$,  $R_{Z'} = \sum_{i={\maxtype} + 1}^K r_i \cdot \algserveone{i}$,  $R_W = \sum_{i=1}^{\maxtype} r_i \cdot \algservetwo{i}$,  $R_{W'} = \sum_{i={\maxtype}+1}^M r_i \cdot \algservetwo{i}$,  $ Z = \sum_{i=1}^{\maxtype} \algserveone{i}$ and $ Z' = \sum_{i={\maxtype} + 1}^K \algserveone{i}$. Suppose that $\maxtype \le M$ and $Z + Z' - n_{\maxtype} \geq 0$. Then, we have 
\[R_W + r_{\maxtype} \cdot \left(W' + Z + Z' - n_{\maxtype} \right)\ge \sum_{i=1}^{\maxtype} r_i \cdot (\nestedvar{i}{{\maxtype}} - \deltan_i) \,,\]
where $(\crr_{\text{nest}}, \{\nestedvar{i}{j}\}_{j \in [K], i \in [j]})$ is the optimal solution to the LP \ref{lp:nested_polytope}.
\end{lemma}
The proof of this lemma is presented at the end of this section.
    
    \item \textbf{Case 1b: ${\maxtype} > M$.}
    For this case, some constraint involving $\bucket{{\maxtype}}$ must be tight and such constraints are of the form 
    \begin{align}
      \sum_{i=1}^{M} \bucketij{i}{1} + \sum_{i=M}^{\ell} \bucket{i} &\leq \nest{\ell} & \ell \in \{{\maxtype}, {\maxtype}+1, \ldots, M\} \label{nested_proof_k_geq_M} \,.
    \end{align}
    Let $\ell \in \{{\maxtype}, {\maxtype}+1, \ldots, M\} $ be an index for which the above is tight, meaning that we have: 
    $ \sum_{i=1}^\ell \algserveone{i} = n_\ell $.
    Given this equality, we can compute a lower bound on our algorithm's reward as follows by considering only the reward from the first row of buckets
    \begin{align*}
        \text{Rew}_{\nestedPolyra, \tau} \left( \{I_t\}_{t \in [T]} \right) &\geq \sum_{i=1}^{\maxtype} r_i \cdot \algserveone{i} + \sum_{i={\maxtype} + 1}^K r_i \cdot \algserveone{i} \\
        &= \sum_{i=1}^{\maxtype} r_i \cdot \algserveone{i} + r_{\maxtype} \cdot \left( \sum_{i={\maxtype} + 1}^K \algserveone{i} \right) + \sum_{i={\maxtype} + 1}^K \Delta_{i, {\maxtype}} \cdot \algserveone{i} \\
        &\geq \sum_{i=1}^{\maxtype} r_i \cdot \deltan_i + \sum_{i={\maxtype} + 1}^K \Delta_{i, {\maxtype}} \cdot \algserveone{i} \\
        &\geq \crr_{\text{nest}} \cdot r_{\maxtype} \cdot C + \sum_{i={\maxtype} + 1}^K \Delta_{i, {\maxtype}} \cdot \algserveone{i} \qquad \text{ by constraint \eqref{nested_lp_revenue_target}}
    \end{align*}
    The second-to-last inequality is true because $\sum_{i=1}^\ell \algserveone{i} = n_\ell$ and the fact that $\ell \in \{\maxtype, \maxtype+1, \ldots, K\}$ 
    give us a lower bound on $\sum_{i=1}^k r_i \cdot \algserveone{i} + r_{\maxtype} \cdot \left( \sum_{i={\maxtype} + 1}^{\ell} \algserveone{i} \right)$ by setting $\algserveone{1}$ to be as large as possible, followed by $\algserveone{2}$, etc. (See footnote \ref{footnote:optimality}.) The last inequality  follows from  
    constraint \eqref{nested_lp_revenue_target} in \ref{lp:nested_polytope} (i.e., the fact that $\crr_\text{nest}  \cdot C \cdot r_j \leq \sum_{i=1}^j \nestedvar{i}{j} \cdot r_i)$ and  constraint \eqref{nested_lp_j_ge_M} that for any  $j \geq M$, $s_i^j = \deltan_{j}$. These two constraints lead to $\sum_{i=1}^k r_i \cdot \deltan_i \geq \crr_\text{nest} \cdot r_k \cdot C$ for any $k\ge M$. The last inequality is the desired result.
\end{itemize}

\textbf{Case 2: }Recall that in this case, we do not have $\algserveone{{\maxtype}} + \algservetwo{{\maxtype}} \geq \numinput{{\maxtype}}$ because we do not have the capacity to serve all the agents in $\bucketij{{\maxtype}}{2}$ (i.e. $\algservetwo{{\maxtype}} < \bucketij{{\maxtype}}{2}$). Clearly this can only happen for ${\maxtype} \leq M$ as accepted impatient agents are always serviced. The fact that we ran out of capacity to serve all of $\bucketij{{\maxtype}}{2}$ implies that
\begin{equation}
    \sum_{i=1}^K \algserveone{i} + \sum_{i={\maxtype}}^M \algservetwo{i} = C
    \label{sum_is_C}
\end{equation}
In other words, the total amount of agents we are serving must be exactly at capacity. Using this, we get the following lower bound on the reward: 
\begin{align*}
    \text{Rew}_{\nestedPolyra, \tau}\left( \{I_t\}_{t \in [T]} \right) &\geq \sum_{i=1}^{\maxtype} r_i \cdot \algserveone{i} + \sum_{i={\maxtype} + 1}^K r_i \cdot \algserveone{i} + \sum_{i={\maxtype}}^M r_i \cdot \algservetwo{i} \\
    &= \sum_{i=1}^{\maxtype} r_i \cdot \algserveone{i} + r_{\maxtype} \cdot \left( \sum_{i={\maxtype}+1}^K \algserveone{i} + \sum_{i = {\maxtype}}^M \algservetwo{i} \right) + \sum_{i={\maxtype} + 1}^K \Delta_{i,{\maxtype}} \cdot \algserveone{i} + \sum_{i={\maxtype}}^M \Delta_{i, {\maxtype}} \cdot \algservetwo{i} \\
    &\geq \sum_{i=1}^{\maxtype} r_i \cdot \deltan_i + r_{\maxtype} \cdot \left( C - n_{\maxtype} \right) + \sum_{i={\maxtype} + 1}^K \Delta_{i,{\maxtype}} \cdot \algserveone{i} + \sum_{i={\maxtype}+1}^M \Delta_{i, {\maxtype}} \cdot \algservetwo{i} \\
    &\geq \sum_{i=1}^{\maxtype} r_i \cdot \deltan_i + r_{\maxtype} \cdot \left( C - n_{\maxtype} \right) + \sum_{i={\maxtype} + 1}^K \Delta_{i,{\maxtype}} \cdot \numinput{i}\,.
\end{align*}
Here, the second-to-last inequality is true because when equation \eqref{sum_is_C} holds, a lower bound on $\sum_{i=1}^{\maxtype} r_i \cdot \algserveone{i} + r_{\maxtype} \cdot \left( \sum_{i={\maxtype}+1}^K \algserveone{i} + \sum_{i = {\maxtype}}^M \algservetwo{i} \right)$ is attained by greedily choosing to put as much weight as possible in $z_1$ followed by $z_2$, etc. Doing so results in setting $z_i = \deltan_i$ for $i \in [k_0]$, and by equation \eqref{sum_is_C} the term in parentheses becomes $C - n_{k_0}$ and this is done by choosing $\algserveone{i} = \deltan_i$ for $i \in [{\maxtype}]$. See footnote \ref{footnote:optimality}. The last inequality is true because by definition of ${\maxtype}$, all types $i$ for which $i > \maxtype$ have the property that our algorithm serves at least $\numinput{i}$ copies of them. 

Given that our goal is inequality \eqref{nested_alg_reward_target} (i.e., $\text{Rew}_{\nestedPolyra, \tau}\left( \{I_t\}_{t \in [T]} \right) \geq \crr_{\text{nest}} \cdot r_{\maxtype} \cdot C + \sum_{i={\maxtype}+1}^K \Delta_{i,{\maxtype}} \cdot \numinput{i}$), all that is left to show is that 
$$ \sum_{i=1}^{\maxtype} r_i \cdot \deltan_i + r_{\maxtype} \cdot \left( C - n_{\maxtype} \right) \geq \crr_{\text{nest}} \cdot r_{\maxtype} \cdot C\,.  $$ 
To show this, we use the following chain of inequalities:
\begin{align*}
    \sum_{i=1}^{\maxtype} r_i \cdot \deltan_i + r_{\maxtype} \cdot \left( C - n_{\maxtype} \right) &\geq \crr_\text{nest} \cdot r_{\maxtype}\cdot C- \sum_{i=1}^{\maxtype} r_i \cdot (\nestedvar{i}{\maxtype} - \deltan_i)+ r_{\maxtype} \cdot \left( C - n_{\maxtype} \right)\\
    &\ge \crr_\text{nest} \cdot r_{\maxtype}\cdot C- r_{\maxtype}\cdot \sum_{i=1}^{\maxtype}   (\nestedvar{i}{\maxtype} - \deltan_i)+ r_{\maxtype} \cdot \left( C - n_{\maxtype} \right)\\
&=\crr_\text{nest} \cdot r_{\maxtype}\cdot C- r_{\maxtype}\cdot \sum_{i=1}^{\maxtype}   \nestedvar{i}{\maxtype}+r_{\maxtype}\nest{\maxtype}+ r_{\maxtype} \cdot \left( C - n_{\maxtype} \right)
\\ &\ge \crr_\text{nest} \cdot r_{\maxtype}\cdot C\,,
\end{align*}
where the first inequality
follows form constraint \eqref{nested_lp_revenue_target} in the LP \ref{lp:nested_polytope} that we have $\sum_{i=1}^k r_i \cdot \deltan_i + \sum_{i=1}^k r_i \cdot (\nestedvar{i}{k} - \deltan_i) \geq \crr_\text{nest} \cdot r_k \cdot C$ for any $k\le M$. The second inequality 
follows from constraint \eqref{nested_lp_j_le_M} that $\deltan_i \leq \nestedvar{i}{j}$  for any  $i \leq j \leq M $. The equality holds because $\sum_{i=1}^{\maxtype} \deltan_i = n_{\maxtype}$, and finally, the last inequality holds because by constraint \eqref{nested_lp_capacity} in the LP \ref{lp:nested_polytope}, $\sum_{i=1}^{\maxtype} s_i^k \leq C$. Thus, we have shown that inequality \eqref{nested_alg_reward_target} holds and we have argued earlier why this implies a $\crr_{\text{nest}}$ CR for this case. 
\end{proof}
\subsubsection{Proof of Lemma \ref{lem:star}}
Here, we will show that $R_W + r_{\maxtype} \cdot \left(W' + Z + Z' - n_{\maxtype} \right)\ge \sum_{i=1}^{\maxtype} r_i \cdot (\nestedvar{i}{{\maxtype}} - \deltan_i)$. To do, we start with making the following two claims about the right and left hands of the aforementioned inequality. 

\textit{Claim 1:} The expression  $R_W + r_{\maxtype} \cdot \left(W' + Z + Z' - n_{\maxtype} \right)
$ is greater than or equal to the optimal objective value of the following optimization problem evaluated at $n_{\maxtype}$, i.e., $R_W + r_{\maxtype} \cdot \left(W' + Z + Z' - n_{\maxtype} \right)\ge \textsc{lower}(n_{\maxtype})$, where
\begin{align}
    \textsc{lower}(x)= \max_{\omega_1, \ldots, \omega_{\maxtype}} \quad & \sum_{i=1}^{\maxtype} r_i \cdot \omega_i + r_{\maxtype} \cdot \left( W' +x - n_{\maxtype} \right) 
        \nonumber 
        \\
        \lplabel[nested_proof_lower_bound_1]{(\textsc{lower}(x))} \text{s.t} \quad & \omega_i \leq \bucketij{i}{2} & i \in [{\maxtype}] \label{first_lp_first_const} \\
        & W'+ x+\sum_{i=1}^{\maxtype} \omega_i \leq C \label{first_lp_second_const} 
        \end{align}
        \textit{Proof of Claim 1:} We show the claim in two steps. In the first step, we show that $R_W 
        + r_{\maxtype} \cdot \left(W' + Z + Z' - n_{\maxtype} \right)\ge \textsc{lower}(Z+Z')$ and in the second step, we show that $ \textsc{lower}(Z+Z')\ge \textsc{lower}(n_{\maxtype})$. This completes the proof.  
        Step 1: To show this step,  we verify  that  $\omega_i = \algservetwo{i}$, $i\in [\maxtype]$, is the optimal solution to problem $\textsc{lower}(Z+Z')$. Recall that $\algservetwo{i}$ is the number of type $i$ agents that are served in $\bucketij{i}{2}$ where we define $\algservetwo{i} = 0$ if $i > M$. From our serving rule, it is clear that
        our algorithm chooses $\algservetwo{i}$ using the $C - W' - Z - Z'$ capacity remaining to maximize $\sum_{i=1}^{\maxtype} r_i \cdot \omega_i$ subject to $\omega_i \leq \bucketij{i}{2}$. This is precisely the LP presented, i.e, $\textsc{lower}(Z+Z')$. (Note that the expression ``$r_{\maxtype} \cdot \left( W' + Z + Z' - n_{\maxtype} \right)$" in the objective value of the LP does not impact its optimal solution as it is not a function of $\omega_i$'s. It is become more clear later why we include this expression in the objective function.)
        
        Step 2: Here, we show that  $\textsc{lower}(Z+Z')\ge \textsc{lower}(n_{\maxtype})$.  To do so, we consider three cases, based on whether or not the second constraints in these two problems (i.e., constraints \eqref{first_lp_second_const}) are binding. Observe that when the second constraint in ($\textsc{lower}(n_{\maxtype})$) is binding (i.e.,  $W' + n_{\maxtype}+\sum_{i=1}^{\maxtype} \omega_i  = C$), the second constraint in ($\textsc{lower}(Z+Z')$) is also binding (i.e.,     
    $W'+ Z + Z'+\sum_{i=1}^{\maxtype} \omega_i = C$). This is because we have  $Z+Z'\ge \nest{\maxtype}$ by our assumption. In the first case, in the optimal solution of  both  ($\textsc{lower}(Z+Z')$) and ($\textsc{lower}(n_{\maxtype})$), the second constraint is not binding. This leads to the optimal solution of $\omega_i^*=\bucketij{i}{2}$ for any $i\in [\maxtype]$ for both problems. It is then easy to see that  the objective value of $\textsc{lower}(n_{\maxtype})$  (i.e., $\sum_{i=1}^{\maxtype} r_i \cdot \bucketij{i}{2} + r_{\maxtype} \cdot W'$) is at most that of $\textsc{lower}(Z+Z')$ (i.e., $\sum_{i=1}^{\maxtype} r_i \cdot \bucketij{i}{2} + r_{\maxtype} \cdot (W'+Z + Z'-\nest{\maxtype})$).  Again recall that $Z+Z'\ge \nest{\maxtype}$.
    Now, consider the second case in which the second constraint in ($\textsc{lower}(n_{\maxtype})$) is not binding while the second constraint in ($\textsc{lower}(Z+Z')$) is binding. The optimal objective value of ($\textsc{lower}(n_{\maxtype})$) is then given by
    \[\sum_{i=1}^{\maxtype} r_i \cdot \bucketij{i}{2} + r_{\maxtype} \cdot W' <  \sum_{i=1}^{\maxtype} r_i \cdot \bucketij{i}{2} + r_{\maxtype} \cdot (C-\nest{\maxtype}- \sum_{i=1}^{\maxtype} \bucketij{i}{2}  ) = \sum_{i=1}^{\maxtype} (r_i-r_{\maxtype}) \cdot \bucketij{i}{2} + r_{\maxtype} \cdot (C-\nest{\maxtype}  )\,,  \]
    where the inequality holds because the second constraint in ($\textsc{lower}(n_{\maxtype})$).  
    Now, let $\omega_{i,1}^*$, $i\in [\maxtype]$ be the optimal solution to ($\textsc{lower}(Z+Z')$). 
    The optimal objective value of ($\textsc{lower}(Z+Z')$) is then given by
    \begin{align*}\sum_{i=1}^{\maxtype} r_i \cdot  \omega_{i,1}^* + r_{\maxtype} \cdot (W'+Z + Z'-\nest{\maxtype}) &= 
    \sum_{i=1}^{\maxtype} r_i \cdot  \omega_{i,1}^* + r_{\maxtype} \cdot (C-Z-Z'- \sum_{i=1}^{\maxtype}    \omega_{i,1}^* +Z + Z'-\nest{\maxtype})\\
    &=   \sum_{i=1}^{\maxtype} r_i \cdot  \omega_{i,1}^* + r_{\maxtype} \cdot (C- \sum_{i=1}^{\maxtype}    \omega_{i,1}^*-\nest{\maxtype})\\
    &=  \sum_{i=1}^{\maxtype} (r_i-r_{\maxtype}) \cdot  \omega_{i,1}^* + r_{\maxtype} \cdot (C-\nest{\maxtype})\,,
    \end{align*}
    where the first equation holds because we assumed that the second constraint in ($\textsc{lower}(Z+Z')$) is binding. Then, since $r_1< r_2<\ldots < r_{\maxtype}$ and $\omega_{i,1}^*\le \bucketij{i}{2}$, we have $ \sum_{i=1}^{\maxtype} (r_i-r_{\maxtype}) \cdot  \omega_{i,1}^* + r_{\maxtype} \cdot (C-\nest{\maxtype}) \ge \sum_{i=1}^{\maxtype} (r_i-r_{\maxtype}) \cdot \bucketij{i}{2} + r_{\maxtype} \cdot (C-\nest{\maxtype}  ) $. Recall that the right hand side is an upper bound on the optimal objective value of ($\textsc{lower}(n_{\maxtype})$), and hence, we obtain the desired result. Now, consider the final case where the second constraint in both ($\textsc{lower}(Z+Z')$) and ($\textsc{lower}(n_{\maxtype})$) are binding and let $\omega_{i,2}^*$, $i\in [\maxtype]$ be the optimal solution to A2. Then, using the previous arguments, we can show that the optimal objective value of ($\textsc{lower}(Z+Z')$) and ($\textsc{lower}(n_{\maxtype})$) are respectively 
    $\sum_{i=1}^{\maxtype} (r_i-r_{\maxtype}) \cdot \omega_{i,2}^* + r_{\maxtype} \cdot (C-\nest{\maxtype} )$ and $\sum_{i=1}^{\maxtype} (r_i-r_{\maxtype}) \cdot  \omega_{i,1}^* + r_{\maxtype} \cdot (C-\nest{\maxtype})$. It is then easy to see that the optimal value of ($\textsc{lower}(Z+Z')$) is greater than that of ($\textsc{lower}(n_{\maxtype})$). This is because we have \begin{align*}\sum_{i=1}^{\maxtype} \omega_{i,1}^* = C-W'-Z-Z', \quad \text{ and } \quad \sum_{i=1}^{\maxtype} \omega_{i,2}^* = C-W'-\nest{\maxtype}\,.
    \end{align*}
 This and the fact that $Z+Z'\ge \nest{\maxtype}$ give us $\sum_{i=1}^{\maxtype} \omega_{i,2}^* > \sum_{i=1}^{\maxtype} \omega_{i,1}^*$. Then, observe that to maximize the objective values of  both ($\textsc{lower}(Z+Z')$) and ($\textsc{lower}(n_{\maxtype})$), we start  to `fill' $\omega_{i}$'s in a decreasing order of index $i\in [\maxtype]$. This results in $\omega_{i, 2}^*\ge \omega_{i,1}^*$, and verifies that  the optimal value of ($\textsc{lower}(Z+Z')$) is greater than that of ($\textsc{lower}(n_{\maxtype})$).

     \textit{Claim 2:}  The expression $\sum_{i=1}^{\maxtype} r_i \cdot (\nestedvar{i}{{\maxtype}} - \deltan_i)$ is less than or equal to  the optimal objective value of the following optimization problem.
      \begin{align}
        \sum_{i=1}^{\maxtype} r_i \cdot (\nestedvar{i}{{\maxtype}} - \deltan_i) \leq \max_{\omega_1, \ldots, \omega_{\maxtype}} \quad &\sum_{i=1}^{\maxtype} r_i \cdot \omega_i \nonumber \\
        \lplabel[nested_proof_upper_bound]{(\textsc{upper})} \text{s.t} \quad & \omega_i \leq \deltan_i & i \in [{\maxtype}] \label{B_first_constraint} \\
        &n_{\maxtype}+ \sum_{i=1}^{\maxtype} \omega_i  \leq C \label{B_second_constraint}
    \end{align}
        \textit{Proof of Claim 2:} To see why the claim holds note that choosing $\omega_i = (s_i^{\maxtype} - \deltan_i)$, $i\in [\maxtype]$, is a feasible solution. This is true because from our LP \ref{lp:nested_polytope}, we know that $s_{i}^{\maxtype} \in [\deltan_i, 2 \cdot \deltan_i]$, which gives constraint \eqref{B_first_constraint} and that $\sum_{i=1}^{\maxtype} s_{i}^{\maxtype} \leq C$, which gives constraint \eqref{B_second_constraint}. The completes the proof of the second claim. 

    \textit{Comparing $\textsc{lower}(n_{\maxtype})$ and \textsc{upper}.} Our goal is to show that the optimal objective value for the former is at least that of the latter. Showing this completes the proof. Before we reason about the optimal solutions to each LP, we first prove a crucial property for our analysis.
    
    \noindent \textit{Claim: } $ W' \geq \sum_{i=1}^{\maxtype} \left( \deltan_i - \bucketij{i}{2} \right) $ \\
    \noindent \textit{Proof: } Starting with the definition of $W'$, we have:
        \begin{align}
            W' &= \sum_{i=\maxtype+1}^M w_i \geq \sum_{i=\maxtype+1}^{\ell_2} w_i \overset{(a)}{=} \sum_{i=\maxtype+1}^{\ell_2} \bucketij{i}{2} = \sum_{i=1}^{\ell_2} \bucketij{i}{2} - \sum_{i=1}^{\maxtype} \bucketij{i}{2} \nonumber \\
            &\overset{(b)}{\geq} n_{\ell_2} - \sum_{i=1}^{\maxtype} \bucketij{i}{2} \overset{(c)}{\geq} n_{\maxtype} - \sum_{i=1}^{\maxtype} \bucketij{i}{2} = \sum_{i=1}^{\maxtype} \left(\deltan_i - \bucketij{i}{2}\right) 
            \label{nested_proof_useful_inequality}
        \end{align}
      where (a) holds because by definition of $\maxtype$, $w_i = \bucketij{i}{2}$ for all $i \geq \maxtype$. Inequality (b) holds by definition $\ell_2 \in \{{\maxtype}, {\maxtype}+1, \ldots , M\}$: we have: $\sum_{i=1}^{\ell_2} \bucketij{i}{2} = n_{\ell_2}$. Inequality (c) holds because $\algservetwo{i}\ge 0$ and $\ell_2\ge \maxtype$, and as a result $\nest{\ell_2}\ge \nest{\maxtype}$. This concludes the proof of the claim. \\
    
    Consider how the two LP's $\textsc{lower}(n_{\maxtype})$ and \ref{nested_proof_upper_bound} are different. The former is transformed into the latter by changing $\bucketij{i}{2}$ to $\deltan_i$ in constraint \eqref{first_lp_first_const} and removing $W'$ from both the objective value and constraint \eqref{first_lp_second_const}. 
    
    Let $\{\omega_i^*\}_{i=1}^{\maxtype}$ be an optimal solution to $\textsc{lower}(n_{\maxtype})$ while let $\{\hat{\omega_i}\}_{i=1}^{\maxtype}$ be an optimal solution to \ref{nested_proof_upper_bound}. Our goal is to prove that $\sum_{i=1}^{\maxtype} r_i \cdot \omega_i^* + r_{\maxtype} \cdot W' \geq \sum_{i=1}^{\maxtype} r_i \cdot \hat{\omega_i}$ which can be done by showing that
    \begin{equation}
        \label{nested_proof_target_equation}
        \sum_{i=1}^{\maxtype} (\hat{\omega_i} - \omega_i^*) \leq W'\,.
    \end{equation}
    
     Since both $\textsc{lower}(n_{\maxtype})$ and \ref{nested_proof_upper_bound} are instances of a continuous knapsack problem, we know that both optimal solutions will have a certain structure. For example, with $\textsc{lower}(n_{\maxtype})$, we have a knapsack with $C - W' - n_{\maxtype}$ capacity and we have access to $\bucketij{i}{2}$ copies of type $i$ for $i \in [\maxtype]$, which achieves a reward of $r_i > 0$. We know that the optimal solution will have one of two characteristics: either (1) we use up all the $C - W' - n_{\maxtype}$ capacity, meaning that constraint \eqref{first_lp_second_const} is binding, or (2) Constraint \eqref{first_lp_second_const} is not binding, but we have used up all $\bucketij{i}{2}$ copies of all types $i \in [\maxtype]$ and so constraint \eqref{first_lp_first_const} is binding for all $i$. We now perform case-work on which of the constraints in $\textsc{lower}(n_{\maxtype})$ is binding, as we just described.
    \begin{itemize}
        \item \textbf{Case 1.} Given that \eqref{first_lp_second_const} is tight for $\textsc{lower}(n_{\maxtype})$ and \eqref{B_second_constraint} is certainly satisfied in \ref{nested_proof_upper_bound}, we have: 
        \begin{align*}
            W' + n_{\maxtype} + \sum_{i=1}^{\maxtype} \omega_i^* = C \geq \sum_{i=1}^{\maxtype} \hat{\omega_i} + n_{\maxtype}
        \end{align*}
        Re-arranging gives us the desired result in \eqref{nested_proof_target_equation}. 
        \item \textbf{Case 2.} In this case, we assume that $\omega_i^* = \bucketij{i}{2}$ for all $i \in [\maxtype]$. This implies that: 
        \begin{align*}
            \sum_{i=1}^{\maxtype} (\hat{\omega_i} - \omega_i^*) = \sum_{i=1}^{\maxtype} (\hat{\omega_i} - \bucketij{i}{2}) \leq \sum_{i=1}^{\maxtype} (\deltan_i - \bucketij{i}{2}) \leq W'\,,
        \end{align*}
        which is the desired result in \eqref{nested_proof_target_equation}. Here, first inequality is true from constraint \eqref{B_second_constraint}, while the second inequality is from our useful property \eqref{nested_proof_useful_inequality} (i.e., $ W' \geq \sum_{i=1}^{\maxtype} \left( \deltan_i - \bucketij{i}{2} \right) $) proved earlier. 
    \end{itemize}
    
    Therefore, we have shown that the optimal objective value of $\textsc{lower}(n_{\maxtype})$ is always at least that of \ref{nested_proof_upper_bound}. This completes our proof. 

\subsection{Proof of Theorem \ref{theorem:bq_polytope_same_cr}}
{\color{black}
\begin{proof}{Proof of Theorem \ref{theorem:bq_polytope_same_cr}}

From Theorem \ref{theorem:polyra_nested_performance}, in order to show $\nestedPolyra^{BQ}$ with nests $n_i^{BQ}$ ($i\in [K]$) achieves a CR of $\crr^{BQ}$ as defined by equation \eqref{equation:bq_cr}, we must show that $\crr^{BQ}$ is the optimal value to the following LP:
\begin{align*}
    \max_{\crr\ge 0, \{s_i^j\}_{j \in [K], i \in [j]}} \quad & \crr \\ 
    \text{s.t.} \quad & \crr \cdot C \cdot r_j \leq \sum_{i=1}^j s_{i}^j \cdot r_i & j \in [K] \\
    &\sum_{i=1}^j s_{i}^j \leq C & j \in [K] \\
    &s_{i}^j \in [\deltan^{BQ}_i, 2 \cdot \deltan^{BQ}_i] & j \leq M, i \in [j] \\
    &s_i^j = \deltan^{BQ}_i & j > M, i \in [j]\,,
\end{align*}
\noindent where $\deltan_i^{BQ} = n_i^{BQ} - n_{i-1}^{BQ}$ for $i\in [K]$ and $n_0^{BQ} = 0$. Note that by our explicit formula for $\crr^{BQ}$ and $n^{BQ}_i$ from Definition \eqref{def:bq_nest}, we compute that $\deltan_i^{BQ} = \crr^{BQ} \cdot \left(1 - \frac{r_{i-1}}{r_i}\right) \cdot C$ where we define $r_0 \defeq 0$.

We observe that the above LP only has exactly $K$ constraints involving $\crr$, namely that $\crr \leq \frac{\sum_{i=1}^j s_i^j \cdot r_i}{C \cdot r_j}$ for all $j \in [K]$. To show that $\crr = \crr^{BQ}$ at optimality, it is sufficient to show the following: 

\begin{itemize}
    \item For all $j \in [K]$, there exists a feasible $(\crr^{BQ}, \{s_i^j\}_{i=1}^j)$ 
    such that $\crr^{BQ} \le \frac{\sum_{i=1}^j s_i^j \cdot r_i}{C \cdot r_j} $. 

    \item There does not exist a feasible solution $(\Gamma, \{s_i^j\}_{i=1}^j)$ for which $\crr > \crr^{BQ}$.  
\end{itemize}

For the first statement, the feasible solution we exhibit for any $j \in [K]$ is $s_i^j = \deltan_i^{BQ}$. This is feasible since $\sum_{i=1}^j s_i^j = \sum_{i=1}^j \deltan_i^{BQ} = n_j^{BQ} \leq n_K^{BQ} = C$. Furthermore, we have: 

\begin{align*}
    \sum_{i=1}^j \deltan_i^{BQ} \cdot r_i = C \cdot \crr^{BQ} \cdot  \sum_{i=1}^j  \cdot \left(1 - \frac{r_{i-1}}{r_i}\right) \cdot r_i =  \crr^{BQ} \cdot r_j \cdot C
\end{align*}
as desired.  Note that the above equation shows that $\crr^{BQ} = \frac{\sum_{i=1}^j s_i^j \cdot r_i}{C \cdot r_j}$ at $s_i^j = \deltan_i^{BQ}$, which verifies the feasibility of the suggested solution. 

For the second statement, consider the inequality $\crr \cdot C \cdot r_K \leq \sum_{i=1}^K s_{i}^K \cdot r_i$ in the above LP. Since we have that $s_i^K = \deltan_i^{BQ}$ by the LP, we see that:  
$$ \sum_{i=1}^K s_{i}^K \cdot r_i = \sum_{i=1}^K \deltan_i^{BQ} \cdot r_i =  \sum_{i=1}^K \crr^{BQ} \cdot \left(1 - \frac{r_{i-1}}{r_i} \right) \cdot r_i \cdot C = \crr^{BQ} \cdot r_K \cdot C\,. $$

Therefore, we get that $\crr \cdot C \cdot r_K \leq \crr^{BQ} \cdot C \cdot r_K$ and hence every feasible solution must have $\crr \leq \crr^{BQ}$, which proves that $\crr^{BQ}$ is the optimal solution.


    
\end{proof}
}

\section{Proof of Statements from Section \ref{section:two_type}}
\label{sec:proof_2types}

Before we present the proof of Theorem \ref{thm:opt_2}, we present a property of all the polytopes in this work that is crucial to showing our CR results. 

\begin{definition}[Downwards Closure] 
\label{def:downward_closed}
A polytope $\feasible$ is said to be downwards closed if for all $\stateB, \stateB' \geq 0$ for which $\stateB \leq \stateB'$ (component-wise), it is the case that $\stateB' \in \feasible$ implies $\stateB \in \feasible$. Equivalently, the converse states that if $\stateB \not \in \feasible$, then $\stateB' \not \in \feasible$.
\end{definition}
This property implies that if $\stateB$ can no longer fit any more type $i \in [K]$ agents at some point in an arrival phase, then all type $i$ agents from that point onward in that arrival phase will be rejected. Conversely, if at the end of the arrival phase, our state $\stateB'$ is able to fit another type $i$ agent, then certainly no type $i$ agents were rejected throughout the arrival phase. 

\subsection{Proof of Theorem \ref{thm:opt_2}} 
\begin{proof}{Proof of Theorem \ref{thm:opt_2}.}
First, we will show that $\frac{2}{3 - r_1/r_2}$ is an upper bound on the CR of any online algorithm.  

Consider the following two input sequences: 
\begin{enumerate}
    \item  $I_1^a =[\underbrace{1, 1, \ldots, 1}_{C}]$, $I_2^a= \emptyset$
    \item  $I_1^b = [\underbrace{1, 1, \ldots, 1}_{C}, \underbrace{2, 2, \ldots, 2}_{C}]$, $I_2^b =[ \underbrace{2, 2, \ldots, 2}_{C}]$
\end{enumerate}
Consider any non-anticipating algorithm $\alg$. Such an algorithm should accept the same number of  agents of type $1$ in both input sequences $a$ and $b$.  Let $x$ be the faction of  agents of type $1$ that   algorithm $\alg$ accepts in the first time period.  Then, in the input sequence $a$, the CR of this algorithm in the first time period, as well as across two time periods,  is $x$.  Now consider how such an algorithm would behave on input $b$. Given that $x$ fraction of  agents of type $1$ is accepted, the maximum possible total reward of this algorithm over the two time periods of input $b$ is given by $ C\cdot(x \cdot r_1 + (2 - x) \cdot r_2)$ while the maximum revenue of the offline solution is $2 C\cdot r_2$, leading to the overall competitive ratio of no more than $(x \cdot r_1 + (2 - x) \cdot r_2)/ (2r_2)$. 

{\color{black}To obtain high reward on both sequences $a$ and $b$ (relative to the offline optimal), we want to maximize $\min(x,(x \cdot r_1 + (2 - x) \cdot r_2)/ (2r_2)) $, which can be done by setting the two equal to each other:}
$$ \frac{ x \cdot r_1 + (2- x) \cdot r_2 }{2 \cdot r_2} = x \quad \Rightarrow \quad x = \frac{2}{3 - r_1/r_2}\,.$$
We have just showed that the cumulative CR is at most $\frac{2}{3 - r_1/r_2}$. Then, considering the fact that the optimal offline reward (without using patience) in both time periods of input $b$ is equal to $C\cdot r_2$, we can conclude that per-period  CR of algorithm $\alg$ is at most $  \frac{2}{3 - r_1/r_2}$. This completes the proof of the upper bound.

We now proceed to proving that regardless of what the arrival sequence is, our polytope $\nestedPolyra^{(2,1)}$ defined in equation \eqref{eq:feasible_two_type} achieves a CR of $\frac{2}{3 - r_1/r_2}$.

Let $\stateB$ be the bucket state at the end of the arrival phase in some period $t$. We must show that the reward collected in the service phase on $\stateB$ is at least $\frac{2}{3 - r_1/r_2}$ of the optimal reward in period $t$ (without using flexibility). We will perform case-work based on whether an incoming  agent of type $k\in \{1, 2\}$ would be accepted or rejected at state $\stateB$. We say that our state $\stateB$ rejects an incoming agent of type $k$ if an additional unit of type $k$ arriving to this state $\stateB$ would be rejected (i.e. no positive fractional amount of that agent can fit in our state $\stateB$). 

\textbf{Case 1: An incoming  agent of type $2$ would be rejected.}  If our state $\stateB$ is unable to accommodate an agent of type $2$, it must be the case that the first row is full (i.e., $\bucketOneOne+\bucketTwo =C$). This implies that  the total reward in this time period is {\color{black}at least}: 
\begin{align*}
    r_1 \cdot \bucketOneOne + r_2 \cdot \bucketOneTwo &= \bucketOneOne\cdot r_1 + (C-\bucketOneOne)\cdot r_2 \\
    &\geq \left(\frac{1}{3 - \frac{r_1}{r_2}} \cdot r_1 + \left(1- \frac{1}{3 - \frac{r_1}{r_2}} \right) \cdot r_2 \right) \cdot C \\
    &= \frac{2}{3 - \frac{r_1}{r_2}} \cdot r_2 \cdot C\,, 
\end{align*}
where the first inequality follows from the fact that $\bucketOneOne\le C\cdot \frac{1}{3 - r_1/r_2}$ and coefficient of $\bucketOneOne$ in the above expression is $C\cdot(r_1-r_2)<0$. Now, considering that the optimal clairvoyant reward in this time period is at most $C \cdot r_2$, we can conclude that the CR of our algorithm in this time period is at least $\frac{2}{3 - r_1/r_2}$, which is the desired result.  

\textbf{Case 2: An incoming  agent of type $2$ would be accepted, but an incoming agent of type $1$ be rejected.} Given that an incoming type $2$ agent would be accepted, we have that $\bucketOneOne + \bucketTwo < C$ as that is the only constraint involving $\bucketTwo$. The fact that an incoming type $1$ agent is rejected means that for each of $\bucketOneOne$ and $\bucketOneTwo$, there is some constraint in $\feasible$ involving them that is tight. Since $\bucketOneOne + \bucketTwo < C$, this implies that we must have $\bucketOneOne=  C\cdot \frac{1}{3 - r_1/r_2}$ and $\bucketOneTwo =C\cdot \frac{1}{3 - r_1/r_2}$. 

More importantly, the fact that an incoming type $2$ agent is accepted implies that no type $2$ agents from $I_t$ were rejected. This is because nested polytopes are downward closed according to Definition \ref{def:downward_closed}. Hence, we conclude that $\bucketTwo$ is precisely the number of type $2$ agents present in $I_t$, which in turn is {\color{black} at least the number of type 2 agents accepted and serviced in the optimal solution} because no type $2$ agents have been rejected. This means that the optimal clairvoyant reward for this time period is no greater than $\bucketTwo \cdot r_2 + (C -\bucketTwo ) \cdot r_1$.  

Recall that our algorithm's serving rule will choose to serve everyone in $\bucketOneOne$ and $\bucketTwo$ (i.e. the first row) and use the remaining space to serve as many agents from $\bucketOneTwo$ as possible. Based on this rule, the following expression exactly computes the reward our algorithm achieves:
$$ \bucketTwo \cdot r_2+ 
 \min\{(C - \bucketTwo), \bucketOneOne+\bucketOneTwo )\} \cdot r_1 \,.$$

We notice that if $C - \bucketTwo \leq \bucketOneOne + \bucketOneTwo$, then our algorithm's reward is exactly the same as the upper bound on the optimal reward, giving us a CR of $1$. Now consider the case of $C - \bucketTwo > \bucketOneOne + \bucketOneTwo$. 
Since we have an expression for our reward as well as an upper bound on the optimal reward, we look at their ratio with a little algebra: 
\begin{align*}
  \frac{ \bucketTwo \cdot r_2+ (\bucketOneOne+\bucketOneTwo) \cdot r_1}{\bucketTwo \cdot r_2 + (C -\bucketTwo ) \cdot r_1} 
  \ge \frac{(\bucketOneOne+\bucketOneTwo) \cdot r_1}{(C -\bucketTwo ) \cdot r_1}  
  = \frac{2 \cdot \frac{C}{3-r_1/r_2}}{C-\bucketTwo} \ge  \frac{2 \cdot \frac{C}{3-r_1/r_2}}{C} \ge  \frac{2}{3-r_1/r_2}\,.
\end{align*}
Here, the first inequality holds because for any $a, b>0$ with $a<b$ and $c>0$, we have $\frac{a+c}{b+c}\ge \frac{a}{b}$, while the first equality holds because $\bucketOneOne=\bucketOneTwo =\frac{C}{3-r_1/r_2}$. 

\textbf{Case 3: Any incoming  agent can be accepted.} {\color{black} In this case, no agent has been rejected by our algorithm. This means that all agents which the optimal offline algorithm choose to service are present in our state $\stateB$. We will serve exactly all the type $2$ agents in $\bucketTwo$ and type $1$ agents in $\bucketOneOne$, and use the remaining $C - \bucketTwo - \bucketOneOne$ capacity to serve other type $1$ agents in $\bucketOneTwo$. This serving rule is equivalent to saying that we will serve all type 2 agents and and as many type $1$ agents as possible, but that is exactly the optimal offline solution as well. Hence in this case, we have a CR of $1$. }
\end{proof}

\section{Proof of Statements from Section \ref{section:three_type}}

\subsection{Full Proof of Theorem \ref{thm:3_type_upper_bound}}
\label{secproofthree} 
\label{section:proof:3_class_upper}
\begin{proof}{Proof of Theorem \ref{thm:3_type_upper_bound}.}
To show the result, we consider the following four input sequences.
\begin{itemize}
     \item  \textbf{Input $a$.}  
     $I_1^a = [ \underbrace{1,1, \ldots,  1}_{2 \cdot C}, \underbrace{2, 2, \ldots,  2}_{C}, \underbrace{3, 3, \ldots,  3}_{C} ],  I_2^a = [ \underbrace{M+1, M+1, , \ldots,  M+1}_{C}, \ldots, \underbrace{3, 3, \ldots,  3}_{C}]$. 
     {\color{black} Note that if $M = 2$, then $I_2^a$ contains only $2 \cdot C$ type $3$ agents. There is no difference in our analysis between having $2 \cdot C$ or just $C$ type 3 agents here because there is no capacity to serve the additional $C$ type 2 agents anyways; we simply choose $2 \cdot C$ to save the casework of having different arrivals for $M = 1$ and $M = 2$. }
     
    \item \textbf{Input $b$.}   $I_1^b = [ \underbrace{1,1, \ldots,  1}_{C}, \underbrace{2, 2, \ldots,  2}_{C}, \underbrace{3, 3, \ldots,  3}_{C} ]$ and
        $I_2^b = [\underbrace{M+1, M+1, \ldots,  M+1}_{C}]$. 
    
    \item \textbf{Input $c$.}    $I_1^c = [\underbrace{1, 1, \ldots, 1}_{C}, \underbrace{2, \ldots, 2}_{C}]$ and 
        $I_2^c = \emptyset$. 
            
    \item \textbf{Input $d$.} $I_1^d = [\underbrace{1, 1, \ldots, 1}_{C}]$ and 
       $ I_2^d = \emptyset$. 
\end{itemize}

Observe that all of the input sequences $\{I_1^b, I_2^b\}, \{I_1^c, I_2^c\}$ and $\{I_1^d, I_2^d\}$ are simply truncated versions of $\{I_1^a, I_2^a\}$ where the input sequence suddenly ends after a certain $C$ copies of a type arrive. In order for any non-anticipating algorithm $\mathcal{A}$ to be $\crr$ competitive, it must be robust towards these sudden truncations of the input sequence.

\textbf{Definitions of Variables. }We now start with a few definitions that describes the  behavior of any non-anticipating algorithm $\mathcal{A}$ against the input sequence $a$, i.e., $\{I_1^a, I_2^a\}$. The $\optupper_{i,t}$'s correspond to the variables in the LP, while the $a_{i,t}$'s are defined for clarity in the arguments to follow. 
\begin{itemize}
    \item $\{\optupper_{i,t}\}_{ i \in [3], t \in [2]}$: The number of type $i$  agents that $\mathcal{A}$ serves at the end of time period $t$ when all the  agents have arrived.  
    
    \item $\{a_{i,t}\}_{i \in [2], t \in [2]}$: The number of type $i\in [2]$  agents that $\mathcal{A}$ accepts in $I_t^a$ for $t\in [2]$. Note that if $I_t^a$ does not contain  agents of type $i$, then $a_{i,t} \defeq 0$. Further, we only define $a_{i,t}$ for $i \in [2]$ because type $3$ is always inflexible, so $a_{3, t}$ would always be equal to $\optupper_{3,t}$. 
\end{itemize}
We show some simple relations between the $a_{i,t}$'s and $\optupper_{i,t}$ that will be used in the arguments to follow. It is easy to verify that
\begin{align}a_{1,1}  = \optOneOne + \optOneTwo\,;\label{align:three_type_a11}
\end{align}
that is, the number of served  agents of type $1$ across two time periods is equal to the number of accepted  agents of type $1$ in the first time period. This is because  agents of type $1$ only appear in the first time period. Similarly, we have 
\begin{align} \label{align:three_type_a21} a_{2,1}  = \sum_{j=1}^{M} \optupper_{2,j} \qquad \text{ and } \qquad a_{2,2} = \begin{cases} 
    \optupper_{2,2} & M = 1 \\
    0 & M = 2 
\end{cases} \,.\end{align} 
To see why, first consider the case of $M=1$. Then, the number of accepted  agents of type $2$ in the first time period, i.e., $a_{2,1}$, is equal to $\sum_{t=1}^{M} \optupper_{2, t}= \optupper_{2,1}$. (Recall that with $M=1$, type $2$ is impatient.) Similarly, the number of accepted  agents of type $2$ in the second time period, i.e., $a_{2,2}$, is equal to $\optupper_{2,2}$. Now, consider the case of $M=2$. The number of accepted  agents of type $2$ in the first time period, i.e., $a_{2,1}$, is equal to $\sum_{j=1}^{M} \optupper_{2,j}= \optupper_{2,1}+\optupper_{2,2}$. In addition, the number of accepted  agents of type $2$ in the second time period, i.e., $a_{2,2}$, is equal to $0$. This is because with $M=2$, under input $a$, no  agents of type $2$ arrive in the second time period. 

Since we assumed that $\mathcal{A}$ achieves at least a $\crr$ fraction of the optimal revenue in each time period for any input sequence $I_t, t \in [T]$, $\mathcal{A}$ must satisfy certain requirements against the 4 input sequences $a,b,c,d$ defined above. 

\textbf{Obtaining the CR of $\crr$ against input $a$.}
Under input $a$, we have that $\textsc{opt}_{\tau}\left( \{I_t^a\}_{t \in [2]} \right) = r_3 \cdot C$ for both $\tau \in [2]$ whereas $\mathcal{A}$ will serve $\optupper_{i,\tau}$ copies of  agent type $i$ for $i \in [3]$. Thus, in order to achieve the CR of $\crr$ in both time periods $t \in [2]$, we must have $\crr \cdot r_3 \cdot C \leq r_1 \cdot \optOnet + r_2 \cdot \optTwot + r_3 \cdot \optupper_{3,t}$,  for $t = 1, 2$. In addition, $\optOnet + \optTwot + \optupper_{3,t} \leq C$ must be true because of our capacity constraint. This leads to constraints \eqref{align:Ia_revenue} and \eqref{align:Ia_capacity} in the LP. 

\textbf{Obtaining the CR of $\crr$ against input $b$.}
This input is only useful when $M=1$. This is because with $M=2$, both inputs $a$ and $b$ lead to the same constraints. Assuming that $M=1$, we now use the fact that algorithm $\alg$ is non-anticipating. Since input $a$ and input $b$ are identical for $t = 1$, $\mathcal{A}$ makes the exact same decisions on both for $t = 1$, while for $t = 2$ we know that $\alg$ accepts $a_{2,2}$ type 2 agents on both inputs. In particular, this means that $\optupper_{1, 2}$ (the number of type $1$  agents that were accepted but not served in the first time period) must be the same under inputs $a$ and $b$. Similarly, $a_{2,2} = \optupper_{2,2}$ (the number of type $2$  agents that were served in the second time period) must be the same under inputs $a$ and $b$. Hence, the total reward in the second time period under input $b$ is $r_1 \cdot \optOneTwo + r_2 \cdot \optupper_{2,2}$. This implies that algorithm $\alg$ obtains the CR of $\crr$ under input $b$ if $\crr \cdot \textsc{opt}_2(\{I_t^b\}_{t\in[2]}) =  \crr \cdot C \cdot r_2 \le r_1 \cdot \optOneTwo + r_2 \cdot \optupper_{2,2}$, giving rise to constraint  \eqref{align:Ib_revenue} in the LP.

\textbf{Obtaining the CR of $\crr$ against input $c$.} Again using the fact that $\alg$ is non-anticipating, $a_{1,1}$ and $a_{2, 1}$ (the number of accepted  agents of type $1$ and $2$, respectively, in the first time period) must be the same under input $a$ and $c$. Now consider input $c$ and observe that under this input, we have $\textsc{opt}_1(\{I_t^c\}_{t \in [2]}) = C \cdot r_2$. This implies that in order to obtain the CR of $\crr$ under input $c$, there must be some way to choose $s_1 \leq a_{1,1}$ type $1$  agents (out of $a_{1, 1}$ type $1$  agents) and $s_2 \leq a_{2,1}$ type 2  agents (out of $a_{2, 1}$ type $2$  agents), such that $r_1 \cdot s_1 + r_2 \cdot s_2 \geq \crr \cdot r_2 \cdot C$ and $s_1 + s_2 \leq C$. We claim that a necessary condition for the existence of such $s_1, s_2$ are the following conditions
\begin{align}
    a_{1,1} \cdot r_1 + a_{2,1} \cdot r_2 \geq \crr \cdot r_2 \cdot C \label{eq:a11}\\
    (C - a_{2,1}) \cdot r_1 + a_{2,1} \cdot r_2 \geq \crr \cdot r_2 \cdot C\,.  \label{eq:a21}
\end{align}
We can see that the existence of $s_1$ and $s_2$ implies the above two conditions because 
\begin{align*}
a_{1,1} \cdot r_1 + a_{2,1} \cdot r_2 &\geq s_1 \cdot r_1 + s_2 \cdot r_2 \geq \crr \cdot r_2 \cdot C \\
(C - a_{2,1}) \cdot r_1 + a_{2,1} \cdot r_2 &\geq (C - s_2) \cdot r_1 + s_2 \cdot r_2 \geq s_1 \cdot r_1 + s_2 \cdot r_2 \geq \crr \cdot r_2 \cdot C
\end{align*}

Substituting  $a_{1,1} = \optupper_{1,1} + \optupper_{1,2}$ and $a_{2,1} = \sum_{t=1}^M \optupper_{2,t}$ in equations \eqref{eq:a11} and \eqref{eq:a21}, we get exactly constraints \eqref{align:Ic_revenue} and \eqref{align:Ic_capacity} in the LP.
 
\textbf{Obtaining the CR of $\crr$ against input $d$.}
Since $\alg$ is non-anticipating, its decisions, and in particular, $a_{1,1}$  (the number of accepted  agents of type $1$ in the first time period), must be the same under input $a$ and $d$. This implies that $\alg$ obtains the CR of $\crr$ under input $d$ if $r_1 \cdot a_{1,1 }\ge \crr \cdot \textsc{opt}_1(\{I_t^d\}_{t \in [2]}) = \crr \cdot C \cdot r_1 $, which is constraint  \eqref{align:Id_revenue} in the LP.
\end{proof} 

\subsection{Properties of the 3-Class Upper Bound}
We present many properties of the 3-class upper bound that will be crucial in proving the optimality of our 3-class polytopes $\feasible^{(1)}$ and $\feasible^{(2)}$. 

\subsubsection{Statement and Proof of Lemma \ref{lemma:3_class_tight_properties}}
\label{sec:lemma:3_class_tight_properties}
\begin{lemma}[Tight Constraints in the 3-Type Upper Bound \ref{3_class_ub}]
    \label{lemma:3_class_tight_properties}
    Let $(\crr_{up}, \{\optupper_{i,t}^*\}_{i \in [3], t \in [2]})$ be any optimal solution to the LP \ref{3_class_ub}, presented in Theorem \ref{thm:3_type_upper_bound}. Then the following five properties must be true:
    \begin{enumerate}[label=(\alph*)]
        \item \label{property:total_is_C} $s_{1,t}^* + s_{2,t}^* + s_{3,t} = C$ for $t \in [2]$ (i.e. constraint \eqref{align:Ia_capacity}) is tight).
        \item \label{property:type_1_cautious} $r_1 \cdot (\optupper_{1,1}^* + \optupper_{1,2}^* ) = \crr_{up} \cdot r_1 \cdot C$ (i.e. constraint \eqref{align:Id_revenue} is always tight).
        \item \label{property:type_2_cautious} $r_1 \cdot \min(\optupper_{1,1}^* + \optupper_{1,2}^*, C - \sum_{t=1}^M \optupper_{2,t}^*) + r_2 \cdot \sum_{t=1}^M \optupper_{2,t}^* = \crr_{up} \cdot r_2 \cdot C $ (i.e. at least one of constraints   \eqref{align:Ic_capacity} and \eqref{align:Ic_revenue} is tight).
        \item \label{property:type_2_caution_M=1} If $M = 1$, then $r_1 \cdot \optupper_{1,2} ^*+ r_2 \cdot \optupper_{2,2}^* = \crr_{up} \cdot r_2 \cdot C$ (i.e. constraint \eqref{align:Ib_revenue} is tight).
        \item \label{property:type_3_cautious} $\crr_{up} \cdot r_3 \cdot C = r_1 \cdot \optOnet^* + r_2 \cdot \optTwot^* + r_3 \cdot \optupper_{3,t}^*$ for $t \in [2]$ (i.e. constraint \eqref{align:Ia_revenue} is tight).
    \end{enumerate}
\end{lemma}
\begin{proof}{Proof of Lemma \ref{lemma:3_class_tight_properties}.}
\label{proof:3_class_tight_properties}
We will first show property \ref{property:total_is_C} holds for \textit{every} optimal solution to the LP \ref{3_class_ub}. Then, for each of properties
\ref{property:type_3_cautious}, \ref{property:type_1_cautious}, \ref{property:type_2_cautious} and \ref{property:type_2_caution_M=1}, we will argue by contradiction that for any optimal solution $(\crr_{up}, \{\optupper_{i,t}^*\}_{i \in [3], t \in [2]})$ for which the corresponding constraint is not tight, we can construct a new feasible solution $(\crr', \{\optupper_{i,t}'\}_{i \in [3], t \in [2]})$ such that $\crr_{up} = \crr'$, but $\optupper_{1,t} + \optupper_{2,t} + \optupper_{3,t} < C$ for some $t \in [2]$, which would imply that $\crr_{up}$ is not the optimal objective value by property \ref{property:total_is_C}.
\begin{enumerate}
    \item [\ref{property:total_is_C}] The main idea behind this proof is that if $s_{1,t}^* + s_{2,t}^* + s_{3,t}^* < C$ for some $t \in [2]$, then we can change the variables $\{s_{i,t}^*\}_{i \in [3], t \in [2]}$ in a way such that all constraints of the LP \ref{3_class_ub} involving $\crr_{up}$ are not tight, implying that $\crr_{up}$ can be increased and contradicting the fact that $\crr_{up}$ is optimal.
    
    Fix the value of $t$ such that $s_{1,t}^* + s_{2,t}^* + s_{3,t}^* + 2 \cdot \delta = C$ for some $\delta > 0$. We construct a new solution $\left( \crr_{up}, \{s_{i,t}'\}_{i \in [3], t \in [2]} \right)$ to \ref{3_class_ub} as follows: 
    
    $$ 
    \begin{array}{rlrlrl}
        s_{1,t}' &= s_{1,t}^* + \delta \quad&\quad s_{2,t}' &= s_{2,t}^* + \delta \quad&\quad s_{3,t}' &= s_{3,t}^*  \\
        s_{1,2-t}' &= s_{1,2-t}^* - \delta/2 \quad & \quad s_{1,t}' &= s_{1,t}^*  + \delta/2 \quad&\quad s_{3,2-t}' &= s_{3,2-t}^* 
    \end{array}
    $$
    
    We now need to check that this new solution $\left(\crr_{up}, \{s_{i,t}'\}_{i \in [3], t \in [2]} \right)$ is feasible and that all constraints involving $\crr_{up}$ are not tight. 
    
    \begin{itemize}
        \item Constraint \eqref{align:Ia_capacity}: By our construction, we have that $s_{1,t}' + s_{2,t}' + s_{3,t}' = C$ and $s_{1,2-t}' + s_{2,2-t}' + s_{3,2-t}' = s_{1,2-t}^* + s_{2,2-t}^* + s_{3,t}^* \leq C$.
        \item Constraint \eqref{align:Ia_revenue}: This constraint is clearly satisfied for $t$ as $s'_{i,t} \geq s_{i,t}^*$ for all $i \in [3]$. For $2 - t$, we have that $r_1 \cdot s_{1, 2-t}' + r_2 \cdot s'_{2, 2-t} + r_3 \cdot s'_{3, 2-t} = r_1 \cdot s_{1, 2-t}^* + r_2 \cdot s^*_{2, 2-t} + r_3 \cdot s^*_{3, 2-t} + \delta \cdot (r_2 - r_1)$ which is strictly greater than $\crr \cdot r_3 \cdot C$ by the fact that $\left(\crr_{up},\{s_{i,t}^*\}_{i\in[3], t \in [2]} \right)$ is feasible. Hence, this constraint is satisfied and not tight.
        \item Constraint \eqref{align:Id_revenue}: This is satisfied and not tight because $s'_{1,t} + s'_{1, 2-t} > s^*_{1,t} + s^*_{1,2-t}$.
        \item Constraint \eqref{align:Ic_capacity}: Notice that we can re-write this constraint in the LP as: 
        \begin{equation}
            \crr \cdot r_2 \cdot C \leq C \cdot r_1 + (r_2 - r_1) \cdot \big( \sum_{t=1}^M s_{2,t} \big)\,. \label{equation:rewritten_Ic_capacity}
        \end{equation}
        Our new solution $\left(\crr_{up}, \{s_{i,t}'\}_{i\in[3], t \in [2]} \right)$ is feasible and makes this constraint not tight because $\sum_{t=1}^M s_{2,t}' > \sum_{t=1}^M s_{2,t}^*$ for both $M = 1$ and $M = 2$. 
        \item Constraint \eqref{align:Ic_revenue}: By the same reasoning as in the argument of constraint \eqref{align:Ia_revenue}, we have that $r_1 \cdot s_{1,\tau}' + r_2 \cdot s_{2,\tau}' > r_1 \cdot s_{1,\tau}^* + r_2 \cdot s_{2,\tau}^*$ for both $\tau = t$ and $\tau = 2 - t$.  
        \item Constraint \eqref{align:Ib_revenue}: This is for the same reason as that of constraint \eqref{align:Ic_revenue}.
    \end{itemize}
    
    Therefore, we have proven that $\left(\crr_{up} , \{s_{i,t}'\}_{i \in [3], t \in [2]}\right)$ is feasible and that all constraints involving $\crr_{up}$ are not tight, which contradicts the optimality of $\crr_{up}$. Hence, \textit{all} optimal solutions must have the property that constraint \eqref{align:Ia_capacity} is tight, i.e. $s_{1,t} + s_{2,t} + s_{3,t} = C$. 
    
    \item[\ref{property:type_1_cautious}] Assume for contradiction that constraint \eqref{align:Id_revenue} is not tight; that is,  our optimal solution has $(\optupper_{1,1}^* + \optupper_{1,2}^* ) > \crr_{up} \cdot r_1 \cdot C$.  We then know that one of $\optupper_{1,t}^* > 0$ for some $t \in [2]$. Fix that particular value of $t$ and construct our new solution $(\crr', \{\optupper_{i,t}'\}_{i \in [3], t \in [2]})$ with the following modifications:
    $$ \optupper_{1,t}' = \optupper_{1,t}^* - \epsilon \qquad \optupper_{2,t}' = \optupper_{2,t}^* + \frac{r_1}{r_2} \cdot \epsilon\,,$$
    where we choose $\epsilon > 0$ sufficiently small such that $r_1 \cdot (\optupper_{1,1}' + \optupper_{1,2}') \geq \crr_{up} \cdot r_1 \cdot C$ (constraint \eqref{align:Id_revenue} is satisfied) and $\optupper_{1,1}' \geq 0$. It is easy to see that $s'_{1,t} + s'_{2,t} + s'_{3,t} < C$ now, which is what we want to for the contradiction. All we need to do now is verify that this new solution $(\crr', \{\optupper_{i,t}'\}_{i \in [3], t \in [2]})$ is feasible by checking the remaining constraints. To do so, we can group constraints \eqref{align:Ia_revenue}, \eqref{align:Ic_capacity}, \eqref{align:Ic_revenue} and  \eqref{align:Ib_revenue} into categories based on the coefficients of $\optupper_{1,t}$ and $\optupper_{2,t}$. 
    Note that all other constraints involving $\optupper_{1,t}$ and $\optupper_{2,t}$ are greater-than-or-equal-to (i.e., $\ge$) inequalities. 
    \begin{itemize}
        \item Constraints \eqref{align:Ia_revenue}, \eqref{align:Ic_revenue} and \eqref{align:Ib_revenue}: these constraints all involve the term $r_1 \cdot \optupper_{1,t} + r_2 \cdot \optupper_{2,t}$ on the right hand side of the $\leq$ inequality. Our new solution has the property that $$ r_1 \cdot \optupper_{1,t}' + r_2 \cdot \optupper_{2,t}' = r_1 \cdot \optupper_{1,t}^* + r_2 \cdot \optupper_{2,t}^*$$ and so all the constraints of this form remain satisfied. 
        \item Constraint \eqref{align:Ic_capacity}: Recall from equation \eqref{equation:rewritten_Ic_capacity} that this constraint can be re-written such that the right hand side of the $\leq$ includes only a term $p \cdot \sum_{\tau = 1}^M \optupper_{2,\tau}$ and our new solution has the property that $\sum_{\tau = 1}^M s_{2, \tau}' \geq \sum_{\tau = 1}^M s^*_{2, \tau}$, meaning that this constraint continues to be satisfied.
    \end{itemize}
    Thus, we have shown a contradiction by property \ref{property:total_is_C} of this Lemma and so it must be the case that every optimal solution has the property that constraint \eqref{align:Id_revenue} is tight.

    \item [\ref{property:type_2_cautious}] Suppose for contradiction that neither constraint  \eqref{align:Ic_capacity} nor  \eqref{align:Ic_revenue} is tight for our solution $(\crr_{up}, \{\optupper_{i,t}^*\}_{i \in [3], t \in [2]})$, meaning that 
    \begin{align}
        r_1 \cdot \left( \optupper_{1,1}^* + \optupper_{1,2}^* \right) + r_2 \cdot \sum_{t=1}^M \optupper^*_{2,t} > \crr_{up} \cdot r_2 \cdot C \label{first_non_binding} \\
        r_1 \cdot \left( C - \sum_{t=1}^M \optupper_{2, t}^* \right) + r_2 \cdot \sum_{t=1}^M \optupper^*_{2,t} > \crr_{up} \cdot r_2 \cdot C\,. \label{second_non_binding}
    \end{align}
    By property \ref{property:type_1_cautious}, we know that $r_1 \cdot (\optupper_{1,1}^* + \optupper_{1,2}^*) = \crr_{up} \cdot r_1 \cdot C < \crr_{up} \cdot r_2 \cdot C$, which means from inequality \eqref{first_non_binding} that $\sum_{t=1}^M \optupper^*_{2,t} > 0$ must be true. This implies that there exists $t \in [2]$ for which $\optupper^*_{2,t} > 0$. Fix this particular value of $t$ and consider a new solution $(\crr', \{\optupper_{i,t}'\}_{i \in [3], t \in [2]})$, where the modifications we make are as follows
    $$ \optupper_{2,t}' = \optupper_{2,t}^* - \epsilon \qquad \optupper_{3,t}' = \optupper^*_{3,t} + \epsilon \cdot \frac{r_2}{r_3}\,. $$
    
    Given that inequalities \eqref{first_non_binding} and \eqref{second_non_binding} are both not binding and $\optupper_{2,t}^*$ appears in both with a positive coefficient, we can find some $\epsilon > 0$ such that our new solution variables $\optupper_{2,t}'$ and $\optupper_{2,t}'$ also satisfy \eqref{first_non_binding} and \eqref{second_non_binding}. Furthermore, we can choose $\epsilon$ small enough such that $\optupper_{2,t}' \geq 0$. Now, we just need to verify that $(\crr', \{\optupper_{i,t}'\}_{i \in [3], t \in [2]})$ is feasible.
    \begin{itemize}
           \item Constraint \eqref{align:Ia_capacity} (i.e., $\optOnet + \optTwot + \optupper_{3,t} \leq C$). By our construction, we have $$\optupper_{1,t}' + \optupper_{2,t}' + \optupper_{3,t}' = \optupper_{1,t}^* + \optupper_{2,t}^* + \optupper_{3,t}^* - \epsilon \cdot \left(1 - \frac{r_2}{r_3} \right) < C \,. $$  
           
        \item Constraint \eqref{align:Ia_revenue} (i.e., $\crr \cdot r_3 \cdot C \leq \optOnet \cdot r_1 + \optTwot \cdot r_2 + \optupper_{3,t} \cdot r_3)$. By our construction, we have $$r_1 \cdot \optupper_{1,t}' + r_2 \cdot \optupper_{2,t}' + r_3 \cdot \optupper_{3,t}'= r_1 \cdot \optupper_{1,t}^* + r_2 \cdot \optupper_{2,t}^* + r_3 \cdot \optupper_{3,t}^*\,.$$ Thus,  this constraint continues to be satisfied. 
        
              \item Constraint \eqref{align:Id_revenue} (i.e., $\crr \cdot r_1 \cdot C \leq(\optOneOne + \optOneTwo) \cdot r_1$) and constraint  \eqref{align:Ib_revenue} (i.e., $\crr \cdot r_2 \cdot C \leq \optOneTwo \cdot r_1 + \optupper_{2,2} \cdot r_2$).   Constraint \eqref{align:Id_revenue} continues to be satisfied by our new solution because $\optupper_{1,1}^*= \optupper_{1,1}'$ and $\optupper_{1,2}^* = \optupper_{1,2}'$. A similar  argument can be applied for constraint  \eqref{align:Ib_revenue}.
 
        \item Constraints  \eqref{align:Ic_capacity} and \eqref{align:Ic_revenue}  are satisfied because we chose $\epsilon$ small enough so that these constraints continue to be satisfied.

    \end{itemize}
    By property \ref{property:total_is_C} of this lemma, we get the desired contradiction, proving that at least one of constraints \eqref{align:Ic_capacity} and \eqref{align:Ic_revenue} must be tight.
    
    \item[\ref{property:type_2_caution_M=1}] For $M = 1$, suppose for contradiction that constraint \eqref{align:Ib_revenue} (i.e., $\crr \cdot r_2 \cdot C \leq \optOneTwo \cdot r_1 + \optupper_{2,2} \cdot r_2$) is not tight for our solution $(\crr_{up}, \{\optupper_{i,t}^*\}_{i \in [3], t \in [2]})$. Similar to our proof for property  \ref{property:type_2_cautious}, we have that $\optupper_{2,2}^* > 0$. This is  because we know that $\optupper_{1,2}^* \leq \crr_{up}\cdot C$ by property \ref{property:type_1_cautious}, and as a result,  it would be impossible to have $r_1 \cdot \optupper_{1,2}^* + r_2 \cdot \optupper_{2,2}^* \geq \crr_{up} \cdot r_2 \cdot C$ without $\optupper_{2,2}^* > 0$. We create our new solution $(\crr', \{\optupper_{i,t}'\}_{i \in [3], t \in [2]})$ with the following modifications 
    $$ \optupper_{2,2}' = \optupper_{2,2}^* - \epsilon \qquad \optupper_{3,2}' = \optupper^*_{3,2} + \epsilon \cdot \frac{r_2}{r_3}\,,$$
    where we choose $\epsilon$ small enough so that our new solution continues to satisfy constraint \eqref{align:Ib_revenue} and such that $\optupper_{2,2}' \geq 0$. We check that this new solution is feasible and therefore optimal. For $M = 1$, the only constraints that involve $\optupper_{2,2}'$ and $\optupper_{3,2}'$ are  \eqref{align:Ia_capacity}, \eqref{align:Ia_revenue}, and \eqref{align:Ib_revenue}. We have already assumed constraint \eqref{align:Ib_revenue} is satisfied, so we just check the remaining two. With respect to constraint \eqref{align:Ia_revenue}  (i.e., $\crr \cdot r_3 \cdot C \leq \optOnet \cdot r_1 + \optTwot \cdot r_2 + \optupper_{3,t} \cdot r_3$), using similar reasoning as in part \ref{property:type_2_cautious}, we see that our construction preserves the value of $r_1 \cdot \optupper_{1,2}^* + r_2 \cdot \optupper_{2,2}^* + r_3 \cdot \optupper_{3,2}^*$ (that is, $r_1 \cdot \optupper_{1,2}^* + r_2 \cdot \optupper_{2,2}^* + r_3 \cdot \optupper_{3,2}^* = r_1 \cdot \optupper_{1,2}' + r_2 \cdot \optupper_{2,2}' + r_3 \cdot \optupper_{3,2}'$), and hence  constraint \eqref{align:Ia_revenue}  still holds. With respect to constraint   \eqref{align:Ia_capacity} (i.e., $\optOnet + \optTwot + \optupper_{3,t} \leq C$), 
     the value of $ \optupper_{1,2}' + \optupper_{2,2}' + \optupper_{3,2}'$ is slightly lower than $ \optupper_{1,2}^* + \optupper_{2,2}^* + \optupper_{3,2}^*$, implying 
    constraint \eqref{align:Ia_capacity} is not tight. This gives us the desired contradiction per property \ref{property:total_is_C} of this lemma.
    
        \item [\ref{property:type_3_cautious}]
  Contrary to our claim, suppose that   constraint \eqref{align:Ia_revenue} is not tight for our solution $(\crr_{up}, \{\optupper_{i,t}^*\}_{i \in [3], t \in [2]})$ for either $t = 1$ or $t = 2$. Fix that particular value of $t$. We claim that $\optupper_{3,t}^* > 0$ because of properties \ref{property:type_2_cautious} and \ref{property:type_2_caution_M=1}.
  From those properties, we get that $r_1 \cdot \optupper_{1,t}^* + r_2 \cdot \optupper_{2,t}^* \leq \crr_{up} \cdot r_2 \cdot C$ for both $t = 1$ and $t = 2$, which means that if we want $r_1 \cdot \optupper_{1,t}^* + r_2 \cdot \optupper_{2,t}^* + r_3 \cdot \optupper_{3,t}^* \geq \crr_{up} \cdot r_3 \cdot C$, then $\optupper_{3,t}^*$ must not be 0. 
    
    Consider a new solution $(\crr', \{\optupper_{i,t}'\}_{i \in [3], t \in [2]})$ where we make the following change 
    $$ \optupper_{3,t}' = \optupper_{3,t}^* - \epsilon\,,$$
    
    where we choose $\epsilon$ small enough such that our new solution continues to satisfy constraint \eqref{align:Ia_revenue} and $\optupper_{3,t}' \geq 0$. (The rest of the variables remain the same.) Our new solution clearly also satisfies constraint \eqref{align:Ia_capacity} (i.e., $\optOnet + \optTwot + \optupper_{3,t} \leq C$) since we have decreased $\optupper_{3,t}^*$ while leaving $\optupper^*_{1,t}$ and $\optupper_{2,t}^*$ unchanged. All other constraints from LP \ref{3_class_ub} continue to be satisfied because they do not involve $\optupper_{3,t}^*$. We get that our new solution is optimal, which gives us a contradiction again because $(\crr', \{\optupper_{i,t}'\}_{i \in [3], t \in [2]})$ is optimal yet does not not have constraint \eqref{align:Ia_capacity} tight. Thus, we get that constraint \eqref{align:Ia_revenue} must be tight for all optimal solutions.    

\end{enumerate} \hfill \Halmos
\end{proof}

\subsubsection{Statement and Proof of Lemma \ref{lemma:3_type_additional_properties}}
\label{sec:lemma:3_type_additional_properties}

The following lemma is used to show that the polytopes $\feasible^{(3, 1)}$ and $\feasible^{(3, 2)}$ are valid as per Definition \ref{def:valid}.

\begin{lemma}[Properties of 3-Type Upper Bound Solution]
\label{lemma:3_type_additional_properties}
Let  $(\crr_{up}, \{\optupper_{i,t}^*\}_{i \in [3], t \in [2]})$ be an optimal solution to \ref{3_class_ub}. Then the following three properties hold.
\begin{enumerate}[label=(\alph*)]
    \item \label{property:revenue_1_and_2} For both $M = 1$ and $M = 2$, every optimal solution must have the property that 
    $$ \Delta_{3,1} \cdot \optupper_{1,t}^* + \Delta_{3,2} \cdot \optupper_{2,t}^* = (1-\crr_{up}) \cdot r_3 \cdot C\,. $$
    \item \label{property:consistent_M=1} For $M = 1$, there exists an optimal solution  such that  $\optOneOne^* \geq \optupper_{1,2}^*\,,$ and  
    \item \label{property:consistent_M=2} For $M = 2$, it is always the case that $C - (\optupper_{2,1}^* + \optupper_{2,2}^*) \geq \frac{1}{2} \left(
    \optupper_{1,1}^* + \optupper_{1,2}^*\right)\,.$
\end{enumerate} 
Here, $\Delta_{i,j} \defeq r_i - r_j$ for any $i, j\in [3]$, $i\ne j$.
\end{lemma}

\begin{proof}{Proof of Lemma \ref{lemma:3_type_additional_properties}.}
\label{proof:3_type_additional_properties} We show each of the three properties separately. 
{\color{white} a}
\begin{enumerate}
    \item [\ref{property:revenue_1_and_2}] 
    By properties \ref{property:total_is_C} and \ref{property:type_3_cautious} of Lemma \ref{lemma:3_class_tight_properties}, we have $\optupper_{1,t}^* + \optupper_{2,t}^* + \optupper_{3,t}^* = C$ for $t\in[2]$ and $\crr_{up} \cdot r_3 \cdot C = r_1 \cdot \optOnet^* + r_2 \cdot \optTwot^* + r_3 \cdot \optupper_{3,t}^*$ for $t \in [2]$. We then immediately get the desired result by multiplying the former by $r_3$ and then subtracting the two equations.
    
    \item [\ref{property:consistent_M=1}] Contrary to our claim, suppose  that $\optOneOne^* < \optupper_{1,2}^*$. Then, we present another optimal solution $(\crr_{up}, \{ \optupper_{i,t}' \}_{i \in [3], t \in [2]})$ for which $\optOneOne' \geq \optupper_{1,2}'$. We construct our new solution $(\crr', \{ \optupper_{i,t}' \}_{i \in [3], t \in [2]})$ to be identical to our old solution $(\crr_{up}, \{ \optupper_{i,t}^*\}_{i \in [3], t \in [2]})$ except with the following changes: 
    $$ \optupper'_{1,t} = \optupper^*_{1,3-t} \qquad \optupper'_{2,t} = \optupper^*_{2,3-t} \qquad \optupper'_{3,1} = \optupper^*_{3,3-t}\quad {t\in [2]}\,.  $$
    Our new solution is constructed by taking the old solution and swapping the first and second rows.. It is clear that $\optOneOne' \geq \optupper_{1,2}'$ now, so we just need to verify feasibility of the new solution. 
    
    Clearly, constraints \eqref{align:Ia_capacity} (i.e., $\optOnet + \optTwot + \optupper_{3,t} \leq C$) and \eqref{align:Ia_revenue} (i.e., $\crr \cdot r_3 \cdot C \leq \optOnet \cdot r_1 + \optTwot \cdot r_2 + \optupper_{3,t} \cdot r_3$) are still satisfied, as these constraints were valid for both $t = 1$ and $t = 2$. In addition, constraint \eqref{align:Id_revenue} (i.e., $\crr \cdot r_1 \cdot C \leq(\optOneOne + \optOneTwo) \cdot r_1$) is still satisfied since it is symmetric between $t = 1$ and $t = 2$. We just need to verify constraints \eqref{align:Ic_capacity}, \eqref{align:Ic_revenue}, and  \eqref{align:Ib_revenue}. For constraint \eqref{align:Ic_capacity}   (i.e., $\crr \cdot r_2 \cdot C \leq \left(C -  \optTwoOne \right) \cdot r_1 +   \optTwoOne \cdot r_2$) and constraint  \eqref{align:Ic_revenue} (i.e.,   
   $\crr \cdot r_2 \cdot C \leq \left(\optOneOne + \optOneTwo \right) \cdot r_1 +  \optTwoOne \cdot r_2 $), we verify them together as follows 
    \begin{align*}
        r_1 \cdot \min(\optupper_{1,1}' + \optupper_{1,2}', C - \optupper_{2,1}') + r_2 \cdot \optupper_{2,1}' &\geq r_1 \cdot \optupper_{1,1}' + r_2 \cdot \optupper_{2,1}' & \text{since $C - \optupper_{2,1}' \ge \optupper_{1,1}'$ by constraint \eqref{align:Ia_capacity}} \\
        &= r_1 \cdot \optupper_{1,2}^* + r_2 \cdot \optupper_{2,2}^* \\
        &\geq \crr_{up} \cdot r_2 \cdot C & \text{by constraint \eqref{align:Ib_revenue}} \,.
    \end{align*}
    
    Finally, for constraint \eqref{align:Ib_revenue} (i.e., $\crr \cdot r_2 \cdot C \leq \optOneTwo \cdot r_1 + \optupper_{2,2} \cdot r_2$), we
   use property \ref{property:revenue_1_and_2} of this lemma which says that $  \Delta_{3,1} \cdot \optOneTwo^* + \Delta_{3,2} \cdot \optTwoTwo^*=\Delta_{3,1} \cdot \optOneOne^* + \Delta_{3,2} \cdot \optTwoOne^*$.  By our definitions, we then have  $\Delta_{3,1} \cdot \optupper_{1,1}' + \Delta_{3,2} \cdot \optupper_{2,1}' = \Delta_{3,1} \cdot \optupper_{1,2}' + \Delta_{3,2} \cdot \optupper_{2,2}'$. Given that $\optupper_{1,1}' \geq \optupper_{1,2}'$, this equality implies that $\optupper_{2,1}' \leq \optupper_{2,2}'$ Taking this equality (i.e., $\Delta_{3,1} \cdot \optupper_{1,1}' + \Delta_{3,2} \cdot \optupper_{2,1}' = \Delta_{3,1} \cdot \optupper_{1,2}' + \Delta_{3,2} \cdot \optupper_{2,2}'$), multiply through by $\frac{r_1}{\Delta_{3,1}}$ and then adding/subtracting a term on both sides, we get
    \begin{equation*}
        r_1 \cdot \optupper_{1,1}' + r_2 \cdot \optupper_{2,1}' +  \left(r_1 \cdot \frac{\Delta_{3,2}}{\Delta_{3,1}} - r_2 \right) \cdot \optupper_{2,1}'= r_1 \cdot \optupper_{1,2}' + r_2 \cdot \optupper_{2,2}' +  \left(r_1 \cdot \frac{\Delta_{3,2}}{\Delta_{3,1}} - r_2 \right) \cdot \optupper_{2,2}'\,.
    \end{equation*}
    We notice that $\left( r_1 \cdot \frac{\Delta_{3,2}}{\Delta_{3,1}} - r_2 \right)$ is negative.\footnote{This simplifies to $r_1 \cdot r_3 - r_1 \cdot r_2 < r_2 \cdot r_3 - r_1 \cdot r_2$ implying that  $r_1 < r_2$, which is true.} This  along with the fact that $\optupper_{2,1}' \leq \optupper_{2,2}'$, we get that 
    \begin{equation*}
        r_1 \cdot \optupper_{1,1}' + r_2 \cdot \optupper_{2,1}' \leq r_1 \cdot \optupper_{1,2}' + r_2 \cdot \optupper_{2,2}'\,.  
    \end{equation*}
    
    We have already shown in the verification of constraints \eqref{align:Ic_capacity} and \eqref{align:Ic_revenue} that $r_1 \cdot \optupper_{1,1}' + r_2 \cdot \optupper_{2,1}' \geq \crr_{up} \cdot r_2 \cdot C$ and so we get the desired result for constraint \eqref{align:Ib_revenue} that $r_1 \cdot \optupper_{1,2}' + r_2 \cdot \optupper_{2,2}' \geq \crr_{up} \cdot r_2 \cdot C$. Thus, we have shown that this new   solution $(\crr', \{\optupper_{i,t}'\}_{i\in[3], t \in [2]})$ is feasible, optimal, and has the desired property. 
    
    \item [\ref{property:consistent_M=2}] Our goal is to show that $C - (\optupper_{2,1}^* + \optupper_{2,2}^*) \geq \frac{1}{2} \left(
    \optupper_{1,1}^* + \optupper_{1,2}^*\right)$ is always true.
    To do so, we will prove a lower bound on $C - (\optupper_{2,1}^* + \optupper_{2,2}^*)$ and an upper bound on $\frac{1}{2} \left( \optupper_{1,1}^* + \optupper_{1,2}^* \right)$. We show that the lower bound on $C - (\optupper_{2,1}^* + \optupper_{2,2}^*)$ is greater and equal to the upper bound on $\frac{1}{2} \left( \optupper_{1,1}^* + \optupper_{1,2}^* \right)$. This gives us the desired result.
    
 \textit{Lower bound on $C - (\optupper_{2,1}^* + \optupper_{2,2}^*)$.}  Here, we show that
 $C - (\optupper_{2,1}^* + \optupper_{2,2}^*) \geq  \frac{(1-\crr_{up}) r_2}{r_2 - r_1} \cdot C$. To present this upper bound, we give a lower bound on $(\optupper_{2,1}^* + \optupper_{2,2}^*)$ using properties  \ref{property:type_1_cautious} and \ref{property:type_2_cautious} of Lemma \ref{lemma:3_class_tight_properties} which state that
    \begin{align*}
        r_1 \cdot (\optupper_{1,1}^* + \optupper_{1,2}^*) &= \crr_{up} \cdot r_1 \cdot C \quad \text{Property \ref{property:type_1_cautious} of Lemma \ref{lemma:3_class_tight_properties}}\\
        r_1 \cdot \min(\optupper_{1,1}^* + \optupper_{1,2}^*, C - \optupper_{2,1}^* - \optupper_{2,2}^* ) + r_2 \cdot (\optupper_{2,1}^* + \optupper_{2,2}^* ) &= \crr_{up} \cdot r_2 \cdot C \quad \text{Property \ref{property:type_2_cautious} of Lemma \ref{lemma:3_class_tight_properties}} \,.
    \end{align*}
    The first equation clearly gives us $\optupper_{1,1}^* + \optupper_{1,2}^* = \crr_{up} \cdot C$. We plug this into the second equation. For the second equation, we perform case-work based on which term of the minimum is taken and solve for $\optupper_{2,1}^* + \optupper_{2,2}^*$ to get:
    \begin{equation*}
        \optupper_{2,1}^* + \optupper_{2,2}^* = \begin{cases}
            \crr_{up} \cdot \left( 1 - \frac{r_1}{r_2} \right) \cdot C & \crr_{up} \cdot \left( 1 - \frac{r_1}{r_2} \right) + \crr_{up} \leq 1 \\
            \frac{\crr_{up} \cdot r_2 - r_1}{r_2 - r_1} \cdot C & \text{otherwise}
        \end{cases}
    \end{equation*}
    We claim that $\optupper_{2,1}^* + \optupper_{2,2}^* \leq \frac{\crr_{up} \cdot r_2 - r_1}{r_2 - r_1} \cdot C$ is always true.  
    This is equivalent to showing $ \crr_{up} \cdot \left( \frac{r_2 - r_1}{r_2} \right) \cdot C \leq \frac{\crr_{up} \cdot r_2 - r_1}{r_2 - r_1} \cdot C$ when $\crr_{up} \cdot \left( 1 - \frac{r_1}{r_2} \right) + \crr_{up} \leq 1$. 
    This is easy to show with some algebra.\footnote{ Note that $ \crr_{up} \cdot \left( \frac{r_2 - r_1}{r_2} \right) \cdot C \leq \frac{\crr_{up} \cdot r_2 - r_1}{r_2 - r_1} \cdot C$ holds if and only if  the following inequality holds 
$       \crr_{up} (r_2^2 - 2 r_2 r_1 + r_1^2) \leq \crr_{up} r_2^2 - r_1r_2 \quad \Leftrightarrow \quad \crr_{up} \leq \frac{1}{2 - \frac{r_1}{r_2}}$
        which is equivalent to the condition $\crr_{up} \cdot \left( 1 - \frac{r_1}{r_2}\right) + \crr_{up} \leq 1$.} Therefore, we have the following lower bound on $C - (\optupper_{2,1}^* + \optupper_{2,2}^*)$, which is the desired result:
    \begin{equation}
    \label{left_hand_side_of_property}
    C - (\optupper_{2,1}^* + \optupper_{2,2}^*) \geq C - \frac{\crr_{up} \cdot r_2 - r_1}{r_2 - r_1} \cdot C = \frac{(1-\crr_{up}) r_2}{r_2 - r_1} \cdot C \,.    \end{equation} 

    \textit{Upper bound on $\frac{1}{2} \cdot \left( \optupper_{1,1}^* + \optupper_{1,2}^* \right)$.} For the upper bound, we have:
    \begin{equation}
    \label{right_hand_side_of_property} \frac{1}{2} \cdot \left( \optupper_{1,1}^* + \optupper_{1,2}^*\right) \leq \frac{1}{2} \cdot \left( \frac{(1-\crr_{up}) \cdot r_3 \cdot C}{r_3 - r_1} + \frac{(1-\crr_{up}) \cdot r_3 \cdot C}{r_3 - r_1} \right) = \frac{(1-\crr_{up}) \cdot r_3 \cdot C}{r_3 - r_1}\,,
    \end{equation}
    where the first inequality is true by property \ref{property:revenue_1_and_2} of this lemma, where we show $ \Delta_{3,1} \cdot \optOneTwo^* + \Delta_{3,2} \cdot \optTwoTwo^* = \Delta_{3,1} \cdot \optOneOne^* + \Delta_{3,2} \cdot \optTwoOne^*= (1-\crr_{up})\cdot C\cdot r_3 $ and hence $s_{1,t}^* \leq \frac{(1 - \crr_{up}) \cdot C \cdot r_3}{\Delta_{3,1}}$ for $t \in [2]$. 
    
   \textit{Comparing the bounds.} Observe the lower bound on  $C - (\optupper_{2,1}^* + \optupper_{2,2}^*)$, presented in  \eqref{left_hand_side_of_property}, is smaller than   the upper bound on $\frac{1}{2} \cdot \left( \optupper_{1,1}^* + \optupper_{1,2}^* \right)$, presented in  \eqref{right_hand_side_of_property}. That is, we have
    $$ \frac{r_2}{r_2 - r_1} \geq \frac{r_3}{r_3 - r_1} \quad \Leftrightarrow \quad -r_1 \cdot r_2 \geq -r_1 \cdot r_3 \quad \Leftrightarrow \quad r_2 \leq r_3\,.$$
    This proves the desired result that $C - (\optupper_{2,1}^* + \optupper_{2,2}^*) \geq \frac{1}{2} \left( \optupper_{1,1}^* + \optupper_{1,2}^*\right)$.
\end{enumerate}
\end{proof}

\subsection{Proof of Statements about the 3-Class Polytopes}
\subsubsection{Statement and Proof of Lemma \ref{lemma:3_type_valid}}
\label{sec:lemma:3_type_valid}
\begin{lemma}[Validity of $\feasibleMOne$] \label{lemma:3_type_valid}
The polytope $\feasibleMOne$ is \consistent{} as per Definition \ref{def:valid}.
\end{lemma}

\begin{proof}{Proof of Lemma \ref{lemma:3_type_valid}.} Recall that 
\begin{align*}  \feasibleMOne = \left\{ (\fbucketOneOne, \fbucketOneTwo, \fbucketTwo, \fbucketThree) \subset \mathbb{R}_+^4: \quad  \begin{array}{cll}
\fbucketOneOne + \fbucketTwo + \fbucketThree &\leq s_{1,1}^* + s_{2,1}^* + s_{3,1}^* & \\
\Delta_{3,1} \cdot \fbucketOneOne + \Delta_{3,2} \cdot \fbucketTwo &\leq \Delta_{3,1} \cdot s_{1,1}^* + \Delta_{3,2} \cdot s_{2,1}^* & \\
\fbucketOnei &\leq \optupper_{1,i}^* & i \in [2] \end{array} \right\}\end{align*}
We verify the two properties of valid polytopes from Definition \ref{def:valid}.

Property (i) is clearly satisfied  because of the constraint $\fbucketOneOne + \fbucketTwo + \fbucketThree \leq s_{1,1}^* + s_{2,1}^* + s_{3,1}^* \leq C$. 
For property (ii), we must show that for any value of $b_{1,2}$ that if such a value were moved to $b_{1,1}$, then no constraints involving $b_{1,1}$ would be broken (assuming $b_2=b_3=0$). The upper bound on $b_{2,1}$ is $s^*_{1,2}$ and so proving this lemma involves showing that (1) $s^*_{1,2} \leq s^*_{1,1}$ and (2) $\Delta_{3,1} \cdot s^*_{1,2}  + \Delta_{3,2} \cdot b_2  \leq (1-\crr_{up}) \cdot r_3 \cdot C$. and (3) $s^*_{1,2} + b_2 + b_3 \leq C$. 
The first condition is a result of property \ref{property:consistent_M=1} of Lemma \ref{lemma:3_type_additional_properties}, the second condition is true because of property \ref{property:revenue_1_and_2} of Lemma \ref{lemma:3_type_additional_properties}, while the third condition is by constraint \eqref{align:Ia_capacity} from \ref{3_class_ub}.

Since $\feasibleMOne$ is not a nested polytope, we also need to show that it is downwards closed as per Definition \ref{def:downward_closed}. This is true because all the constraints limit a positive linear combination of $b_{1,1}, b_{1,2}, b_2$ and $b_3$ to be no more than some constant.
\end{proof}

\subsubsection{Proof of Theorem \ref{thm:opt_alg_M=1}}
\begin{proof}{Proof of Theorem \ref{thm:opt_alg_M=1}.} 
Throughout the proof, we take advantages of properties of the optimal solution to LP  \ref{3_class_ub}, which are presented in a series of lemmas (Lemmas \ref{lemma:3_class_tight_properties} and \ref{lemma:3_type_additional_properties}). 
Per property \ref{property:total_is_C} of Lemma \ref{lemma:3_class_tight_properties}, any optimal solution $(\crr_{up}, \{\optupper_{i,t}^*\}_{i \in [3], t \in [2]})$ to LP \ref{3_class_ub} must have the property that constraint \eqref{align:Ia_capacity} is tight for both $t = 1$ and $t = 2$. That is, $\optupper_{1,t}^* + \optupper_{2,t}^* + \optupper_{3,t}^* = C$, $t\in [2]$. Hence, 
the constraints of $\feasibleMOne$ can be written as follows:
\begin{align}
   \fbucketOneOne + \fbucketTwo + \fbucketThree & \leq C \label{feas:M=1_capacity}\\
    \Delta_{3,1} \cdot \fbucketOneOne + \Delta_{3,2} \cdot \fbucketTwo & \leq (1- \crr_{up}) \cdot r_3 \cdot C \label{feas:M=1_type12} \\
   \fbucketOnei & \leq \optupper_{1,i}^* & i \in [2] \label{feas:M=1_type1}
\end{align}

Consider an arbitrary input sequence $\{I_t\}_{t \in [T]}$ and a particular time period $\tau \in [T]$. Our goal is to show: 
\begin{equation}
    \frac{\text{Rew}_{\feasibleMOne, \tau}(\{I_t\}_{t \in [T]})}{\textsc{opt}_{\tau}(\{I_t\}_{t \in [T]})}\geq \crr_{up}\,, \label{equation:3_class_goal}
\end{equation} 
Here, with a slight abuse of notation, $\text{Rew}_{\feasibleMOne, \tau}(\{I_t\}_{t \in [T]})$ is the reward of the \po{}  algorithm with feasible polytope $\feasibleMOne$ in time period $\tau $. Recall that in our model, reward is earned whenever an agent is served as opposed to accepted, implying that all of $\text{Rew}_{\feasibleMOne, \tau}(\{I_t\}_{t \in [T]})$ is earned by agents served in time period $\tau$. We focus on $I_\tau$ and define the following notation for our analysis:
\begin{itemize}
    \item Let $\numinput{1}, \numinput{2}, \numinput{3}$ be the number of type $1$, $2$ and $3$ agents appearing in $I_\tau$ respectively. 
    \item Let $\numopt{1}, \numopt{2}, \numopt{3}$ be the number of type $1$, $2$ and $3$ agents respectively that the optimal offline benchmark serves out of $I_\tau$ (which cannot utilize partially patient agents). In other words, we have 
    \begin{equation}
        \textsc{opt}_{\tau}(\{I_t\}_{t \in [T]}) = r_1 \cdot \numopt{1} + r_2 \cdot \numopt{2} + r_3 \cdot \numopt{3}\,, \label{equation:opt_revenue}
    \end{equation}
   where $y_i \leq \numinput{i}$ for $i \in [3]$. 
    \item Let $\stateB = (\bucketOneOne, \bucketTwo, \bucketThree, \bucketOneTwo)$ be the final number of agents assigned to each bucket at the end of the input sequence $I_\tau$. By definition of our \po{} algorithm and by Lemma \ref{lemma:3_type_valid}, we know that $\stateB \in \feasibleMOne$. (Lemma \ref{lemma:3_type_valid} shows that $\feasibleMOne$ is a valid polytope.)
\end{itemize}

We show the result based on the following case-work, where in case 1, we assume that an incoming type $3$ agent would be rejected. That is, an incoming  type $3$ agent could not be fitted to the final state $\stateB$; see Section \ref{sec:polyra} for details. In case 2, we assume that an incoming type $2$ agent would be rejected while an incoming type $3$ agent would be accepted, and so on. 

\textbf{Case 1: Incoming type $3$ {would be} rejected.} 
If an incoming type $3$ agent would be rejected, then it must be the case $\bucketOneOne + \bucketTwo + \bucketThree = C$ because if we had $\bucketOneOne + \bucketTwo + \bucketThree < C$, then it would be possible to add a non-zero amount of an incoming type $3$ agent to the bucket  $\bucketThree$ while still satisfying all the constraints of $\feasibleMOne$. Given that $\bucketOneOne + \bucketTwo + \bucketThree = C$, we know that our serving rule will simply serve all agents in the first row. Therefore, we can write the reward of our algorithm as:
\begin{align*}
    \text{Rew}_{\feasibleMOne, \tau} (\{I_t\}_{t \in [T]}) &= r_1 \cdot \bucketOneOne + r_2 \cdot \bucketTwo + r_3 \cdot \bucketThree \\
    &= r_1 \cdot \bucketOneOne + r_2 \cdot \bucketTwo + r_3 \cdot (C - \bucketOneOne - \bucketTwo) \\    
    &= r_3 \cdot C - \Delta_{3,1} \cdot \bucketOneOne - \Delta_{3,2} \cdot \bucketTwo & \text{by definition of $\Delta_{i,j}$} \\
    &\geq r_3 \cdot C - (1- \crr_{up}) \cdot r_3 \cdot C  & \text{by the second constraint in $\feasibleMOne$, i.e.,  \eqref{feas:M=1_type12}} \\
    &= \crr_{up} \cdot r_3 \cdot C \\
    &\geq \crr_{up} \cdot \textsc{opt}_{\tau}(\{I_t\}_{t \in [T]})\,.
\end{align*}
The final inequality is true because the most reward the optimal algorithm  can achieve in a single time period is $C \cdot r_3$. The last inequality is the desired result. Note that the only assumption we really needed to make this case work is $\bucketOneOne + \bucketTwo + \bucketThree = C$, for which an incoming type $3$ agent being rejected is a sufficient condition. Thus, for the remainder of this proof, we assume that $\bucketOneOne + \bucketTwo + \bucketThree < C$.

\textbf{Case 2: Incoming type $3$ would be accepted, but an incoming type $2$ would be rejected.} The fact that an incoming type $3$ agent at the end of $I_\tau$ could be accepted means that no type $3$ agents in $I_\tau$ were rejected at all. By the design of our polytope $\feasibleMOne$, we see that if a constraint involving a particular bucket is tight, then that constraint will remain tight as the variables $\bucketOneOne, \bucketOneTwo, \bucketTwo$ and $\bucketThree$ only increase as we accept more agents. This observation gives us the important relationship, that $\bucketThree = \numinput{3}$. (Recall $\numinput{i}$ is the number of type $i$ agents appearing in $I_{\tau}$) as all type $3$ agents from $I_\tau$ have been accepted into $\bucketThree$. This along with the fact that $\numinput{3} \geq \numopt{3}$ imply the following upper bound on the optimal revenue:
\begin{align}
    \textsc{opt}_{\tau}(\{I_t\}_{t \in [T]}) &= r_1 \cdot \numopt{1} + r_2 \cdot \numopt{2} + r_3 \cdot \numopt{3} \nonumber \\
    &\leq r_2 \cdot (C - \numopt{3}) + r_3 \cdot \numopt{3} & \text{since $\numopt{1} + \numopt{2} + \numopt{3} \leq C$} \\
    &\leq r_2 \cdot (C - \bucketThree) + r_3 \cdot \bucketThree & \text{since } \numopt{3} \leq \numinput{3} = B_3  \,. \label{equation:type_2_rejected_opt}
\end{align} 
(Recall that $y_i$, $i\in [3]$, is the number of type $i$ that the optimal offline benchmark serves out of $I_\tau$.)  
The fact that an incoming agent of type $2$ would be rejected implies that some constraint in $\feasibleMOne$ involving $\bucketTwo$ holds with equality. Since we already assumed that $\bucketOneOne + \bucketTwo + \bucketThree < C$, the only other possible constraint is \eqref{feas:M=1_type12}, implying that \begin{equation}
    \Delta_{3,1} \cdot \bucketOneOne + \Delta_{3,2} \cdot \bucketTwo = (1- \crr_{up}) \cdot r_3 \cdot C \,. \label{equation:type_2_equation_tight}
\end{equation} 
With this equation in hand, our goal will be to show the following inequality
\begin{equation}
   \label{equation:type_2_rejected_desired}
\text{Rew}_{\feasibleMOne, \tau}(\{I_t\}_{t \in [T]}) \geq  \crr_{up} \cdot r_2 \cdot C + \Delta_{3,2}  \cdot \bucketThree\,.
\end{equation} 
Showing this inequality completes the proof for this case because the CR can be bounded as follows:
    \begin{align*}
        \frac{\text{Rew}_{\feasibleMOne, \tau}(\{I_t\}_{t \in [T]})}{\textsc{opt}_{\tau}(\{I_t\}_{t \in [T]})} &\geq \frac{\crr_{up} \cdot r_2 \cdot C  + \Delta_{3,2} \cdot \bucketThree}{r_1 \cdot \numopt{1} + r_2 \cdot \numopt{2} + r_3 \cdot \bucketThree} \qquad\text{by inequality 
        \eqref{equation:type_2_rejected_desired}}
        \\
        &\geq \frac{\crr_{up} \cdot r_2 \cdot C  + \Delta_{3,2} \cdot \bucketThree}{r_2 \cdot (C -\bucketThree) + r_3 \cdot \bucketThree} \qquad\text{by inequality  \eqref{equation:type_2_rejected_opt}}\\
        &= \frac{\crr_{up} \cdot r_2 \cdot C  + \Delta_{3,2} \cdot \bucketThree}{r_2 \cdot C + \Delta_{3,2} \cdot \bucketThree} \\
        &\geq \frac{\crr_{up} \cdot r_2 \cdot C} {r_2 \cdot C }\\
        &= \crr_{up} \,.
    \end{align*}
    where the last inequality is true because of the fact that if $a < b$ and $c > d$, then $\frac{a + c}{b+ d} > \frac{a}{b}$. The last equality is the desired result. 

To show inequality \eqref{equation:type_2_rejected_desired}, we consider the following two cases: (a) $\bucketOneOne = 0$, and (b) $\bucketOneOne > 0$.
\begin{itemize}
\item \textbf{Case 2a: $\bucketOneOne = 0$.} 
Using equation \eqref{equation:type_2_equation_tight}, we get that $B_2 = \frac{1}{\Delta_{3,2}} \cdot (1 - \crr_{up}) \cdot r_3 \cdot C$. By property \ref{property:revenue_1_and_2} of Lemma \ref{lemma:3_type_additional_properties}, we have $\Delta_{3,1} \cdot \optupper_{1,t}^* + \Delta_{3,2} \cdot \optupper_{2,t}^* = (1-\crr_{up}) \cdot r_3 \cdot C$, which gives us 
\begin{align}
B_2 = \frac{1}{\Delta_{3,2}} \cdot \left(\Delta_{3,1} \cdot \optOneOne^* + \Delta_{3,2} \cdot \optupper_{2,1}^*\right) \geq \optOneOne^* + \optupper_{2,1}^*\,. \label{eq:B2_1}
\end{align}
A lower bound on $\text{Rew}_{\feasibleMOne, \tau}(\{I_t\}_{t \in [T]})$ can be achieved by just considering the revenue from $\bucketOneOne, \bucketTwo$, and $\bucketThree$, which we are guaranteed to serve.
\begin{align*}
        \text{Rew}_{\feasibleMOne, \tau} (\{I_t\}_{t \in [T]}) &\geq r_1 \cdot \bucketOneOne + r_2 \cdot \bucketTwo + r_3 \cdot \bucketThree \nonumber \\
        &\geq r_1 \cdot 0 + r_2 \cdot (\optOneOne^* + \optupper_{2,1}^*) + r_3 \cdot B_3 & \text{by inequality \eqref{eq:B2_1}}\\
        &\ge r_1 \cdot \optOneOne^* + r_2 \cdot \optupper_{2,1}^* + r_3 \cdot B_3 \\
        &\geq \crr_{up} \cdot r_2 \cdot C + r_3 \cdot B_3& \text{by constraint \eqref{align:Ib_revenue} in LP \ref{3_class_ub}} \\
        &\geq \crr_{up} \cdot r_2 \cdot C + \Delta_{3,2} \cdot B_3\,.
    \end{align*}
The last inequality is the desired result (see inequality \eqref{equation:type_2_rejected_desired}). 
\item \textbf{Case 2b: $\bucketOneOne > 0$.} 
For this case, we will take advantage of the fact that in our \po{} algorithms, we prioritize assigning incoming partially patient agents to buckets in the second row over the first row. In our three type problem with $M = 1$, we know that if $\bucketOneOne > 0$, then it must mean that $\bucketOneTwo$ is full. Applying this reasoning to this particular case gives us the assumption that $\bucketOneTwo = \optOneTwo^*$.  

    Recall that all agents in the first row (i.e., $\bucketOneOne, \bucketTwo, \bucketThree$) have already been allocated the resource and we use the leftover $C - \bucketOneOne - \bucketTwo - \bucketThree$ resources to serve as many agents in $\bucketOneTwo$ as possible. Therefore, we can write our reward as follows
    \begin{equation}
        \text{Rew}_{\feasibleMOne, \tau} (\{I_t\}_{t \in [T]}) = r_1 \cdot \bucketOneOne + r_1 \cdot \min \left(\bucketOneTwo, C - \bucketOneOne - \bucketTwo - \bucketThree \right) + r_2 \cdot \bucketTwo + r_3 \cdot \bucketThree \, .
        \label{equation:thereward}
    \end{equation}
    With respect to ``$\min$" in the above equation note that  $\bucketOneTwo$ is the number of type $1$ agents remaining after serving all the agents in the first row, while $C - \bucketOneOne - \bucketTwo - \bucketThree$ is the capacity remaining. We will now show that choosing $\bucketOneOne = \optupper_{1,1}^*$ and $\bucketTwo = \optupper_{2,1}^*$ gives an upper bound on equation \eqref{equation:thereward}. To see this, recall the following formula which relates $\bucketOneOne$ and $\bucketTwo$, which is obtained by solving for $\bucketTwo$ in terms of $\bucketOneOne$ in equation \eqref{equation:type_2_equation_tight}:
     \begin{align}\label{eq:B2}\bucketTwo = \frac{1}{\Delta_{3,2}} \big((1 - \crr_{up}) \cdot r_3 \cdot C_3 - \Delta_{3,1} \cdot \bucketOneOne \big)\,.\end{align}
     
     We plug this value of $\bucketTwo$ into equation \eqref{equation:thereward} and we consider how the resulting equation changes with $\bucketOneOne$. The coefficient of $\bucketOneOne$ depends
     whether $\min \left(\bucketOneTwo, C - \bucketOneOne - \bucketTwo - \bucketThree \right)$ is $\bucketOneTwo$ or $ C - \bucketOneOne - \bucketTwo - \bucketThree$:
     \begin{itemize}
         \item If $\bucketOneTwo \leq C - \bucketOneOne - \bucketTwo - \bucketThree $, by replacing $\bucketTwo$ by the expression in equation \eqref{eq:B2},   the coefficient of $\bucketOneOne$ in  equation \eqref{equation:thereward} is 
         $$ r_1 + r_2 \cdot \frac{\Delta_{3,1}}{\Delta_{3,2}} = - r_3 \cdot \frac{\Delta_{2,1}}{\Delta_{3,2}} < 0\,.$$
         Here, we used some simple algebra to get the first equality. 
         \item If $\bucketOneTwo > C - \bucketOneOne - \bucketTwo - \bucketThree $, by replacing $\bucketTwo$ by the expression in equation \eqref{eq:B2},  the coefficient of $\bucketOneOne$ is 
         $$ r_1 - r_1 + r_1 \cdot \frac{\Delta_{3,1}}{\Delta_{3,2}} - r_2 \cdot \frac{\Delta_{3,1}}{\Delta_{3,2}} = -(r_2 - r_1) \cdot \frac{\Delta_{3,1}}{\Delta_{3,2}} < 0\,. $$
     \end{itemize}
     
     \medskip
     This shows that the coefficient of $\bucketOneOne$ in equation \eqref{equation:thereward} is always negative, meaning that a lower bound is given by setting $\bucketOneOne$ as large as possible. By the last constraint in $\feasibleMOne$, i.e., 
     equation  \eqref{feas:M=1_type1} with $i = 1$, we see that $\bucketOneOne \leq \optvar{1}{1}^*$, so a lower bound is obtained by choosing $\bucketOneOne = \optvar{1}{1}^*$. This also implies that $\bucketTwo = \optvar{2}{1}^*$. This is because we know (1) from equation \eqref{equation:type_2_equation_tight} that $\Delta_{3,1} \cdot \bucketOneOne + \Delta_{3,2} \cdot \bucketTwo = (1- \crr_{up}) \cdot r_3 \cdot C$ and (2)
     from property \ref{property:revenue_1_and_2} of Lemma \ref{lemma:3_type_additional_properties} that 
     $\Delta_{3,1} \cdot \optupper_{1,1}^* + \Delta_{3,2} \cdot \optupper_{2,1}^* = (1-\crr_{up}) \cdot r_3 \cdot C$. 
     
    At this point, we have concluded that choosing $\bucketOneOne = \optupper_{1,1}^*$ and $\bucketTwo = \optupper_{2,1}^*$ gives an upper bound on the reward. We now piece everything together by using equation \eqref{equation:thereward}, substituting $\bucketOneOne = \optOneOne^*, \bucketOneTwo = \optOneTwo^*$ and $\bucketTwoOne = \optupper_{2,1}^*$ to get: 
    \begin{align*}
        \text{Rew}_{\feasibleMOne, \tau} (\{I_t\}_{t \in [T]}) &= r_1 \cdot \min \left(\bucketOneOne + \bucketOneTwo, C - \bucketTwoOne - \bucketThree \right) + r_2 \cdot \bucketTwoOne + r_3 \cdot \bucketThree \\
        &= r_1 \cdot \min \left(\optOneOne^* + \optOneTwo^*, C - \optupper_{2,1}^* - \bucketThree \right) + r_2 \cdot \optupper_{2,1}^* + r_3 \cdot \bucketThree \\
        &\geq r_1 \cdot \min \left(\optOneOne^* + \optOneTwo^* - \bucketThree, C - \optupper_{2,1}^* - \bucketThree \right) + r_2 \cdot \optupper_{2,1}^* + r_3 \cdot \bucketThree \\
        &\geq r_1 \cdot \min \left(\optOneOne^* + \optOneTwo^*, C - \optupper_{2,1}^* \right) + r_2 \cdot \optupper_{2,1}^* + \Delta_{3,1} \cdot \bucketThree \\
        &\geq \crr_{up} \cdot C \cdot r_2 + \Delta_{3,2} \cdot \bucketThree \, . \qquad \text{by constraints \eqref{align:Ic_revenue}, \eqref{align:Ic_capacity} and $\Delta_{3,1} \geq \Delta_{3,2}$}
        \label{equation:thereward}
    \end{align*}
    This completes the proof of Case 2 as we have shown the desired result from equation \eqref{equation:type_2_rejected_desired}. 
\end{itemize}

\textbf{Case 3: Incoming type $2$ and $3$ are accepted, but incoming type $1$ is rejected.} Using a similar argument as above, we get that $\bucketTwo = \numinput{2} \geq \numopt{2}$ and $\bucketThree = \numinput{3} \geq \numopt{3}$ as no type $2$ or type $3$ agents in $I_\tau$ have been rejected. Furthermore, this means that constraints \eqref{feas:M=1_capacity} and \eqref{feas:M=1_type12} of $\feasibleMOne$ are not tight.  Therefore, in order for an incoming type $1$ to be rejected, it must be the case that constraints \eqref{feas:M=1_type1} is tight for any $i = \{1,2\}$, implying that $\bucketOneOne = \optOneOne^*$ and $\bucketOneTwo = \optOneTwo^*$. Hence, we have 
\begin{align*}
    \text{Rew}_{\feasibleMOne, \tau} (\{I_t\}_{t \in [T]}) &= r_1 \cdot \min \left(\bucketOneOne + \bucketOneTwo, C - \bucketTwo - \bucketThree \right) + r_2 \cdot \bucketTwo + r_3 \cdot \bucketThree \\ 
    &= r_1 \cdot \min \left(\optOneOne^* + \optOneTwo^*, C - \bucketTwo - \bucketThree \right) + r_2 \cdot \bucketTwo + r_3 \cdot \bucketThree \\
    &\geq r_1 \cdot \min \left(\crr_{up} \cdot C, C - \bucketTwo - \bucketThree \right) + r_2 \cdot \bucketTwo + r_3 \cdot \bucketThree &\text{by constraint \eqref{align:Id_revenue}} \\
    &\geq r_1 \cdot \min \left(\crr_{up} \cdot C, C - \bucketTwo - \bucketThree \right) + (r_2-r_1) \cdot \bucketTwo +(r_3-r_1) \cdot \bucketThree + r_1 ( \bucketTwo+ \bucketThree)\\
    &\geq r_1 \cdot \min \left(\crr_{up} \cdot C, C \right) + \Delta_{2,1} \cdot \bucketTwo + \Delta_{3,1} \cdot \bucketThree \\
    &\geq r_1 \cdot \crr_{up} \cdot C + \Delta_{2,1} \cdot \bucketTwo + \Delta_{3,1} \cdot \bucketThree
\end{align*}
Therefore, we have the desired inequality.

\textbf{All incoming agent types are accepted.} Since no incoming agents are rejected, our \po{} algorithm with feasible set $\feasibleMOne$ has accepted all agents in $I_{\tau}$ and will serve all the type $2$ and $3$ agents along with as many type $1$ agents as possible. This is clearly at least as much reward as what the clairvoyant optimal algorithm would serve, so we achieve a CR of $1$. 
\end{proof}

\subsubsection{Proof of Theorem \ref{theorem:three_type_M=2}}

\begin{proof}{Proof of Theorem \ref{theorem:three_type_M=2}}
Recall that $\feasibleMTwo \defeq \nestedPolyra\left(\nest{1}, \nest{2}, \nest{3} \right)$, where 
\begin{equation*}
    \nest{1} = \frac{1}{2} \cdot (\optupper_{1,1}^* + \optupper_{1,2}^* ), \qquad 
    \nest{2} = \frac{1}{2} \cdot ( \optupper_{1,1}^* + \optupper_{1,2}^* + \optupper_{2,1}^* +\optupper_{2,2}^*  ), \qquad
    \nest{3} =C.   
\end{equation*}

 For any $i\in [3]$, define $\deltan_{i} = \nest{i}-\nest{i-1}$, where we set $\nest{0} =0$.
    By property \ref{property:total_is_C} of Lemma \ref{lemma:3_class_tight_properties}, which states that $\optupper_{1,t}^* + \optupper_{2,t}^* + \optupper_{3,t}^* = C$, $t\in [2]$, we have
    $\deltan_3 = n_3 - n_2 =\frac{1}{2} \cdot \left( \optupper_{3,1}^* +\optupper_{3,2}^*  \right)$, resulting in the following values of $\deltan_1, \deltan_2, \deltan_3$
\begin{equation*}
    \deltan_1 = \frac{1}{2} \left( \optupper_{1,1}^* + \optupper_{1,2}^* \right), \qquad
    \deltan_2 = \frac{1}{2} \left( \optupper_{2,1}^* + \optupper_{2,2}^* \right), \qquad 
    \deltan_3 = \frac{1}{2} \left( \optupper_{3,1}^* + \optupper_{3,2}^* \right)\,.
\end{equation*}    
    We now apply Theorem \ref{theorem:polyra_nested_performance}, which gives an LP that calculates the CR of any particular instance of the nested polytope. We use that LP to calculate ${\crr}_{\text{nest}} $, the CR of $\nestedPolyra\left(\nest{1}, \nest{2}, \nest{3} \right)$ as follows
\begin{align}
   {\crr}_{\text{nest}} \defeq \max_{\crr, \{\nestedvar{i}{j}\}_{j \in [3], i \in [j]}} \quad &\crr \nonumber \\
    \text{s.t.} \quad & \crr \cdot C \cdot r_1 \leq \nestedvar{1}{1} \cdot r_1 \label{nested_M=2_type_1_rev} \\
    & \crr \cdot C \cdot r_2 \leq \nestedvar{1}{2} \cdot r_1 + \nestedvar{2}{2} \cdot r_2 \label{nested_M=2_type_2_rev} \\
    & \crr \cdot C \cdot r_3 \leq \nestedvar{1}{3} \cdot r_1 + \nestedvar{2}{3} \cdot r_2 + \nestedvar{3}{3} \cdot r_3 \label{nested_M=2_type_3_rev}\\
    & \deltan_1 \leq \nestedvar{1}{1}, \nestedvar{1}{2} \leq 2 \cdot \deltan_1 \label{nested_M=2_type_1_ub} \\
    & \deltan_2 \leq  \nestedvar{2}{2} \leq 2 \cdot \deltan_2  \label{nested_M=2_type_2_ub} \\
    & \nestedvar{i}{3} \leq \deltan_{i} & i \in [3] \label{nested_M=2_type_3_ub} \\
    & \sum_{i=1}^j \nestedvar{i}{j} \leq C & j \in [3] \label{nested_M=2_capacity} \\
    & \nestedvar{i}{j} \geq 0 \,, \quad \crr \geq 0 & j \in [3], i \in [j]
\end{align}
In order to show that a \po{} algorithm with feasible region $\feasibleMTwo$ achieves a $\crr_{up}$ CR, it would be sufficient to show that $ {\crr}_{\text{nest}}  \geq \crr_{up}$. To do so, we start with the optimal solution to the upper bound $(\crr_{up}, \{\optupper_{i, t}^*\}_{i \in [3], t \in [2]})$ and use it to create a feasible instance of the aforementioned  LP for which $\crr = \crr_{up}$. We set the variables $\nestedvar{i}{j}$ as follows: 
\begin{align}
    \nestedvar{1}{1} = 2 \cdot \deltan_1, \qquad
    \nestedvar{1}{2} = \min(C - 2 \cdot \deltan_2, 2 \cdot \deltan_1) , \qquad
    \nestedvar{2}{2} = 2 \cdot \deltan_2, \qquad
    \nestedvar{i}{3} = \deltan_i,  i\in [3] \label{nested_proof_the_assignment}
\end{align}

We verify that such an assignment is feasible. We have that constraints \eqref{nested_M=2_type_1_rev}, \eqref{nested_M=2_type_2_rev} and \eqref{nested_M=2_type_3_rev} are satisfied by looking at the corresponding constraints \eqref{align:Id_revenue},   \eqref{align:Ic_revenue}, and  \eqref{align:Ia_revenue} respectively from the upper bound. To see why, consider  constraint \eqref{nested_M=2_type_1_rev}. This constraint  holds if 
$\crr_{up} \cdot C \cdot r_1 \leq \nestedvar{1}{1} \cdot r_1$, which by our construction, it is equivalent to  
$\crr_{up} \cdot C \cdot r_1 \leq (\optupper_{1,1}^* + \optupper_{1,2}^* ) \cdot r_1$. Then, clearly $\crr_{up} \cdot C \cdot r_1 \leq (\optupper_{1,1}^* + \optupper_{1,2}^* ) \cdot r_1$ holds by constraint  \eqref{align:Id_revenue} in \ref{3_class_ub}.
Next consider constraint  \eqref{align:Ic_revenue}. This constraint holds if $\crr_{up} \cdot C \cdot r_2 \leq \nestedvar{1}{2} \cdot r_1 + \nestedvar{2}{2} \cdot r_2$. By our construction, this is the same as:
\[\crr_{up} \cdot C \cdot r_2 \leq \min\big(C - 2 \cdot \deltan_2, 2 \cdot \deltan_1\big) \cdot r_1 + 3\cdot \deltan_2 \cdot r_2 = 
\min(C - \left( \optupper_{2,1}^* + \optupper_{2,2}^* \right), \left( \optupper_{1,1}^* + \optupper_{1,2}^* \right)) \cdot r_1 + \left( \optupper_{2,1}^* + \optupper_{2,2}^* \right) \cdot r_2
\]
The above inequality holds per constraints \eqref{align:Ic_capacity} and \eqref{align:Ic_revenue} in \ref{3_class_ub}. To show constraint \eqref{nested_M=2_type_3_rev} holds, note that this constraint is equivalent to 
\[\crr \cdot C \cdot r_3 \leq \nestedvar{1}{3} \cdot r_1 + \nestedvar{2}{3} \cdot r_2 + \nestedvar{3}{3} \cdot r_3 = \frac{1}{2}\left( \optupper_{1,1}^* + \optupper_{1,2}^* \right) \cdot r_1 + \frac{1}{2} \left( \optupper_{2,1}^* + \optupper_{2,2}^* \right)\cdot r_2 + \frac{1}{2} \left( \optupper_{3,1}^* + \optupper_{3,2}^* \right)\cdot r_3\,. \]
The above inequality holds because of constraint \eqref{align:Ia_revenue} in \ref{3_class_ub}.

Now, consider constraint \eqref{nested_M=2_type_1_ub}. For both $\nestedvar{1}{1}$ and $\nestedvar{1}{2}$, the upper bound of $2 \cdot \deltan_1$ is immediately satisfied, while the lower bound of $\deltan_1$ is trivially satisfied by $\nestedvar{1}{1}$. For the lower bound on $\nestedvar{1}{2} =  \min(C - 2 \cdot \deltan_2, 2 \cdot \deltan_1)$ if $2 \cdot \deltan_1$ is taken in the minimum, then we are done. Otherwise, if $C - 2 \cdot \deltan_2$ is the minimum, then the lower bound is also satisfied because $$C - 2 \cdot \deltan_2 = C - (\optupper_{2,1}^* + \optupper_{2,2}^*) \ge  \frac{1}{2} \left(
    \optupper_{1,1}^* + \optupper_{1,2}^*\right) = \deltan_1 \,.$$ 
where the  inequality is true by property \ref{property:consistent_M=2} from Lemma \ref{lemma:3_type_additional_properties}.
For constraint \eqref{nested_M=2_type_2_ub} and \eqref{nested_M=2_type_3_ub}, they are all trivially satisfied by our assignment in equation \eqref{nested_proof_the_assignment}. Finally, we argue that constraint \eqref{nested_M=2_capacity} is satisfied. For $j=1$, constraint \eqref{nested_M=2_capacity} required us to show $\nestedvar{1}{1} = 2\cdot \deltan_1 = \optupper_{1,1}^* + \optupper_{1,2}^* \le C$, which is true because of property \ref{property:type_1_cautious} from Lemma \ref{lemma:3_class_tight_properties}, which states that $r_1 \cdot (s_{1,1}^* + s_{1,2}^*) = r_1 \cdot \crr_{up} \cdot C$. For $j = 2$, constraint \eqref{nested_M=2_capacity} we need to show that $\nestedvar{1}{2} + \nestedvar{2}{2} \leq C$, which is true by the definition of $\nestedvar{1}{2}$ and $\nestedvar{2}{2}$ from equation \eqref{nested_proof_the_assignment}. 
For $j = 3$, constraint \eqref{nested_M=2_capacity} required us to show that  
\[\nestedvar{1}{3}+\nestedvar{2}{3}+\nestedvar{3}{3} =  \deltan_1+\deltan_2+\deltan_3 = \frac{1}{2} \left( \optupper_{1,1}^* + \optupper_{1,2}^* + \optupper_{2,1}^* + \optupper_{2,2}^* +\optupper_{3,1}^* + \optupper_{3,2}^* \right) \leq \frac{1}{2} \cdot 2 \cdot C
\,,\]
where the inequality holds because of constraint \eqref{align:Ia_capacity} from LP \ref{3_class_ub}.  

We have thus shown that the assignment of $\nestedvar{i}{j}$ from \eqref{nested_proof_the_assignment}, along with $\crr = \crr_{up}$ leads to a feasible solution, meaning that ${\crr}_{\text{nest}} \geq \crr_{up}$. This allows us to conclude by Theorem \ref{theorem:polyra_nested_performance} that this particular instance of the nested algorithm as given by \eqref{nested_proof_the_assignment} has a CR at least $\crr_{up}$, and hence is optimal. 
\end{proof}
\vspace{1em}

\section{Upper Bounds for \texorpdfstring{$K > 3$}{K > 3}}
\label{appendix:not_tight_example}

We will show by example that for $K > 3$, our intuitive overlapping worst-case arrival sequences from Theorem \ref{thm:3_type_upper_bound} do not lead to an attainable CR.  
The crux of our upper bound constructions have been the choice of the worst-case arrival sequence and the resulting collection of truncations that it generates. For $K=2$ and $3$, by presenting algorithms whose CR matches exactly those upper bounds, we showed that the upper bound is tight/attainable.

For $K> 3$ types, our upper bound construction can easily be generalized, but we will see from the following example that doing so may not lead to a tight upper bound. To show this, we present \textit{alternative} worst-case arrival sequences under which we may get a tighter upper bound than the bound obtained from extending our worst-case arrival sequences in Theorem \ref{thm:3_type_upper_bound}.

\begin{example}[Overlapping arrival sequences do not lead to a tight upper bound]
  \label{example:natural_choice}
  Suppose that we have $K=4$ types with rewards $r_1 < r_2 < r_3 < r_4$ and $M = 2$ (i.e., types $1$ and $2$ are partially patient, while types $3$ and $4$ are impatient). For notational convenience, we choose $C = 3$. If we use the same reasoning as before, our worst-case arrival sequence $\{I_1, I_2\}$ is
  \[\begin{array}{ll}
      I_1 = \{ 1, 1, 1, \: 2, 2, 2, \:3, 3, 3, \:4, 4, 4 \} \qquad & I_2 = \{ 3, 3, 3,\: 4, 4, 4  \}
    \end{array}\]
  and its collection of truncations is
  \[
    \begin{array}{ll}
      I_1^1 = \{ 1, 1, 1\} \qquad                                    & I_2^1 = \{  \}         \\
      I_1^2 = \{ 1, 1, 1, \: 2, 2, 2\} \qquad                        & I_2^2 = \{  \}         \\
      I_1^3 = \{ 1, 1, 1, \: 2, 2, 2, \:3, 3, 3\} \qquad             & I_2^3 = \{  \}         \\
      I_1^4 = \{ 1, 1, 1, \: 2, 2, 2, \:3, 3, 3, \:4, 4, 4 \} \qquad & I_2^4 = \{  \}         \\
      I_1^5 = \{ 1, 1, 1, \: 2, 2, 2, \:3, 3, 3, \:4, 4, 4 \} \qquad & I_2^5 = \{ 3, 3, 3  \} \\
    \end{array}
  \]

  We present two alternative choices of worst-case arrival sequences as follows:

  \begin{itemize}
    \item \textbf{Removing type(s)}. Consider using the following worst-case arrival sequence
          \[\begin{array}{ll}
              \overline{I_1} = \{ 1, 1, 1, \:3, 3, 3, \:4, 4, 4 \}, \quad & \overline{I_2} = \{ 3, 3, 3,\: 4, 4, 4  \} \\
            \end{array}\]
          The difference between $\{\overline{I_1}, \overline{I_1}\}$ and $\{I_1, I_2\}$ is that this new worst-case arrival sequence (and its truncations) do not send any type $2$ agents. This worst-case arrival sequence leads to a tighter upper bound when $r_2$ is close to $r_3$. Consider an extreme case where $r_2 = r_3$ (or are arbitrarily close). If that is the case, then an adversary should not send any type $2$ agents because type $2$ agents are partially patient, while type $3$ agents are not, yet they achieve (nearly) the same reward. In general for $K > 4$, for some reward values $r_1, \ldots, r_K$ we can construct a tighter worst-case arrival sequence by choosing an index $j \in [M]$ and removing types $j, j+1, \ldots, M$ from the original worst-case arrival sequence.

    \item \textbf{Considering three time periods, rather than two.} Consider the following worst-case arrival sequence.
          $$
            \begin{array}{lll}
              \tilde{I_1} = \{ 1, 1, 1, \: 2, 2, 2, \:3, 3, 3, \:4, 4, 4 \}, & \tilde{I_2} = \{ 1, 1, 1, \: 2, 2, 2, \:3, 3, 3, \:4, 4, 4 \}, & \tilde{I_3} = \{ 3, 3, 3, 4, 4, 4 \} \\
            \end{array}
          $$

          The difference between $\{\tilde{I_1}, \tilde{I_2}, \tilde{I_3}\}$ and $\{I_1, I_2\}$ is that this new worst-case arrival sequence sends agents in ascending order of reward over three time periods rather than two. Let $\crr_{up}$ be the upper bound given by considering the original worst case arrival sequence $\{I_1, I_2\}$ while $\tilde \crr$ is the upper bound when we consider $\{\tilde{I_1}, \tilde{I_2}, \tilde{I_3}\}$ as the worst case. Notice that $\tilde{I_1} = I_1$ and so any non-anticipating algorithm must behave the same in the first time period against both worst-case arrival sequences.

          Suppose $\alg$ is an algorithm that is $\crr_{up}$ competitive. After accepting and serving some subset of the 12 agents in $I_1$, $\alg$ begins $t = 2$ with say $c_1$ unserved type $1$ agents and $c_2$ unserved type $2$ agents (i.e. the accepted partially patient agents from $t = 1$ that we did not serve). Under our original second time period arrival sequence $I_2$, we enforce that this commitment of $c_1$ type $1$ and $c_2$ type $2$ agents does not preclude $\alg$ from achieving a $\crr_{up}$ CR when the worst case arrival sequences continues as $I_2 = \{3, 3, 3,\: 4, 4, 4\}$. However, it may be the case that if the arrival sequence were to continue as $\tilde{I_2}$ and $\tilde{I_3}$, then this commitment of $c_1$ type 1 agents and $c_2$ type 2 agents makes it impossible for any algorithm to be $\crr_{up}$ competitive anymore against all the truncations of $\{\tilde{I_2}, \tilde{I_3}\}$: perhaps we can only hope to be $\tilde \crr < \crr_{up}$ competitive. In other words, depending on the commitment $c_1, c_2$, it is possible that starting in that state, $\{\tilde{I_2}, \tilde{I_3}\}$ more accurately depicts the worst-case scenario than $\{I_2\}$. In general, it is unclear if there is an upper bound to the number of time periods we need to consider in order to get the tightest upper bound.
  \end{itemize}

  As a concrete numerical example, if we take the case where $r_1 = 1, r_2 = 3, r_3 = 4, r_4 = L$, then as $L \to \infty$ our old upper bound gives $\frac{9}{20} = 0.45$ whereas taking the tighter of the two new worst-case arrival sequences shown above yields $\frac{15}{34} \approx 0.4411 < 0.45$.

The upper bounds created by these new worst-case arrival sequences may be only tighter than our original upper bound only when $K > 3$ because for $K = 2$ and $K = 3$, there are too few types for some of these nuances to show. These new worst-case arrival sequences may be tighter when there are tensions between multiple partially patient and impatient types. With $K = 2$ or $K = 3$, since there is always either just $1$ partially patient type or $1$ impatient type, we do not see such nuances.
\end{example}

{
\color{black}

\section{Proof of Theorem \ref{theorem:our_alg_beats_bq}}
\label{appendix:nested_vs_bq}
\begin{proof}{Proof of Theorem \ref{theorem:our_alg_beats_bq}}
    Our goal here is to prove that for $M \geq 1$, the optimal nested polytope (given by Definition \ref{def:optimal_nested_algorithm}) achieves a CR which is strictly greater than the CR of the BQ nested \po{} algorithm, given by $\crr^{BQ}$. Fixing some rewards $r_1 < \ldots < r_K$, we define the following collection of CR's $ \crr_{1}^*, \ldots, \crr_{K-1}^*$ where $\crr_{m}^*$ is the CR of the optimal nested \po{} algorithm when $M = m$, which is computed by solving the LP in Definition \ref{def:optimal_nested_algorithm} for $M = m$. 
    We will show that 
    \begin{equation}
        \label{ordering_of_cr}
        \crr^{BQ} < \crr_{1}^* \leq \crr_{2}^* \ldots \leq \crr_{K-1}^* \, ,
    \end{equation}
    which is the desired result.

    \noindent 
    Let us start by rewriting the LP in Definition \ref{def:optimal_nested_algorithm} by indexing the variables with $m$: 
    \begin{align*}
        \crr_{m}^* \defeq \max_{\crr_{m}, \{\deltan_{i,m}\}_{i=1}^K,\{ s_{i,m}^{j}\}_{j \in [K], i \in [j]}} \quad & \crr_{m} \\
        \text{s.t.} \quad & \crr_{m} \cdot r_j \cdot C \leq \sum_{i=1}^j s_{i,m}^{ j} \cdot r_i & j \in [K] \\ 
        & \sum_{i=1}^j s_{i,m}^{j} \leq C & j \in [K]\\
        & \deltan_{i,m} \leq s_{i,m}^{ j} \leq 2 \cdot \deltan_{i,m}  & j < m, i \in [j] \\
        & s_{i,m}^{j} = \deltan_{i,m} & j > m, i \in [j]\\
        &\crr_{m}\ge 0, \deltan_{i,m}\ge 0 & i\in [K]\\
        & s_{i,m}^{ j} \ge 0 & j\in [K], i\in [j]\,.
    \end{align*}

    Here, $\crr^*_m$ is the CR of the optimal nested algorithm instance when the rewards are $r_1, \ldots, r_K$ and the first $M = m$ types are flexible. The corresponding optimal solution $\deltan_{i,m}^*$ is the optimal marginal nest sizes for type $i$ when $M = m$. To show our desired result from \eqref{ordering_of_cr}, we will first show that $\crr^{BQ} < \crr_{1}^*$ and then show that for any $m \in [K-2]$ we have $\crr_{m}^* \leq \crr_{m+1}^*$. 

    \textbf{Step 1.} To show that $\crr^{BQ} < \crr_{1}^*$, we present an explicit construction of a solution to the LP for $M = 1$ which achieves a strictly higher objective value than $\crr^{BQ}$. Recall from the definition of the BQ nested Polytope (Definition \ref{def:bq_nest}) that: 
    $$ \crr^{BQ} = \left( K - \sum_{i=1}^{K} \frac{r_{i-1}}{r_{i}} \right)^{-1} $$
    and that we can compute the optimal (marginal) nest sizes $\deltan_i^{BQ} = n_{i}^{BQ} - n_{i-1}^{BQ}$:
    $$ 
    \deltan_i^{BQ} = \crr^{BQ} \cdot \left(1 - \frac{r_{i-1}}{r_{i}} \right) \cdot C\,.
    $$
    where $r_0 \defeq 0$.
    
    To construct a solution to the LP for $\crr_1^*$ with a better CR than $\crr^{BQ}$, we choose the following solution and then argue its feasibility: 
    \begin{align}
    \label{contructing_n1}
        \deltan_{i,1} &=
        \begin{cases}
            \frac{C}{C'} \cdot \frac{1}{2} \deltan^{BQ}_1 & i = 1 \\
            \frac{C}{C'} \cdot \left( \deltan^{BQ}_2 + \frac{1}{2} \deltan^{BQ}_1 \cdot \frac{r_1}{r_2} \right) & i = 2 \\
            \frac{C}{C'} \cdot \deltan^{BQ}_i & \text{otherwise}
        \end{cases}, \qquad 
        s^{j}_{i,1} = 
        \begin{cases}
            \min(2 \cdot \deltan_{i,1}, C) & i = j = 1 \\
            \deltan_{i, 1} & i \in [j], j > 1 
        \end{cases}, \qquad  
        \crr_1 = \frac{C}{C'} \cdot \crr^{BQ}
    \end{align}
    where $C' \defeq C \cdot \left( \frac{1}{2} \deltan_{1}^{BQ} + \deltan_2^{BQ} + \frac{1}{2} \deltan_1^{BQ} \cdot \frac{r_1}{r_2} + \sum_{i=3}^K \deltan_i^{BQ}\right)$ and is chosen so that the $\deltan_{i,1}$'s sum up to $C$.  Note that since $\deltan_1^{BQ} > 0$ and $\sum_{i=1}^K \deltan_i^{BQ} = C$, we have that $C' < C$ and so $\crr_1 > \crr^{BQ}.$

    We claim that the above choices of variables leads to a feasible solution to the LP defining $\crr_1^*$. 

    \begin{itemize}
        \item We first show $s_{i,1}^j$ is feasible. For $j = 1$, we need to show that then we have $\deltan_{i,1} \leq s_{i,1}^{1} \leq 2 \cdot \deltan_{i,1}$, where $s_{i,1}^{1} = \min(2 \cdot \deltan_{i,1}, C)$. This is clear if $2 \cdot \deltan_{i,1} < C$. Otherwise, if $C \leq 2 \cdot \deltan_{i,1}$, then we just need to show that $\deltan_{i,1} \leq C$. Then,  clearly we have $\deltan_{i,1} \leq C$ because  $\deltan_{i,1}$ is chosen such that $\sum_{i=1}^K \deltan_{i,1} = C$.
        For $j > 1$ we set $s_{i,1}^{j} = \deltan_{i,1}$ explicitly.
        
        \item $\sum_{i=1}^j s_{i, 1}^{j} \leq C$. For $j = 1$, this holds because $s_{i,1}^{1} = \min(2 \cdot \deltan_{i,1}, C)$. For $j > 1$, we have that $\sum_{i=1}^j s_{i, 1}^{j} = \sum_{i=1}^j \deltan_{i,1} \leq \sum_{i=1}^K \deltan_{i,1} = C$.
        
        \item $\sum_{i=1}^j s_{i, 1}^{j} \cdot r_i \geq \crr_1 \cdot r_j \cdot C$. For $j = 1$, we need to show that $\min(2 \cdot \deltan_{i,1}, C) \cdot r_1 \geq \crr_1 \cdot r_1 \cdot C$. This is clearly true when the minimum takes on the value $C$. Otherwise, we have that $2 \cdot \deltan_{i,1}\cdot r_1 = \frac{C}{C'} \cdot \deltan_1^{BQ} \cdot r_1 = \frac{C}{C'} \cdot \crr^{BQ} \cdot r_1 \cdot C = {\crr_1} \cdot r_1 \cdot C$ as desired.

        Otherwise, for $j > 1$, we have:
            \begin{align*}
                \sum_{i=1}^j r_i \cdot s_{i,1}^{j} &= \sum_{i=1}^j r_i \cdot \deltan_{i,1}\\
                &= r_1 \cdot \deltan_{1,1} + r_2 \cdot \deltan_{2,1} + \sum_{i=3}^j r_i \cdot \deltan_{i,1} \\ 
                &= \frac{C}{C'} \cdot \left[ r_1 \cdot \deltan_1^{BQ} + r_2 \cdot \deltan_2^{BQ} + \sum_{i=3}^j r_i \cdot \deltan_i^{BQ} \right] \\
                &\geq \frac{C}{C'} \cdot  \crr^{BQ} \cdot r_j \cdot C \\
                &= \crr_{1} \cdot r_j \cdot C \,.
            \end{align*}
    \end{itemize}

    \noindent Hence, we have shown that the LP for $\crr_1^{*}$ must have an optimal value at least $\crr_1^{*} > \crr_1 = \crr^{BQ} \cdot \frac{C}{C'} > \crr^{BQ}$.




    \textbf{Step 2.} To show that $\crr_m^{*} \leq \crr_{m+1}^{*}$ for any $m \in [K-2]$, we claim that any feasible solution to the LP for $\crr_{m}^{*}$  is also a feasible solution to $\crr_{m+1}^{*}$, achieving an objective value at least $\crr_{m}^{*}$. More let $\Gamma_m^*, \{\deltan_{i,m}^*\}, \{s_{i, m}^{j*} \}_{j \in [K], i \in [j]}$ be the optimal solution to the LP for $\crr_m^*$ (it can actually be any feasible solution, not necessarily optimal). We construct a new feasible solution for the LP for $\crr^*_{m+1}$ as follows:
    \begin{align*}
        \crr_{m+1} &= \crr_{m}^{*} \\
        \deltan_{i, m+1} &= \deltan_{i.m}^{*} & i \in [K] \\
        s_{i, m+1}^{ j} &= s_{i, m}^{ j*} & j \in [K], i \in [j]
    \end{align*}
    
    We claim that this solution is feasible. Comparing the LP for $\crr_{m}^{*}$ and $\crr_{m+1}^{*}$, we see that the only constraint which differs is the one for $j = m+1$. For $j = m+1$, in the LP for $\crr_{m}^{*}$, we have the constraint $ s_{i, m}^{ m+1} = \deltan_{m+1, m}$ while in the LP for $\crr_{m+1}^{*}$, we have the constraint $\deltan_{m+1, m+1} \leq s_{i, m+1}^{m+1} \leq  2 \cdot \deltan_{m+1, m+1}$. Therefore, we have any solution to the LP for $\crr_{m}^{*}$ is feasible for the LP to $\crr_{m+1}^{*}$ which proves that $\crr_{m}^{*} \leq \crr_{m+1}^{*}$. 
    \Halmos
    
\end{proof}
}
\section{Proof of Theorem \ref{theorem:main_optimality_gap}} \label{sec:theorem:main_optimality_gap}
We prove this statement through the following steps: 

\begin{enumerate}
    \item Presenting a loose, but explicit upper bound $\overline{\crr}_{up}$ on the CR of any non-anticipating deterministic algorithm (Lemma \ref{thm:loose_upper_bound}).
    \item An explicit instance of a nested polytope with nest sizes $(\overline{\deltan_i})_{i=1}^K$ that may not be the optimal nested polytope, but achieves a CR of $\crr_{nest}$ where $\crr_{nest} \leq \crr^*$ (Definition \ref{lem:bar_gamma}).
    \item Showing that $\crr_{nest} \geq 0.8 \cdot \overline{\crr}_{up}$ (Lemma \ref{thm:optimality_gap}), which also implies that $\crr^* \geq 0.8 \cdot \overline{\crr}_{up}$ and proves Theorem \ref{theorem:main_optimality_gap}. 
\end{enumerate}

\subsection{Step 1: A Loose Upper Bound.} 
The following lemma presents $M+1$ upper bound LPs, denoted by $\crr_{\lpp, m}$, $m\in \{0, 1, \ldots, M\}$. To construct $\crr_{\lpp, m}$, we consider similar overlapping worst-case  arrival sequences as in the proof of Theorem \ref{thm:3_type_upper_bound}. However, the worst-case arrival sequences used in constructing $\crr_{\lpp, m}$ chooses to only send types $1, 2, \ldots, m, M+1, \ldots, K$ and omitting types $m+1, \ldots, M$. An adversary may choose to do this depending on how close $r_{m+1}, \ldots, r_M$ are to $r_{M+1}$. For example, consider the case where $r_{M} \approx r_{M+1}$. Then, it is better for an adversary to omit type $M$ agents and send only type $M+1$ agents since their reward are similar, but the former is partially patient while the latter is impatient. By omitting some partially patient types with high rewards, we can potentially construct tighter upper bounds.

\begin{lemma} [Upper Bound LPs] 
\label{lemma:upper_bound_lp_m}
For any rewards $r_1 < \ldots< r_K$ with the first $M \in [K-1]$ types partially patient and any $m \in \{0, 1, \ldots, M\}$, the following LP gives an upper bound on the CR of any deterministic online algorithm: 
\begin{align}
    \crr_{\lpp, m} \defeq \max \quad \crr \nonumber \\
    \text{s.t.} \quad & r_i \cdot \crr \cdot C \leq \sum_{j=1}^i r_j \cdot s_j & i \in \{1, 2, \ldots, m\} \label{sup_before_m} \\
    & 2 \cdot r_i \cdot \crr \cdot C \leq 2 \cdot \sum_{j=1}^{m-1} r_{j} \cdot s_j + \sum_{j=M}^{i} r_j \cdot s_j & i \in \{ M+1, \ldots, K-1\} \label{sup_after_m} \\
    & 2 \cdot r_K \cdot \crr \cdot C \leq \sum_{j=1}^{m-1} r_{j} \cdot s_j + \sum_{j=M}^{K} r_j \cdot s_j \label{sup_last_type} \\
    & \sum_{i=1}^K s_i \leq 2 \cdot C \label{sup_sum_constraint}\\
    & \crr, s_i \geq 0 & i \in [K] \nonumber 
\end{align}

Furthermore, $\crr_{\lpp}$ as defined below, is an upper bound on the CR of any online algorithm 
\begin{equation}
    \crr_{\lpp}\, \defeq  \min_{m \in \{0, 1, \ldots, M\}} \crr_{\lpp, m}\,.
\end{equation}
\end{lemma}

While $\crr_{LP}$ is certainly an upper bound on the CR of any online algorithm, it is quite difficult to analyze since it is the minimum of $M+1$ terms, each of which is the solution to a non-trivial LP. As a result, we define the following closed-form upper bound which is weaker than $\crr_{LP}$ but much easier to analyze .

\begin{lemma}[A Closed-Form Upper Bound] 
\label{thm:loose_upper_bound} 
Define the following quantity $\overline{\crr}_{up}$:

\begin{equation}
    \label{crrup_def}
    \overline{\crr}_{up} \defeq \min \left( \frac{1}{G} \, ,\frac{2}{2\cdot G + M - \sum_{i=1}^{M+1} \frac{r_{i-1}}{r_i} - \frac{r_M}{r_{M+1}} + \frac{r_M}{r_K}} \right)
\end{equation}
where $G$ is defined as follows: 
\begin{equation}
    \label{eq:G} 
    G \defeq K - M - \sum_{i=M+2}^K \frac{r_{i-1}}{r_{i}}\,.
\end{equation}

For any $K$ types of agents with rewards $r_1 < \ldots < r_K$ and the the first $M \in [K-1]$ partially patient, we have that $\overline{\crr}_{up} \geq \crr_{LP}$, where $\crr_{LP}$ is defined in Lemma \ref{lemma:upper_bound_lp_m}. Thus, $\overline{\crr}_{up}$ is also a valid upper bound on the CR of any deterministic online algorithm
\end{lemma}

While $\crr_{LP}$ is defined to be the minimum of $M+1$ terms, each of which is the solution to a LP, $\overline{\crr}_{up}$ is the minimum of 2 terms, which roughly correspond to $\crr_{LP, 0}$ and $\crr_{LP, M}$. It turns out that intuitively considering just these two scenarios (sending no patient types or sending all the patient types), results in a tight enough upper bound for our subsequent analysis.

\subsection{Step 2: Near-optimal Nested Polytope}
Here, we construct an explicit near-optimal instance of the nested polytope. We show that  a \po{} algorithm with this  near-optimal nested polytope  achieves at least 80\% of the optimal CR, by comparing its CR to $\overline{\crr}_{up}$ defined earlier. Note that the CR of this nested \po{} algorithm   is a lower bound on the CR of our optimized nested \po{} algorithm in Definition \ref{def:optimal_nested_algorithm}. The near-optimal nested \po{} algorithm uses the polytope $\nestedPolyra(\overline{\deltan_1}, \ldots, \overline{\deltan_K})$ where the nest sizes $\overline{\deltan_i}$, $i\in [K]$ is defined in the following definition. We will shortly discuss how this near-optimal nested \po{} algorithm is related to our optimized nested \po{} algorithm in Definition \ref{def:optimal_nested_algorithm}.  

\begin{definition} [Near-Optimal Nested Algorithm Construction] \label{lem:bar_gamma}
Define
\begin{align} 
    \overline \crr = 
    \frac{2}{2 \cdot G + M - \sum_{i=1}^{M+1} \frac{r_{i-1}}{r_i}}\,
    \label{definition:gamma_bar}\,,
\end{align}
where $G$ is defined in equation \eqref{eq:G}. Further, define the following nest sizes $\{\overline{\deltan_i}\}_{i=1}^K$
\begin{equation}
\label{definition:nested_instance}
\overline{\deltan}_i = 
\begin{cases} 
\frac{1}{2} \cdot \overline \crr \cdot \left( 1 - \frac{r_{i-1}}{r_{i}} \right) \cdot C & i \in [M] \\
\overline \crr  \cdot \left( 1 - \frac{1}{2} \frac{r_{i-1}}{r_{i}} \right) \cdot C & i = M+ 1 \\
\overline \crr  \cdot \left( 1 - \frac{r_{i-1}}{r_{i}} \right) \cdot C & i= M+2, \ldots, K
\end{cases}
\end{equation} where $r_0 \defeq 0$. Define $\crr_{nest}$ to be the CR of this nested \po{} algorithm with polytope $\nestedPolyra(\overline{n_1}, \ldots, \overline{n_K})$ (recall that $\overline{n}_i = \overline\deltan_1 + \ldots + \overline\deltan_i$). It is easy to verify that $\overline{n}_K = C$, so we have constructed a valid polytope.
\footnote{ To see why  $\overline{n}_K = C$, note that by definition,  
$\sum_{i=1}^K \overline{\deltan}_i = \sum_{i=1}^{M} \overline{\deltan}_i + \overline{\deltan}_{M+1} + \sum_{i=M+2}^{K} \overline{\deltan}_i$, which can be written as 
\begin{align*} \overline{\crr} \cdot C \cdot \Bigg[ \underbrace{\frac{M}{2} - \frac{1}{2} \cdot \sum_{i=1}^{M} \frac{r_{i-1}}{r_i} - \frac{1}{2} \frac{r_M}{r_{M+1}} }_{
\frac{1}{2} \cdot \left( M - \sum_{i=1}^{M+1} \frac{r_{i-1}}{r_{i}} \right)
} + \underbrace{1 + (K-M-1) - \sum_{i=M+2}^{K} \frac
{r_{i-1}}{r_i}}_{G}\Bigg] = C\,.
\end{align*}

} 
\end{definition}

\subsection{Step 3: The Competitive Ratio of the Near-Optimal Nested Algorithm}
The following the main result of this section that completes the proof of Theorem \ref{theorem:main_optimality_gap}.

\begin{lemma}[Near-Optimality of $\nestedPolyra(\overline{n}_1, \ldots, \overline{n}_K)$]
\label{thm:optimality_gap}
Consider the near-optimal nested \po{} algorithm with feasible polytope $\nestedPolyra(\overline{n}_1, \ldots, \overline{n}_K)$, where the nest sizes $\overline{n}_i$, $i\in[K]$ are defined in Definition \ref{lem:bar_gamma}. Let
 $\crr_{\text{nest}}$ be the CR of this algorithm. Then,  $\crr_{\text{nest}}$ can be lower bounded as follows:
$$\crr_{\text{nest}} \geq 0.8 \cdot\overline{\crr}_{\up} \,,$$
where $\overline{\crr}_{\up}$, defined in Theorem \ref{thm:loose_upper_bound}, is an upper bound on the CR of any non-anticipating algorithm.  
\end{lemma}
To show that $\crr_{\text{nest}} \geq 0.8 \cdot\overline{\crr}_{\up}$, we crucially take advantage of the intermediate quantity $\overline{\crr}$, defined in equation \eqref{definition:gamma_bar}. We show that the following two inequalities hold
\begin{equation*}
    \frac{\crr_{\text{nest}}}{\overline{\crr}} \geq f_1(G) \qquad {\text and } \qquad \frac{\overline{\crr}}{\overline{\crr}_{\up}} \geq f_2(G) 
\end{equation*}
for some functions $f_1, f_2$ which will be defined in Lemmas \ref{lemma:appx_one_of_two} and \ref{lemma:appx_two_of_two} in the appendix. The ratio ${\crr_{\text{nest}}}/{\overline{\crr}}$ measures how close the CR of our nested algorithm ($\crr_{\text{nest}}$) is to $\overline{\crr}$, an optimistic CR of $\nestedPolyra(\overline{\deltan_1}, \ldots, \overline{\deltan_K})$. Furthermore, the ratio ${\overline{\crr}}/{\overline{\crr}_{\up}}$ measures how well this optimistic CR approximates the CR of any algorithm, not necessarily nested. We highlight that as one of our novel technical contributions, the lower bounds $f_1, f_2$ we provide for the aforementioned ratios are both function of the variable $G = K - M - \sum_{i=M+2}^{K} \frac{r_{i-1}}{r_{i}}$, rather than being constants. (Recall that $1/G$ is an upper bound on the CR of any online algorithm in the situation when you have $K - M$ types with rewards $r_{M+1}, \ldots, r_{K}$ all of which are impatient.)
Bounding these ratios as a function of $G$ is necessary as $\min_G f_1(G) \approx 0.816$ and  $\min_G f_2(G) = 0.8$, which implies that simply bounding the two ratios with constants would not be sufficient. 
We show in Lemma \ref{lemma:appx_combine} that $f_1(G) \cdot f_2(G) \ge 0.8$ for all $G$, which completes the proof. The statements of Lemmas \ref{lemma:appx_one_of_two}, \ref{lemma:appx_two_of_two} and \ref{lemma:appx_combine} and their proofs  can be found in Section \ref{sec:optimality_gap_proofs}.

\section{Proofs of the Lemmas Used in the Proof of Theorem \ref{theorem:main_optimality_gap}}

{\color{black}
\subsection{Proof of Lemma \ref{lemma:upper_bound_lp_m}
}
\begin{proof}{Proof of Lemma \ref{lemma:upper_bound_lp_m} }
Fix a particular value of $m$. The argument here is to use the arrival sequence which sends the first $m$ out of $M$ flexible types and all of the $K-M$ inflexible types. We start out with a \textit{primary} input sequence and we consider truncated versions of it. Define our primary input sequence $\{I_1, I_2\}$ as follows: 
\begin{align*}
    I_1 &= \{ \underbrace{1, \ldots, 1}_{C},  \underbrace{2, \ldots, 2}_{C}, \ldots \underbrace{m, \ldots, m}_{C}, \underbrace{M+1, \ldots, M+1}_{C} \ldots,  \underbrace{K, \ldots, K}_{C} \} \\
    I_2 &= \{ \underbrace{M+1, \ldots, M+1}_{C},  \underbrace{M+2, \ldots, M+2}_{C}, \ldots,  \underbrace{K, \ldots, K}_{C} \}
\end{align*}
$I_1$ involves $C$ copies of agent types $1$ through $m$ out of the $M$ flexible types, followed by $C$ copies each of agent types $M+1$ through $K$ all of which are inflexible. $I_2$ involves $C$ copies of each of the inflexible types $M+1$ through $K$. We define  truncated versions of this input sequence, denoted by $\{I_1^{(i,j)}, I_2^{(i,j)}\}$ for $i \in \{1, 2, \ldots, m, M+1, \ldots, K\}$ and $j \in \{1, 2\}$, which consist of sending up to and including the $C$ agents of type $i$ in period $j$ (we only define $\{I_1^{(i,j)}, I_2^{(i,j)}\}$ when type $i$ agents appear in $I_j$). For example: 
\begin{align*}
    I_1^{(1,1)} &= \{\underbrace{1, \ldots, 1}_{C}\}, & I_2^{(1,1)} &= \emptyset \\
    I_1^{(M+1, 1)} &= \{\underbrace{1, \ldots, 1}_{C}, \ldots, \underbrace{M+1, \ldots, M+1}_{C} \}  & I_2^{(M+1, 1)} &= \emptyset  \\
    I_1^{(M+1, 2)} &= I_1 & I_2^{(M+1, 2)} &= \{\underbrace{M+1, \ldots, M+1}_{C}\}
\end{align*}


Consider any algorithm $\alg$ that claims to be $\crr_{\lpp, m}$ competitive and how it behaves when faced with our primary input sequence $\{I_1, I_2\}$. Let $a_{i,t}$ be the number of type $i$ agents that $\alg$ accepts from $I_t$ and define $s_i = a_{i,1} + a_{i,2}$ to be the total number of type $i$ agents accepted in $\{I_1, I_2\}$. Note that for $t=2$, since type $i \in [m]$ agents do not appear in $I_2$, we have that $a_{i,2} = 0$. In order for $\alg$ to be truly $\crr_{\lpp, m}$ competitive, it must achieve this CR against not only $\{I_1, I_2\}$, but also $\{I_1^{(i,j)}, I_2^{(i, j)}\}$. For each of these input sequences, we consider what must be true about the $a_{i,t}$'s in order for $\alg$ to be $\crr_{\lpp, m}$ competitive against them. \\

Consider input sequences given by the bullets below. In each bullet, we present a collection of arrival sequences and state that if any algorithm wants to be $\crr_{LP, m}$ competitive overall, then it needs to satisfy certain constraints. 

\begin{itemize}
    \item \textbf{Sequences $\{I_1, I_2\}$.} The optimal clairvoyant revenue for each of $I_1$ and $I_2$ is $r_K \cdot C$. In order for our algorithm to achieve an $\crr_{\lpp,m}$ CR against that, a necessary condition is that the total reward of all the agents we accept is at least $2 \cdot \crr_{\lpp, m} \cdot r_K$:
    \[ \sum_{t=1}^2 \sum_{i=1}^K r_i \cdot a_{i,t} \geq 2 \cdot \crr_{\lpp, m} \cdot r_K\,.  \]
    The left hand side can also be written as $\sum_{i=1}^K r_i \cdot s_i$. This results in constraint \eqref{sup_last_type} from the LP. 
    \item \textbf{Sequences $\{I_1^{(i,1)}, I_2^{(i, 1)}\}$ for $i \in \{1, 2\ldots, m\}$.} 
    For such values of $i$, note that $I_2^{(i, 1)} = \emptyset$ and so we focus on how $\mathcal{A}$ behaves on $I_1^{(i,1)}$. The optimal algorithm achieves a reward of $C \cdot r_i$ on $I_1^{(i,1)}$ and so in order for $\alg$ to be $\crr_{\lpp,m}$ competitive, certainly a necessary condition is that the total revenue from the agents we have accepted from $I_1^{(i,1)}$ must be at least $\crr_{\lpp,m } \cdot r_i \cdot C$.  Since $I_1^{(i,1)}$ and $I_1$ are identical for the first $C \cdot i$ agents, when $\alg$ is run on $I_1^i$, it must have the same behavior as on the first $C \cdot i$ agents of $I_1$, meaning $\alg$ accepts $a_{j,1}$ copies of type $j$ for $j \in [i]$. Therefore, the following must necessarily be true for $\alg$ to truly be $\crr_{\lpp}$ competitive: 
    \[ \sum_{j=1}^i r_j \cdot a_{j,1} \geq \crr_{\lpp, m} \cdot r_i \cdot C \]
    Since $a_{j,2} = 0$ for $j \in [M]$ (no flexible agents appear in the second time period), we can replace $a_{j,1}$ with $s_{j}$ and this gives constraint \eqref{sup_before_m} in the LP.  
    
    \item \textbf{Sequences $\{I_1^{(i,1)}, I_2^{(i,1)}\}$ and $\{I_1^{(i,2)}, I_2^{(i,2)}\}$ for $i \in \{M+1, \ldots, K-1\}$}. Here, we consider these two input sequences together because they are both truncated after $C$ copies of type $i$ (inflexible) agents have arrived, meaning that the optimal revenue for the time period that the truncation happens is $C \cdot r_i$. For $\{I_1^{(i,1)}, I_2^{(i,1)}\}$, the same reasoning as the $i \in [M]$ case applies, so we must have that 
     \begin{equation}
        \label{sup_first}
         \sum_{j=1}^i r_j \cdot a_{j,1} \geq \crr_{\lpp, m} \cdot r_i \cdot C
     \end{equation}
    For $\{I_1^{(i,2)}, I_2^{(i,2)}\}$, we claim that the following is a \textit{upper bound} on the  total reward of the agents that $\alg$ has accepted at the end of $I_2^{(i,2)}$ 
    \begin{equation}
        \label{eq:sup_second}
        \sum_{j=1}^m r_j \cdot a_{j,1} + \sum_{j=M+1}^i r_j \cdot a_{j,2}\,.
    \end{equation} 
    At the end of $I_2^{(i,2)}$, the agents that $\alg$ has accepted and not yet serviced can only contain partially patient agents from period 1 (i.e. $I_1^{(i,2)}$) and agents that arrived in $I_2^{(i,2)}$, all of which are inflexible. Up to $a_{j,1}$ agents of type $j \in [m]$ can still be waiting for service, and up to $a_{j,2}$ agents of type $j \geq M+1$. The total reward of all such agents is by \eqref{eq:sup_second} above, that is certainly an upper bound on the reward that an algorithm can service at the end of $I_2^{(i,2)}$. 
    In order for an algorithm to be $\crr_{\lpp,m}$ competitive, the quantity in \eqref{eq:sup_second} above must also be at least $\crr_{\lpp,m} \cdot r_i \cdot C$. Adding up the inequality \eqref{sup_first} with the fact that expression \eqref{eq:sup_second} must be at least $\crr_{\lpp,m} \cdot r_i \cdot C$ yields 
    \begin{align*}
    &\:\sum_{j=1}^i r_j \cdot a_{j,1} + \sum_{j=1}^m r_j \cdot a_{j,1} + \sum_{j=M+1}^i r_j \cdot a_{j,2} \\
    =\:& 
    \left(\sum_{j=1}^m r_j \cdot a_{j,1}  + \sum_{j=M+1}^i r_j \cdot a_{j,1}\right)+ \sum_{j=1}^m r_j \cdot a_{j,1} + \sum_{j=M+1}^i r_j \cdot a_{j,2} \\
     =&\: 2\sum_{j=1}^m r_j \cdot a_{j,1}  + \sum_{j=M+1}^i r_j \cdot (a_{j,1}+a_{j,2}) \\
     =&\:2 \cdot \sum_{j=1}^m r_j \cdot s_j + \sum_{j=M+1}^i r_j \cdot s_j\\
     \geq&\: 2 \cdot \crr_{\lpp,m} \cdot C\,.
    \end{align*}
    This gives us constraint \eqref{sup_after_m} in the LP. 
\end{itemize}

Finally, we know that since all the agents in $\{I_1, I_2\}$ must be served within these 2 time periods, the total number of agents accepted cannot exceed $2 \cdot C$. This gives us constraint \eqref{sup_sum_constraint}. Thus, we have verified that  $\crr_{\lpp,m}$, $m \in \{0,1, \ldots , M\}$ is an upper bound on the CR of any non-anticipating algorithm $\alg$.

\end{proof}

\subsection{Proof of Lemma \ref{thm:loose_upper_bound}}

\begin{proof}{Proof of Lemma \ref{thm:loose_upper_bound}}

  In this proof, we will greatly leverage Lemma \ref{lemma:closed_form} (presented in Section \ref{sec:closed-form_lemma}), which gives us closed form expressions for $\crr_{LP, 0}$ and $\crr_{LP, M}$ under some mild conditions. For any $m \in \{0, 1, \ldots, M\}$, recall that  $\crr_{LP, m}$ is the optimal objective value to the LP defined in Lemma \ref{lemma:upper_bound_lp_m}, whereas $\crr_{LP, m}^*$ are the closed form expressions that correspond to $\crr_{LP, m}$ under a mild assumptions given in Lemma \ref{lemma:closed_form}. 

  First, we will show that we can equivalently write $\overline{\crr}_{up}$ in terms of these closed form expressions: 
  \begin{equation}
        \label{equivalent_closed_form}
      \overline{\crr}_{up} = \min(\crr_{LP, 0}^*, \crr_{LP, M}^*)\,.
  \end{equation}
 To see this, recall the definition of $\crr_{LP, m}^*$ for any $m \in \{0, 1, \ldots, M\}$: 
  $$ 
   \crr_{LP, m}^* \defeq  2 \cdot \left[\frac{r_m}{r_K} + 2 \cdot \left(1 - \frac{r_m}{r_{M+1}} \right) + \sum_{j=1}^m \left(1 - \frac{r_{j-1}}{r_j}\right) + 2 \cdot \sum_{j=M+2}^{K}  \left( 1 - \frac{r_{j-1}}{r_j} \right)\right]^{-1}\,.
  $$
  For $m = 0$, we get that $$\crr_{LP, 0}^* = 2 \cdot \left[ 2 + 2 \cdot \sum_{j=M+2}^{K} \left( 1 - \frac{r_{j-1}}{r_j} \right) \right]^{-1} = \left[1 + (K - M - 1) - \sum_{j=M+2}^{K} \frac{r_{j-1}}{r_j} \right]^{-1} = \frac{1}{G}\,.$$ 

  For $m = M$, we have: 
  \begin{align*}
        \crr_{LP, M}^* &=  2 \cdot \left[\frac{r_M}{r_K} + 2 \cdot \left(1 - \frac{r_M}{r_{M+1}} \right) + \sum_{j=1}^M \left(1 - \frac{r_{j-1}}{r_j}\right) + 2 \cdot \sum_{j=M+2}^{K}  \left( 1 - \frac{r_{j-1}}{r_j} \right)\right]^{-1} \\
&=  2 \cdot \left[\frac{r_M}{r_K} - \frac{r_M}{r_{M+1}} + \sum_{j=1}^M \left(1 - \frac{r_{j-1}}{r_j}\right) - \frac{r_M}{r_{M+1}} + \underbrace{2 + 2 \cdot \sum_{j=M+2}^{K}  \left( 1 - \frac{r_{j-1}}{r_j} \right)}_{=2 G, \text{ from above}}\right]^{-1} \\
            &= 2\left[ 2\cdot G + M - \sum_{i=1}^{M+1} \frac{r_{i-1}}{r_i} - \frac{r_M}{r_{M+1}} + \frac{r_M}{r_K} \right]^{-1}\,.
    \end{align*}
 Therefore, we have shown equation \eqref{equivalent_closed_form}.

  Our goal in this proof is to show that $\overline{\crr}_{up} \geq \crr_{LP}$, where so far  we have shown that $\overline{\crr}_{up} = \min(\crr_{LP, 0}^*, \crr_{LP, M}^*)$. We will perform some case-work depending on whether $\overline{\crr}_{up}$ takes on the value $\crr_{LP, 0}^*$ or $\crr_{LP, M}^*$. 

  \textbf{Case 1: $\overline{\crr}_{up} = \crr_{LP, 0}^*$.} From Lemma \ref{lemma:closed_form}, we have that $\crr_{LP, 0}^* = \crr_{LP, 0}$ holds if the condition $\frac{1}{r_0} - \frac{2}{r_{M+1}} + \frac{1}{r_{K}} \geq 0$ is true. However, since $r_0$ is defined to be 0, this is always true and thus $\overline{\crr}_{up} = \crr_{LP, 0}^* = \crr_{LP, 0} \geq \crr_{LP}$, where the last inequality is true since $\crr_{LP}$ is defined to be the minimum of $M+1$ terms, which includes $\crr_{LP, 0}$ (and also $\crr_{LP, M}$).

  \textbf{Case 2: $\overline{\crr}_{up} = \crr_{LP, M}^*$.} Similar to Case 1, if the condition $\frac{1}{r_M} - \frac{2}{r_{M+1}} + \frac{1}{r_K} \geq 0$ holds, then $\crr_{LP, M}^* = \crr_{LP, M}$ and we are done by the same reasoning as Case 1. Otherwise, we define $\tilde{m}$ as follows: 
  $$ \tilde{m} \defeq \max \left\{m\in \{0, 1, 2,\ldots, M\} \mid \frac{1}{r_m} - \frac{2}{r_{M+1}} + \frac{1}{r_K} \geq 0 \right\}\,. $$
  Observe that $\tilde m$ is well defined because at $m=0$, we have $\frac{1}{r_m} - \frac{2}{r_{M+1}} + \frac{1}{r_K}\ge 0$ (recall that $r_0=0$). By our definition of $\tilde{m}$, we have: 
  \begin{align}
      \frac{1}{r_{\tilde{m}}} - \frac{2}{r_{M+1}} + \frac{1}{r_K} &\geq 0 \label{tilde_m_satisfies} \\
    \frac{1}{r_{m}} - \frac{2}{r_{M+1}} + \frac{1}{r_K} &< 0 \quad \forall m \in \{\tilde{m} + 1, \ldots, M\} \label{other_m_not_satisfy}
  \end{align}
  Now, consider any $m \in \{\tilde{m} + 1, \ldots, M\}$. We claim that $\crr_{LP, m}^* \geq \crr_{LP, m-1}^*$. To see this, let us compare the closed form expressions for $\crr_{LP, m}^*$ and $\crr_{LP, m-1}^*$:
    \begin{align*}
        \crr_{LP, m}^* &= 2 \cdot \left[\frac{r_m}{r_K} + 2 \cdot \left(1 - \frac{r_{m}}{r_{M+1}} \right) + \sum_{j=1}^{m} \left(1 - \frac{r_{j-1}}{r_j}\right) + 2 \cdot \sum_{j=M+2}^{K}  \left( 1 - \frac{r_{j-1}}{r_j} \right)\right]^{-1} \\
        \crr_{LP, m-1}^* &= 2 \cdot \left[\frac{r_{m-1}}{r_K} + 2 \cdot \left(1 - \frac{r_{m-1}}{r_{M+1}} \right) + \sum_{j=1}^{m-1} \left(1 - \frac{r_{j-1}}{r_j}\right) + 2 \cdot \sum_{j=M+2}^{K}  \left( 1 - \frac{r_{j-1}}{r_j} \right)\right]^{-1} \\
  \end{align*}

  Showing that $\crr_{LP, m}^* \geq \crr_{LP, m-1}^*$ is equivalent to showing that 

$$ 
\frac{r_m}{r_K} - 2 \cdot \frac{r_{m}}{r_{M+1}}  + \left(1 - \frac{r_{m-1}}{r_m}\right)  \leq \frac{r_{m-1}}{r_K} - 2 \cdot \frac{r_{m-1}}{r_{M+1}} \, .
$$

Simplifying the above to one side gives: 
$$ 
 (r_m - r_{m-1}) \cdot \left[ \frac{1}{r_K} - \frac{2}{r_{M+1}} - \frac{1}{r_m} \right] \leq 0 \, ,
$$
which is true since $r_{m} > r_{m-1}$ and the expression in square brackets is negative by \eqref{other_m_not_satisfy}. Thus, 
\begin{align*}
    \overline{\crr}_{up} = \crr_{LP, M}^* \geq \crr_{LP, M-1}^* \geq \ldots \geq \crr_{LP, \tilde{m}}^* = \crr_{LP, \tilde{m}} \geq \crr_{LP}\,.
\end{align*}
Here, the second-to-last inequality is true by applying Lemma \ref{lemma:closed_form} and equation \eqref{tilde_m_satisfies}, while the last inequality is true because $\crr_{LP}$ is defined to be a minimum of $M+1$ terms, one of which is $\crr_{LP, \tilde{m}}$. This completes the proof.

\end{proof}


    

\subsection{Lemma \ref{lemma:closed_form} and its Proof} \label{sec:closed-form_lemma}
\begin{lemma}
\label{lemma:closed_form}
Suppose we have rewards $0 =r_0<r_1 < \ldots < r_K$ with the first $M \in [K-1]$ types partially patient and any $m \in \{0,1, \ldots, M\}$ such that $\frac{1}{r_m} - \frac{2}{r_{M+1}} + \frac{1}{r_K} \geq 0$. 
Then, a closed form solution to the LP for $\crr_{\lpp, m}$ is given by: 
\begin{equation}
\label{closed_form_lp_solution}
    \crr^*_{\lpp, m} \defeq \frac{2}{\frac{r_m}{r_K} + 2 \cdot \left(1 - \frac{r_m}{r_{M+1}} \right) + \sum_{j=1}^m \left(1 - \frac{r_{j-1}}{r_j}\right) + 2 \cdot \sum_{j=M+2}^{K}  \left( 1 - \frac{r_{j-1}}{r_j} \right)}\,.
\end{equation}
\end{lemma}

\begin{proof}{Proof of Lemma \ref{lemma:closed_form}} 
Fix a value of $m$. For brevity, we will use $\crr^*$ to refer to $\crr^*_{\lpp, m}$ as defined by equation \eqref{closed_form_lp_solution}. To show that $\crr^*$ is the optimal solution to the LP for $\crr_{\lpp, m}$, we will present an explicit primal and dual solution and verify that complementary slackness is satisfied. For the sake of brevity, we take $C = 1$. Note that we can replace each $s_i$ with $s_i \cdot C$ ad all the $C$'s would cancel out in each constraint. Therefore, the objective value of the LP does not depend on $C$. The primal dual pair is:  

{ \scriptsize 
\begin{align}
\text{max} \quad  & \crr & \text{maximize} \quad & 2 \cdot t & \nonumber \\
\text{s.t.} \quad & r_i \cdot \crr \leq \sum_{j=1}^i r_j \cdot s_j & \text{s.t.} \quad &\lambda_i \geq 0 & i \in \{1, \ldots, m\} \label{eq:primal1} \\
& 2 \cdot r_i \cdot \crr \leq 2 \cdot \sum_{j=1}^{m} r_{j} \cdot s_j + \sum_{j=M+1}^{i} r_j \cdot s_j && \lambda_i \geq 0 & i \in \{M+1, \ldots,K-1\} \label{eq:primal2} \\
&  2 \cdot r_K \cdot \crr \leq \sum_{j=1}^{m
} r_{j} \cdot s_j + \sum_{j=M+1}^{K} r_j \cdot s_j && \lambda_K \geq 0 \label{eq:primal3} &  \\
&  \sum_{j=1}^K s_j \leq 2  && t \geq 0 & \label{eq:primal4} \\
& s_i \geq 0 && t \geq r_i \cdot \left( \sum_{j=i}^m \lambda_i + 2 \cdot \sum_{j=M+1}^{K-1} \lambda_i + \lambda_K \right) & i \in \{1, \ldots, m\} \label{eq:dual1}\\
& s_i \geq 0 &&t \geq r_i \cdot \left( \sum_{j=i}^K \lambda_j \right) & i \in \{M+1, \ldots, K\} \label{eq:dual2} \\
& \crr \geq 0 && \sum_{j=1}^m r_j \cdot \lambda_j + 2 \cdot \sum_{j=M+1}^K r_j \cdot \lambda_j \geq 1 & \label{eq:dual3} 
\end{align}}

Consider the following closed form for the primal/dual pair.

{ \scriptsize
\begin{align*}
    \crr^* &= \frac{2}{L^*}  & t^* &= \frac{1}{L^*} \\
    s_i^* &= 
    \begin{cases} 
        \crr^* \cdot \left( 1 - \frac{r_{i-1}}{r_i} \right) & i \in \{1,  \ldots, m\} \\
        2 \cdot \crr^* \cdot \left( 1 - \frac{r_{m}}{r_{M+1}} \right) & i = M+1 \text{ and } M+1 < K \\
        2 \cdot \crr^* \cdot \left( 1 - \frac{r_{i-1}}{r_i} \right) & i \in \{M+2, \ldots, K-1\} \\
        2 \cdot \crr^* \cdot \left( 1 - \frac{r_{K-1}}{r_K} + \frac{1}{2} \cdot \frac{r_m}{r_K} \right) & i = K \text{ and } M+1 < K \\
        2 \cdot \crr^* \cdot \left( 1 - \frac{1}{2} \cdot \frac{r_m}{r_K} \right) & i = K \text{ and } M+1 = K 
    \end{cases} 
    &
    \lambda^*_i &= 
    \begin{cases} 
        t^* \cdot \left( \frac{1}{r_i} - \frac{1}{r_{i+1}} \right) & i \neq m \\
        t^* \cdot \left( \frac{1}{r_m} - \frac{2}{r_{M+1}} + \frac{1}{r_K} \right) & i = m
    \end{cases}
\end{align*}
}

where 
$$ 
L^* \defeq \frac{r_m}{r_K} + 2 \cdot \left(1 - \frac{r_m}{r_{M+1}} \right) + \sum_{i=1}^m \left(1 - \frac{r_{i-1}}{r_i}\right) + 2 \cdot \sum_{i=M+2}^{K}  \left( 1 - \frac{r_{i-1}}{r_i} \right) \,.
$$
Note that we define $r_{0} = 0$ and $r_{K+1} = \infty$ (and so $\frac{1}{r_{K+1}} = 0$). 

It is clear that the proposed solutions above achieve the same objective value, so we just need to verify that they are each feasible. 

\textbf{Primal LP Feasibility.}  We examine each constraint separately.  
\begin{itemize}
    \item Constraint \ref{eq:primal1}: For any $i\in \{1, 2,\ldots, m\}$, we have 
    \begin{align*}
        \sum_{j=1}^i r_j \cdot s_j^* &= \crr^* \cdot \sum_{j=1}^i r_j \cdot \left(1 - \frac{r_{i-1}}{r_i} \right) = \crr^* \cdot r_i \,.
    \end{align*}

    \item Constraint \eqref{eq:primal2}: We verify this constraint by expanding the left hand side as follows:
    \begin{align*}
        2 \cdot \sum_{j=1}^{m} r_{j} \cdot s_j^* + \sum_{j=M+1}^{i} r_j \cdot s_j^* &= 2 \cdot \crr^* \cdot r_m + 2 \cdot \crr^* \cdot \sum_{j=M+1}^{i} r_i \cdot \left(1 - \frac{r_{i-1}}{r_i} \right) \\
        &= 2 \cdot \crr^* \cdot r_m + 2 \cdot \crr^* (r_i - r_m) \\
        &= 2 \cdot \crr^* \cdot r_i\,.
    \end{align*}
    Note that if $M+1 = K$, then this constraint does not exist in the LP, so we do not need to consider that case. 

    \item Constraint \eqref{eq:primal3}: 
    We consider two cases based on whether $M+1 = K$ or $M+1 < K$. For the case when $M+1 = K$, the right hand side of constraint \eqref{eq:primal3} can be written as:
    \begin{align*}
        \sum_{j=1}^m r_j \cdot s_j^* + r_K \cdot s_K^* = \crr^* \cdot r_m + \crr^* \cdot (2 \cdot r_K - r_m) = 2 \cdot \crr^* \cdot r_K
    \end{align*}

    For the case when $M+1 < K$, we have: 
    \begin{align*}
        \sum_{j=1}^m r_j \cdot s_j^* + \sum_{j=M+1}^K r_j \cdot s^*_j &= \crr^* \cdot r_m + 2 \cdot \crr^* \cdot (r_{K-1} - r_m) + 2 \cdot \crr^* \cdot (r_K - r_{K-1} + 1/2 \cdot r_m) = 2 \cdot \crr^* \cdot r_K\,.
    \end{align*}

    \item Constraint \eqref{eq:primal4}: For the case where $M+1 = K$, we have: 
    \begin{align*}
        \sum_{j=1}^K s_j^* = \crr^* \cdot \left[ \sum_{j=1}^m \left(1 - \frac{r_{j-1}}{r_j} \right) + 2  \cdot \left(1 - \frac{1}{2} \cdot \frac{r_{m}}{r_K} \right) \right] = 2\,.
    \end{align*}
    The term inside the square brackets above is exactly $L^*$ when $M +1 = K$ (in the definition of $L^*$, the last summation does not exist since $M+2 > K$). 

    For the case where $M+1 < K$, we have: 
    \begin{align*}
        \sum_{j=1}^K s_j^* &= \crr^* \cdot \left[\sum_{j=1}^m \left(1 - \frac{r_{i-1}}{r_i} \right) + 2 \cdot \left(1 - \frac{r_m}{r_{M+1}} \right) + 2 \cdot \sum_{j=M+2}^{K-1} \left(1 - \frac{r_{i-1}}{r_i} \right) + 2 \cdot \left(1 - \frac{r_{K-1}}{r_K} + \frac{1}{2} \frac{r_m}{r_K} \right) \right] = 2\,.
    \end{align*}

    Again, we see that the term in square brackets is exactly $L^*$. 

    \item \textbf{Positivity}. All of our solution values $s_i^*, \crr^*$ are non-negative because $r_i / r_j < 1$ for any $i < j$.

\end{itemize}

\textbf{Dual LP Feasibility. } We examine each of the constraints separately. 
\begin{itemize}
    \item Constraint \ref{eq:dual1}: We expand the inside of the parentheses on the left hand side as:
    \begin{align*}
        \sum_{j=i}^m \lambda_i^* &+ 2 \sum_{M+1}^{K-1} \lambda_i^* + \lambda_K \\
        &= 
        t^* \left( \frac{1}{r_i} - \frac{1}{r_{m}} \right) + t^* \left( \frac{1}{r_m} - \frac{2}{r_{M+1}} + \frac{1}{r_K} \right) + 2 t^* \left(\frac{1}{r_{M+1}} - \frac{1}{r_K} \right) + t^* \left(\frac{1}{r_K} - \frac{1}{r_{K+1}}\right) = \frac{t^*}{r_i}\,.
    \end{align*}

    \item Constraint \eqref{eq:dual2}: It is easy to verify that:
    \begin{align*}
        \sum_{j=i}^K \lambda_j^* = t^* \left(\frac{1}{r_i} - \frac{1}{r_{K+1}} \right) = \frac{t^*}{r_i}\,.
    \end{align*}

    \item Constraint \eqref{eq:dual3}: The left  side of the inequality can be written as: 
    \begin{align*}
        \sum_{j=1}^m r_j \cdot \lambda_j^* + 2 \cdot \sum_{j=M+1}^K r_j \cdot \lambda_j^*
        =& t^* \sum_{j=1}^{m-1} \left(1 - \frac{r_j}{r_{j+1}} \right) + t^* \left(1 - 
2 \cdot \frac{r_m}{r_{M+1}} + \frac{r_m}{r_K} \right) + 2 \cdot t^* \sum_{j=M+1}^{K} \left(1 - \frac{r_j}{r_{j+1}} \right) \\
=\, & t^*\left[\sum_{j=1}^{m} \left(1 -  \frac{r_{j-1}}{r_j}\right) - 1  + \left(1 - 
2 \cdot \frac{r_m}{r_{M+1}} + \frac{r_m}{r_K} \right) + 2 \sum_{j=M+2}^{K}   \left(1 - \frac{r_{j-1}}{r_{j}} \right) + 2 \right]\,.
    \end{align*}
    
    The term in brackets above is exactly $L^*$ and so the above simplifies to $1$, which satisfies the constraint.
    \item \textbf{Positivity}. All of the variables are non-negative due to the fact that $r_i < r_j$ for $i < j$ (and hence $\frac{1}{r_i} - \frac{1}{r_j} > 0$), as well as the assumption in the lemma statement that $\frac{1}{r_m} - \frac{2}{r_{M+1}} + \frac{1}{r_K} \geq 0$.
    
\end{itemize}

Since we have exhibited a feasible primal and dual pair which achieve the same objective value, this proves that $\crr^*_{\lpp, m}$ indeed has the closed form given in equation \eqref{closed_form_lp_solution}.

\end{proof}
}

\subsection{Proof of the Lemmas Used in the Proof of Lemma \ref{thm:optimality_gap}}
\label{sec:optimality_gap_proofs}

Here, we first present the three lemmas used in the proof of Lemma \ref{thm:optimality_gap}. We then proceed to showing them. 

\begin{lemma}[$\crr_{\text{nest}}$ vs. the Upper Bound on the CR of any Nested Alg. ($\overline{\crr}$)]
\label{lemma:appx_one_of_two} Let $\bar \crr$ be the quantity defined in equation \eqref{definition:gamma_bar}. Then, 
\begin{equation}
     \frac{\crr_{\text{nest}}}{\overline{\crr}} \geq f_1(G) \defeq
    \begin{cases}
        1 & 2 \cdot \overline{n}_M \leq C \\
        1 - \frac{1}{2} e^{-2 \cdot G + 1} & 2 \cdot \overline{n}_M > C 
    \end{cases} 
    \label{equation:f1}
\end{equation}
where $G$ is defined in equation \eqref{eq:G}. 
\end{lemma}

\begin{lemma}[Upper Bound on the CR of any Nested Alg. ($\overline{\crr}$) vs. that of any Alg. ($\overline{\crr}_{up}$)]
\label{lemma:appx_two_of_two}
Let $\bar \crr$ be the quantity defined in equation \eqref{definition:gamma_bar} and $\overline{\crr}_{up} = \min \left( \crr_{\lpp}, \frac{1}{G} \right)$, defined in equation \eqref{eq:G}, be an upper bound on the CR of any non-anticipating algorithm. Then, 
\begin{equation}
\frac{\overline{\crr}}{\overline{\crr}_{up}} \geq f_2(G) \defeq  \begin{cases}
        1 - \frac{1 - \max(2 - G, 0)}{2 \cdot G + 1 - \max(2-G, 0)} & 2 \cdot \overline{n}_M \leq C \\
        1 - \frac{1 - \max(2-G, 0)}{4 \cdot G - 2} & 2\cdot \overline{n}_M > C
    \end{cases} \label{equation:f2}
\end{equation}
where $G$ is defined in equation \eqref{eq:G}.
\end{lemma}

\begin{lemma}[Combining $f_1(\cdot)$ and $f_2(\cdot)$]
\label{lemma:appx_combine}
For any value of $G \in (1, K-M)$, we have that $f_1(G) \cdot f_2(G) \geq 0.8$, where $G$ is defined in equation \eqref{eq:G}.
\end{lemma}

With these 3 lemmas, we have proved the desired result for Lemma \ref{thm:optimality_gap} that $\crr_{nest} \geq 0.8 \cdot \overline{\crr}_{up}$, which also proves Theorem \ref{theorem:main_optimality_gap} that $\crr^* \geq 0.8 \cdot \overline{\crr}_{up}$ (since $\crr^* \geq \crr_{nest}$).

\subsubsection{Proof of Lemma \ref{lemma:appx_one_of_two}}
\begin{proof}{Proof of Lemma \ref{lemma:appx_one_of_two}.}
The goal of this lemma is to show that our instance of the nested polytope, which is constructed using $\overline \crr$ and $\overline{\deltan_i}$ has a CR, denoted by $\crr_{\text{nest}}$, which is not too far from $\overline \crr$ itself. Recall that 

\begin{align*} 
    \overline \crr &= 
    \frac{2}{2 \cdot G + M - \sum_{i=1}^{M+1} \frac{r_{i-1}}{r_i}} &
{\overline{\deltan_i}} &= 
\begin{cases} 
\frac{1}{2} \cdot \overline \crr \cdot \left( 1 - \frac{r_{i-1}}{r_{i}} \right) \cdot C & i \in [M] \\
\overline \crr  \cdot \left( 1 - \frac{1}{2} \frac{r_{i-1}}{r_{i}} \right) \cdot C & i = M+ 1 \\
\overline \crr  \cdot \left( 1 - \frac{r_{i-1}}{r_{i}} \right) \cdot C & i= M+2, \ldots, K
\end{cases}
\end{align*} To show this, we will use the LP \ref{lp:nested_polytope} from Theorem \ref{theorem:polyra_nested_performance}, which tells us how to compute the CR of any arbitrary instance of the nested polytope with given values of $(\overline{\deltan_1}, \ldots, \overline{\deltan_K})$. We would like to show that this LP has an objective value at least $f_1(G) \cdot \overline{\crr}$. To do so, we will a feasible solution $(\crr, (s_i^j)_{j\in [K], i\in [j]})$ to the LP with $\crr = f_1(G) \cdot \overline \crr$. Such a feasible solution needs to satisfy constraints of LP \ref{lp:nested_polytope}, i.e., constraints  \eqref{nested_lp_revenue_target}, \eqref{nested_lp_capacity},  \eqref{nested_lp_j_le_M}, and \eqref{nested_lp_j_ge_M}.

Since our function $f_1(G)$ is piecewise, we consider two cases based on the relationship between $2 \cdot n_M$ and $C$.  Recall that 
\begin{equation*}
 f_1(G) \defeq
    \begin{cases}
        1 & 2 \cdot \overline{n_M} \leq C \\
        1 - \frac{1}{2} e^{-2 \cdot G + 1} & 2 \cdot \overline{n_M} > C 
    \end{cases} 
\end{equation*}

\textbf{Case 1: $2 \cdot \overline{n_M} \leq C$}: For this case, we see that $f_1(G) = 1$, which means that we need to exhibit a collection of feasible $\{s_i^j\}_{j\in[K], i \in [j]}$ such that all constraints in \ref{lp:nested_polytope} are satisfied with $\crr = \overline{\crr}$. We consider two sub-cases based on whether $j \leq M$ or $j > M$.

\begin{itemize}
    \item Case 1a: $j \leq M$. In this case, constraint \eqref{nested_lp_j_le_M} requires $s_i^j \in [\overline{\deltan_i}, 2 \cdot \overline{\deltan_i}]$. We choose the solution $s_i^j = 2 \cdot \overline{\deltan_i}$. Since $\sum_{i=1}^j s_i^j = \sum_{i=1}^j 2 \cdot \overline{\deltan_i} = 2 \cdot \overline{n_j} \leq 2 \cdot \overline{n_M} \leq C$, we know that the capacity constraint \eqref{nested_lp_capacity} is also satisfied. For $j \leq M$, the only other constraint is the revenue requirement \eqref{nested_lp_j_le_M}, which we verify as follows: 
    \begin{align}
        \sum_{i=1}^j r_i \cdot s_i^j &= 2 \cdot \sum_{i=1}^j r_i \cdot \frac{1}{2} \cdot \overline \crr \cdot \left( 1 - \frac{r_{i-1}}{r_i} \right) \cdot C \nonumber \\
        &= \overline \crr \cdot C \cdot \sum_{i=1}^j r_i - r_{i-1} \nonumber \\
        &= \overline \crr \cdot r_j \cdot C\,. \label{appx_jleqM}
    \end{align}
    The first equation holds because by our definition and  nest sizes, we have $s_i^j = 2\cdot\deltan_i = \frac{1}{2}\cdot \bar \crr \cdot (1-r_{i-1}/r_i)$.
    
    \item Case 1b: $j > M$. Constraint \eqref{nested_lp_j_ge_M} forces us to choose $s_{i}^j = \deltan_i$. This clearly satisfies the capacity constraint \eqref{nested_lp_capacity} and so the only constraint left is the revenue requirement \eqref{nested_lp_revenue_target}:
    \begin{align*}
        \sum_{i=1}^j r_i \cdot s_i^j &= \overline \crr \cdot C \cdot \left[ \frac{1}{2} \cdot \sum_{i=1}^M r_i \cdot \left( 1 - \frac{r_{i-1}}{r_i} \right)  + r_{M+1} \cdot \left(1 - \frac{1}{2} \frac{r_{M}}{r_{M+1}} \right)+ \sum_{i=M+2}^j r_i \cdot \left( 1 - \frac{r_{i-1}}{r_i} \right) \right] \\
        &= \overline \crr \cdot C \cdot \left[ \frac{1}{2} \cdot r_M + \left(r_{M+1} - \frac{1}{2} r_M \right) + (r_j - r_{M+1}) \right]\\
        &=  \overline \crr \cdot r_j \cdot C\,.
    \end{align*}
\end{itemize}
Note that only in the $j \leq M$ case did we use the assumption that $2 \cdot n_M \leq C$. This means that for Case 2, the same argument holds for $j > M$ and we only need to work with $j \leq M$. 

\textbf{Case 2: $2 \cdot \overline{n_M} > C$ and $j \leq M$}. Our goal is this case is to prove that for all $j \in [M]$, we have can find a choice of $s_i^j$ for $i \in [j]$ such that $ f_1(G) \cdot \overline \crr \leq \sum_{i=1}^j r_i \cdot s_i^j $. Fix a particular value of $j \in [M]$. If $2 \cdot \overline{n_j} \leq C$, then we are done because setting $s_i^j = 2 \cdot \overline{\deltan}_i$ for $i \in [j]$ is feasible, and we can apply the same set of equalities as in \eqref{appx_jleqM}. 

Otherwise, assume that $2 \cdot \overline{n_j} > C$ and so setting $s_i^j = 2 \cdot \overline{\deltan_i}$ for $i \in [j]$ is not feasible as it breaks the capacity constraint on $s_i^j$ (i.e. constraint \eqref{nested_lp_capacity} in \ref{lp:nested_polytope}). Since we can no longer  choose $s_i^j = 2 \cdot \overline{\deltan_i}$ for all $i \in [j]$, we choose to \textit{greedily} set them so that $\sum_{i=1}^j r_i \cdot s_i^j$ is maximized. This involves  first setting all the $s_i^j$'s , $i\in [j]$, to be $\overline{\deltan_i}$ (the lower bound of $s_i^j$) and then using the remaining $C - \sum_{i=1}^j \overline{\deltan_i} = C- \overline{n_j}$ capacity to greedily make increase each of $s_{j}^j, s_{j-1}^j, s_{j-2}^j, \ldots, s_{1}^j$ by as much as possible up to $2 \cdot \overline{\deltan_i}$ in that order. More formally, let $\ell$ be defined as follows: 
$$
\ell \defeq \max \left\{ k \in [j] \: : \: \overline{n_j} + \sum_{i'= k}^j \overline{\deltan_{i'}} > C\right\}\,. 
$$
Note that $\ell$ is well defined because we assumed that $2 \cdot n_j > C$ so certainly the index $1$ is in the set and hence we are not taking the maximum of an empty set. We choose our $s_i^j$ as follows:

\begin{equation}
    s_i^j = \begin{cases}
        \overline{\deltan_i} & i < \ell \\
        \overline{\deltan_i} + C - n_j - \sum_{i'=\ell+1}^j \overline{\deltan_{i'}} & i = \ell \\
        2 \cdot \overline{\deltan_i} & j \ge i > \ell
    \end{cases}
\end{equation}
For $i > \ell$, our construction chooses $s_i^j = 2 \cdot \overline{\deltan_i}$ (the upper bound of each $s_i^j$), while for $i < \ell$, we choose $s_i^j$ to be $\overline{\deltan_i}$ (the lower bound for $s_i^j$). For $i = \ell$, we choose $s_\ell^j$ to use up the remaining capacity so that $\sum_{i=1}^j s_i^j = C$ is satisfied. {To see why this is true, note that
\begin{align*}
    \sum_{i=1}^j s_i^j &= \sum_{i=1}^{\ell - 1} \overline{\deltan_i} + \overline{\deltan_{\ell}} + C - n_j - \sum_{i=\ell+1}^j \overline{\deltan_i} + 2 \cdot \sum_{i=\ell+1}^j \overline{\deltan_i} \\
    &= \sum_{i=1}^j \overline{\deltan_i} + C - n_j = C\,.
\end{align*}
}
This shows that constraint  \eqref{nested_lp_capacity} is satisfied by our choice of $s_i^j$. Now, we verify that $s_i^j \in [\deltan_i, 2 \cdot \deltan_i]$ (constraint \eqref{nested_lp_j_le_M}). This is clear for $i \neq \ell$. For $i = \ell$, we need to show that $C - \overline{n_j} - \sum_{i'=\ell+1}^j \deltan_{i'} \in [0, \deltan_\ell]$. The upper bound of $\deltan_{\ell}$ is true by definition of $\ell$. Recall that by definition of $\ell$, we have  that $\overline{n_j} + \sum_{i=\ell}^j \deltan_i > C$, which implies that $C - \overline{n_j} - \sum_{i'=\ell+1}^j \overline{\deltan_{i'}} \leq \overline{\deltan_\ell}$. For the lower bound of $0$, we consider two small cases. If $\ell = j$, then we see that $C - \overline{n_j} - \sum_{i'=j+1}^j \overline{\deltan_{i'}} = C - \overline{n_j} \geq 0$. Otherwise, if $\ell < j$, then what we want to show (i.e. $C - \overline{n_j} - \sum_{i=\ell+1}^j  \overline{\deltan_{i'}} \geq 0$) is true because $\ell$ was chosen to be the \textit{maximum} of the set, meaning that $\ell + 1 \in [j]$ has the property that $\overline{n_j} + \sum_{i=\ell+1}^j \overline{\deltan_i} \leq C$. This is exactly what we want to show.

Having verified constraints \eqref{nested_lp_capacity} and \eqref{nested_lp_j_le_M}, we now proceed to show that for this fixed value of $j$, we have $\sum_{i=1}^j r_i \cdot s_i^j \geq f_1(G) \cdot \overline \crr\cdot C$.
\begin{align*}
    \sum_{i=1}^j r_i \cdot s_i^j &= \sum_{i=1}^{\ell-1} r_i \cdot \overline{\deltan_i} + r_\ell \cdot \left( C - n_j - \sum_{i=\ell+1}^j \overline{\deltan_i} \right) + 2\cdot \sum_{i=\ell+1}^j r_i \cdot \overline{\deltan_i} \\
    &= \frac{1}{2} \cdot \overline \crr \cdot r_j \cdot C + r_\ell \cdot (C - \overline{n_j} - (\overline{n_j} - \overline{n_\ell})) + \frac{1}{2} \cdot \overline \crr \cdot (r_j - r_{\ell}) \cdot C \\
    &= \overline \crr \cdot r_j \cdot C + r_\ell \cdot \left(C - 2 \overline{n_j} + \overline{n_\ell} - \frac{1}{2} \cdot \overline \crr \cdot C \right) \,.
\end{align*}
By definition of $f_1(G)$,  our goal is to show that $\frac{\sum_{i=1}^j r_i \cdot s_i^j}{\overline \crr \cdot r_j \cdot C} \geq 1 - \frac{1}{2} e^{-2 \cdot G + 1}$.  Dividing the above equation through by $\overline \crr \cdot r_j \cdot C$, it would be sufficient to have 

$$ \frac{r_\ell}{\overline \crr \cdot r_j \cdot C} \cdot \left(\frac{1}{2} \cdot \overline \crr \cdot C - C + 2\cdot \overline{n_j} - \overline{n_\ell} \right) \leq \frac{1}{2}  e^{-2 \cdot G + 1}\,.  $$ 
Using the fact that $\overline{n_k} = \frac{1}{2} \cdot \overline \crr \cdot C \cdot \sum_{i=1}^k  \left( 1 - \frac{r_{i-1}}{r_i} \right)$ for any $k \in [M]$, we re-write the left hand side of the above as: 

$$ \frac{1}{2} \frac{r_\ell}{r_j} \cdot \left(1 -\frac{2}{\overline{\crr}} + 2 \cdot \sum_{i=1}^j \left(1 - \frac{r_{i-1}}{r_i} \right) - \sum_{i=1}^\ell \left(1 - \frac{r_{i-1}}{r_i} \right) \right) = \frac{1}{2} \frac{r_\ell}{r_j} \cdot \left(1 -\frac{2}{\overline{\crr}} + 2j - \ell - 2 \cdot \sum_{i=1}^j \frac{r_{i-1}}{r_i} + \sum_{i=1}^{\ell} \frac{r_{i-1}}{r_i} \right)\,. $$

Using the definition of $\overline \crr =
             \frac{2}{2 \cdot G + M - \sum_{i=1}^{M+1} \frac{r_{i-1}}{r_i}}$  from equation \eqref{definition:gamma_bar}, we have that that the above expression is upper bounded by
\begin{align*}
    &\: \frac{1}{2} \frac{r_\ell}{r_j} \cdot \left(1 - 2 \cdot G - M + \sum_{i=1}^{M+1} \frac{r_{i-1}}{r_i} + 2j - \ell - 2 \cdot \sum_{i=1}^j \frac{r_{i-1}}{r_i} + \sum_{i=1}^{\ell} \frac{r_{i-1}}{r_i} \right) \\
    = \:& \frac{1}{2} \frac{r_\ell}{r_j} \cdot \left(1 - 2 \cdot G - M + 2j - \ell - \sum_{i=\ell+1}^{j} \frac{r_{i-1}}{r_i} + \sum_{i=j+1}^{M+1} \frac{r_{i-1}}{r_i} \right) \\
    = \:& \frac{1}{2} \frac{r_\ell}{r_j} \cdot \Big(1 - 2 \cdot G - M + 2j - \ell - \sum_{i=\ell+1}^{j} \frac{r_{i-1}}{r_i} + \underbrace{\sum_{i=j+1}^{M+1} \frac{r_{i-1}}{r_i} - (M-j+1)}_{\leq 0} + (M-j+1) \Big) \\
    \leq \: &\frac{1}{2} \frac{r_\ell}{r_j} \cdot \Big(2 - 2 \cdot G + j-\ell - \sum_{i=\ell+1}^j \frac{r_{i-1}}{r_i} \Big) \\ 
    \leq \: & \frac{1}{2} e^{-2 \cdot G + 1}\,. 
\end{align*}
In the last step, we applied Lemma \ref{lemma:auxilliary_lemma}, stated below, with $x_i = \frac{r_{\ell + i - 1}}{r_{\ell + i}}$ for $x \in [j- \ell]$, $c = 2 - 2 \cdot G + j - \ell$ and $n = j - \ell$. That gives us the desired result. 
\end{proof}

\begin{lemma}
\label{lemma:auxilliary_lemma}
Consider the following multivariate function: 
$$ f(x_1, x_2, \ldots, x_n) = \prod_{i=1}^n x_i \cdot \left(c - \sum_{i=1}^n x_i \right)\,, $$
where $x_i \in (0,1)$ for $i \in [n]$. Then, we have
$$ f(x_1, \ldots, x_n) \leq e^{-(n+1-c)}\,. $$
\end{lemma}

\begin{proof}{Proof of Lemma \ref{lemma:auxilliary_lemma}.}
Let $s = \sum_{i=1}^n x_i$. We assume that $c - s > 0$.  Otherwise the statement is trivially true. For any fixed value of $s$, by the inequality of arithmetic and geometric mean (AM–GM)\footnote{The inequality of arithmetic and geometric means states that for any non-negative real numbers $x_1, \ldots x_n$, the following is true: 
$ \frac{x_1 + \ldots + x_n}{n} \geq \sqrt[n]{x_1 \cdots x_n} $,
where the left side is the arithmetic mean while the right side is the geometric mean. 
}, we have that
$$ \sqrt[n]{\prod_{i=1}^n} x_i \leq \frac{s}{n}\,, $$
which implies that 
$$ f(x_1, \ldots, x_n) \leq  \left( \frac{s}{n}\right)^n \cdot \left( c - n \cdot \frac{s}{n} \right)\,. $$
If we define $x \defeq \frac{s}{n}$, then we have that $f(x_1, \ldots, x_n) \leq g(x)$ where $g(x) = x^n \cdot (c - n \cdot x)$ for $x \in (0,1)$. Our assumption that $c - s > 0$ now carries over and becomes $c - n \cdot x > 0$ implying that $x \in \left( 0, \frac{c}{n} \right)$. Our goal now is to show that $g(x) \leq e^{-(n+1-c)}$  for any $x$ in that interval. To do so, we observe that the zeros of $g$ are precisely $x = 0$ and $x = \frac{c}{n}$ and it is clear that $g(x)$ is positive for all $x \in \left( 0, \frac{c}{n} \right)$. This means that $g(x)$, a polynomial in $x$, must have a local maximum in this interval. By a simple derivative calculation, we see that the only critical point of $g$ in that interval is at $x^* = \frac{c}{n+1}$ and so it must be a local maximum. Therefore, we have
\begin{align*}
    g(x) \leq g(x^*) &= \left(  \frac{c}{n+1} \right)^{n+1} = \left(  1 - \frac{n+1-c}{n+1} \right)^{n+1} \leq e^{-(n+1-c)} \,.
\end{align*}
This completes the proof. 
\end{proof}
\subsubsection{Proof of Lemma \ref{lemma:appx_two_of_two}}
\begin{proof}{Proof of Lemma \ref{lemma:appx_two_of_two}.}
We begin by one useful property that we will use multiple times in this proof. 

\textit{Claim: } Recall that $G \defeq K - M - \frac{r_{M+1}}{r_{M+2}} - \frac{r_{M+2}}{r_{M+3}} - \ldots - \frac{r_{K-1}}{r_K}$ as per equation \eqref{eq:G}. For any value of $G$, whose domain is $(1, K-M)$, we have:
\begin{equation}
    \label{optimality_gap_useful_property}
    \frac{r_{M+1}}{r_K} \geq \max(2-G, 0)\,.
\end{equation}
\textit{Proof of the claim: }If $2-G \leq 0$ (i.e. $G \geq 2$), then the statement is trivially true as $\frac{r_{M+1}}{r_K} \geq 0$. Thus,  we only focus on the case where $G \in (1,2)$. Recall that the definition of $G$ is $K - M - \frac{r_{M+1}}{r_{M+2}} - \frac{r_{M+2}}{r_{M+3}} - \ldots - \frac{r_{K-1}}{r_K}$ and note that $G$ is $K-M$ minus the sum of $K-M-1$ terms, each of which is between $0$ and $1$ exclusive. In addition, we notice that the \textit{product} of these $K - M - 1$ terms is exactly $\frac{r_{M+1}}{r_K}$. 
\begin{align*}
    \frac{r_{M+1}}{r_K} &= \frac{r_{M+1}}{r_{M+2}} \cdots \frac{r_{K-1}}{r_K} \\
    &\geq 2 - G\,.
\end{align*}
This inequality is true because when we fix the sum of the $K-M-1$ terms to be $K - M - G$ and each term is in $(0,1)$ and $G \in (1,2)$, then the minimum value of their product occurs when all but one of them is $1$ for a sum of $K - M - 2$ and the last one is $2-G$. This is formally proven in Lemma \ref{calculus_exercise}, stated below and proven in Section \ref{proof:calculus_exercise}

\begin{lemma}
\label{calculus_exercise} 
Given $x_1, \ldots, x_n \in (0,1)$ such that $\sum_{i=1}^n x_i = n- 1+\epsilon$ for some $\epsilon \in (0,1)$, we have that $\prod_{i=1}^n x_i \leq \epsilon$. 
\end{lemma}

Having presented the useful property, we now show the result: Here we would like to compare two quantities $\overline{\crr}_{up}$ and $\overline \crr$ that both have closed form solutions from equations \eqref{crrup_def} and \eqref{definition:gamma_bar}.
Our goal here is to show that 
\begin{align*}
\frac{\overline{\crr}}{\overline{\crr}_{up}} \geq f_2(G) \defeq  \begin{cases}
        1 - \frac{1 - \max(2 - G, 0)}{2 \cdot G + 1 - \max(2-G, 0)} & 2 \cdot n_M \leq C \\
        1 - \frac{1 - \max(2-G, 0)}{4 \cdot G - 2} & 2\cdot n_M > C
    \end{cases}
\end{align*}

\textbf{Case 1: $2 \cdot n_M < C$.}
We split our analysis into sub-cases based on the minimum in the definition of $\overline{\crr}_{up}$.  (Recall that $\overline{\crr}_{up}= \min \left( \crr_{\lpp}, \frac{1}{G} \right)$.)
\begin{itemize}
    \item Case 1a: $\crr_{\lpp} \leq \frac{1}{G}$. Recall that $ \crr_{\lpp}
    = \frac{2}{2 \cdot G + M - \sum_{i=1}^{M+1} \frac{r_{i-1}}{r_i} - \frac{r_M}{r_{M+1}} + \frac{r_M}{r_K}}$ and $G = K - M - \sum_{i=M+2}^{K} \frac{r_{i-1}}{r_{i}}$.
Hence, the condition $\crr_{\lpp} \leq \frac{1}{G}$ can be written as 
\begin{equation}
\label{inequality:the_condition}
   2 \cdot G \leq 2 \cdot G + M - \sum_{i=1}^{M+1} \frac{r_{i-1}}{r_i} -  \cdot \frac{r_M}{r_{M+1}} + \frac{r_M}{r_K} \quad \Rightarrow \quad  M - \sum_{i=1}^{M+1} \frac{r_{i-1}}{r_i} \geq \frac{r_M}{r_{M+1}} - \frac{r_M}{r_K}\,.
\end{equation}
Also, recall that $\overline \crr = \frac{2}{2 \cdot G + M - \sum_{i=1}^{M+1} \frac{r_{i-1}}{r_i}}$. We then bound our desired expression as follows:
\begin{align}
    \frac{\overline \crr }{\crr_{\lpp}} &= 
    \frac{
        2 \cdot G + M - \sum_{i=1}^{M+1} \frac{r_{i-1}}{r_i} - \frac{r_M}{r_{M+1}} + \frac{r_M}{r_K}
    }{
        2 \cdot G + M - \sum_{i=1}^{M+1} \frac{r_{i-1}}{r_i}
    }\nonumber \\
    &= 1 - \frac{
        \frac{r_M}{r_{M+1}} - \frac{r_M}{r_K}
    }{
        2 \cdot G + M - \sum_{i=1}^{M+1} \frac{r_{i-1}}{r_i}
    } \nonumber \\
    & \geq 1 - \frac{
        \frac{r_M}{r_{M+1}} - \frac{r_M}{r_K}
    }{
        2 \cdot G +\frac{r_M}{r_{M+1}} - \frac{r_M}{r_K} 
    } \nonumber \\
    & \geq 1 - \frac{
        1 - \frac{r_{M+1}}{r_K}
    }{
        2 \cdot G + 1 - \frac{r_{M+1}}{r_K}
    }\,. \label{repeated_inequlaity}
\end{align}
Here, in the first inequality, we used inequality \eqref{inequality:the_condition} in order to give an upper bound to $\sum_{i=1}^{M+1}\frac{r_{i-1}}{r_{i}}$ in the denominator. The second inequality is true because we see that $\frac{r_M}{r_{M+1}} - \frac{r_M}{r_K}$ appears in both the numerator and denominator of our fraction. We multiply both of those terms by $\frac{r_{M+1}}{r_M} > 1$, increasing the numerator and denominator of the fraction by the same amount, making it larger and closer to $1$. We apply the useful property in equation \eqref{optimality_gap_useful_property}, and immediately get our desired result:
$$ \frac{\overline \crr}{\crr_{\lpp}} \geq  1 - \frac{
        1 - \frac{r_{M+1}}{r_K}
    }{
        2 \cdot G + 1 - \frac{r_{M+1}}{r_K}
    }  \geq  1 - \frac{
        1 - \max(2-G, 0)
    }{
        2 \cdot G + 1 - \max(2-G, 0)
    }  $$

\item \textbf{Case 1b: $\crr_{\lpp} > \frac{1}{G}$.}
In this case, the negation of inequality \eqref{inequality:the_condition} is true, which allows us to bound the desired expression in the following way
\begin{align*}
   \frac{\overline \crr }{\crr_{\lpp}} =  \frac{\overline{\crr}}{1/G} &= \frac{2 \cdot G}{2 \cdot G + M - \sum_{i=1}^{M+1} \frac{r_{i-1}}{r_i}} \\
    &= 1 - \frac{M - \sum_{i=1}^{M+1} \frac{r_{i-1}}{r_i}}{2 \cdot G + M - \sum_{i=1}^{M+1} \frac{r_{i-1}}{r_i}} \\
    &\geq 1 - \frac{\frac{r_M}{r_{M+1}} - \frac{r_M}{r_K}}{2 \cdot G + \frac{r_M}{r_{M+1}} - \frac{r_M}{r_K}}\,.
\end{align*}
The last inequality is obtained by applying the inverse of inequality \eqref{inequality:the_condition} and noticing that by replacing $M - \sum_{i=1}^{M+1} \frac{r_{i-1}}{r_i}$ in both the numerator and denominator with a larger term, we make the fraction term larger. This resulting expression is exactly the same as the second-to-last expression in inequality \eqref{repeated_inequlaity}. Thus, we conclude that regardless of whether we are in Case 1a or Case1b, the same bound form Case 1a holds. 
\end{itemize}
This concludes the proof for Case 1. Notice that we did not use the fact that $2 \cdot n_M \leq C$. The statements we make are actually true for any relationship between $2 \cdot n_M$ and $C$. In case 2, we show that if $2 \cdot n_M > C$, then we can strengthen our bounds. 

\textbf{Case 2: } $2 \cdot n_M > C$. We use the fact that $n_M = \frac{1}{2} \cdot \overline{\crr} \cdot \sum_{i=1}^M \left(1 -  \frac{r_{i-1}}{r_{i}} \right) \cdot C$ and $\overline{\crr} = \frac{2}{2 \cdot G + M - \sum_{i=1}^{M+1} \frac{r_{i-1}}{r_i}}$ to simplify this condition into the following 
\[2 \cdot n_M > C \quad \Rightarrow \quad \overline{\crr} \cdot \sum_{i=1}^M \left(1 -  \frac{r_{i-1}}{r_{i}}\right) > 1 \quad \Rightarrow 2 \cdot \left( M - \sum_{i=1}^M \frac{r_{i-1}}{r_i} \right) > 2 \cdot G + M - \sum_{i=1}^{M+1} \frac{r_{i-1}}{r_i}\,.\] 

This is further simplified as 
\begin{equation}
    \label{2nMge_C_condition}
    M - \sum_{i=1}^{M+1} \frac{r_{i-1}}{r_i} > 2 \cdot G - 2 \cdot \frac{r_M}{r_{M+1}} \,.
\end{equation} 
Using inequality \eqref{2nMge_C_condition}, we construct a lower bound as follows (note that we do not perform case-work on $\crr_{\lpp}$. Rather we just use the fact that $\crr_{\lpp} \leq \overline{\crr}$): 
\begin{align*}
    \frac{\overline \crr }{\crr_{\lpp}} &\geq
    \frac{
        2 \cdot G + M - \sum_{i=1}^{M+1} \frac{r_{i-1}}{r_i} - \frac{r_M}{r_{M+1}} + \frac{r_M}{r_K}
    }{
        2 \cdot G + M - \sum_{i=1}^{M+1} \frac{r_{i-1}}{r_i}
    } \\
    &= 1 - \frac{
        \frac{r_M}{r_{M+1}} - \frac{r_M}{r_K}
    }{
        2 \cdot G + M - \sum_{i=1}^{M+1} \frac{r_{i-1}}{r_i}
    } \\
    & \geq 1 - \frac{
        \frac{r_M}{r_{M+1}} - \frac{r_M}{r_K}
    }{
        2 \cdot G + 2 \cdot G - 2 \cdot \frac{r_M}{r_{M+1}}
    } \\
    & = 1 - \frac{
        \frac{r_M}{r_{M+1}} \left( 1 - \frac{r_{M+1}}{r_K} \right)
    }{
        4 \cdot G  - 2 \cdot \frac{r_M}{r_{M+1}}
    } \\
    & \geq 1 - \frac{
        1 - \frac{r_{M+1}}{r_K}
    }{
        4 \cdot G  - 2
    }\,. 
\end{align*}
The first inequality is true by inequality \eqref{2nMge_C_condition}, while for the last inequality we take $\frac{r_M}{r_{M+1}} = 1$ to give a lower bound. This is a valid lower bound because of the fact that the function $h(x) = \frac{x \cdot \left( 1 - \frac{r_{M+1}}{r_K} \right)}{4\cdot G - 2 \cdot x}$ for $x\in(0,1)$ is increasing as $x$ approaches the discontinuity point $x = 2 \cdot G > 2$ (as $G > 1$). Therefore, we give a lower bound on $1 - h(x)$ by choosing $x = 1$. Now, we use our useful property that $\frac{r_{M+1}}{r_K} \geq \max(2-G, 0)$ to get our desired result

$$ 1 - \frac{
        1 - \frac{r_{M+1}}{r_K}
    }{
        4 \cdot G  - 2
    }
    \geq 
    1 - \frac{1 - \max(2-G, 0)}{4 \cdot G - 2}\,.
$$

\end{proof}

\subsubsection{Proof of Lemma \ref{lemma:appx_combine}}
\begin{proof}{Proof of Lemma \ref{lemma:appx_combine}.}
Recall that 
\begin{equation*}
  f_1(G) =
    \begin{cases}
        1 & 2 \cdot n_M \leq C \\
        1 - \frac{1}{2} e^{-2 \cdot G + 1} & 2 \cdot n_M > C 
    \end{cases}   
\end{equation*}
and 
\begin{align*}
 f_2(G) =  \begin{cases}
        1 - \frac{1 - \max(2 - G, 0)}{2 \cdot G + 1 - \max(2-G, 0)} & 2 \cdot n_M \leq C \\
        1 - \frac{1 - \max(2-G, 0)}{4 \cdot G - 2} & 2\cdot n_M > C
    \end{cases}\,.
\end{align*}

For $2 \cdot n_M \leq C$, we have that $f_1(G) \cdot f_2(G) = 1 - \frac{1 - \max(2 - G, 0)}{2 \cdot G + 1 - \max(2-G, 0)} $. This function is increasing for $G > 2$ and decreasing for $G \in (1,2)$, meaning that its minimum value is at $G =2$ for a value of $\frac{4}{5}$. 

For $2 \cdot n_M < C$, we consider the function 

$$ f_1(G) \cdot f_2(G) = \left( 1 - \frac{1}{2} e^{-2 \cdot G + 1} \right) \left( 1 - \frac{1 - \max(2-G, 0)}{4 \cdot G - 2} \right)\,.$$

We claim that the minimum of this function occurs at $G = 2$, for which the function value is $\left( 1 - \frac{1}{2} e^{-3} \right) \cdot \frac{5}{6} \approx 0.8125 > \frac{4}{5}$. For $G > 2$, it is easy to see that $f_1(G) \cdot f_2(G)$ is an increasing function. For $G < 2$, we see that $f_1(G) \cdot f_2(G)$ has critical points at $G \approx 1.1394$ and $G \approx 1.5496$ for which the function values are $0.8139$ and $0.8158$, both of which are significantly larger than $\frac{4}{5}$. Therefore, we have proven that $f_1(G) \cdot f_2(G) \geq \frac{4}{5}$ is true for any value of $G$. 
\end{proof}

\subsubsection{Proof of Lemma \ref{calculus_exercise}}
\label{proof:calculus_exercise}
\begin{proof}{Proof of Lemma \ref{calculus_exercise}.}
Here, we show that given $x_1, \ldots, x_n \in (0,1)$ such that $\sum_{i=1}^n x_i = n- 1+\epsilon$ for some $\epsilon \in (0,1)$, we have that $\prod_{i=1}^n x_i \leq \epsilon$. To do so,
we consider the following minimization problem, where we relax $x_i \in (0,1)$ to $x_i \in [0,1]$. Clearly this new minimization problem's objective is a lower bound on that of our original problem. 
\begin{align*}
    \min_{x_1, \ldots, x_n} \quad & \prod_{i=1}^n x_i \\
    \text{s.t} \quad & \sum_{i=1}^k x_i = n - 1 + \epsilon \\
    & x_i \in [0, 1] 
\end{align*}
We claim that the minimum above occurs when all but one of the $x_i$'s are equal to $1$ and the last is equal to $\epsilon$. To see this, we first notice that given that $\epsilon > 0$, any feasible solution cannot choose $x_i = 0$ for any $i \in [n]$ for otherwise there is no way for the remaining $n-1$ of the $x_i$'s to add up to something greater than $n-1$. Now, consider any optimal solution $(x_i^*)_{i=1}^n$ to the above minimization problem. For any two indices $i \neq j$, we claim that one of $x_i^*$ or $x_j^*$ must be equal to $1$. Suppose for contradiction that there exists indices $i \neq j$ such that $0 < x_i^* \leq x_j^* < 1$. Then, we can increase $x_j^*$ by $\delta > 0$ and decrease $x_i^*$ by $\delta$ and this results in a smaller objective value because: 

$$ x_i^* \cdot x_j^* \geq  x_i^* \cdot x_j^* + (x_i^* - x_j^*) \cdot \delta - \delta^2 = (x_i^* + \delta)(x_j^* - \delta)\,,$$
and all other terms $x_k^*$ for $k \not \in \{i,j\}$ remain the same. Given that $0 < x_i^* < x_j^* < 1$, it is always possible to find such a $\delta > 0$ so that the $(x_i^* + \delta)$ and $(x_j^* - \delta)$ are both in $[0,1]$. Notice that their sum is the same as $x_i^* + x_j^*$ meaning that this new solution is feasible. Therefore, we get a contradiction and so it must be the case that one of $x_i^*$ or $x_j^*$ is equal to $1$. Applying this argument to all pairs of indices in $[n]$ gives us the conclusion that all but one of the $x_1^*, \ldots, x_n^*$'s is equal to $1$, which shows our desired result. 
\end{proof}

\section{Additional Figures for Numerical Studies}

{\color{blue}In Figure \ref{sim:by_M_2}, we present the average TR-CR and FA-CR values over \(250\) problem instances. These instances consider \(M\) ranging from \(0\) to \(9\), \(\sigma\) values of \(0.1\), \(0.5\), and \(1.0\), and a fixed \(K = 10\). The figure also includes the theoretical lower bounds for FA-CR, represented as \(\crr^*\) and \(\crr^{BQ}\). These findings align with the observations from Figure \ref{sim:by_M}.

Figure \ref{sim:cdf_1} illustrates the empirical cumulative distribution function (CDF) for the \(250\) problem instances discussed in Section \ref{sec:numerical_studies}. The CDF is evaluated for the selected configurations: \( (M=4, \sigma=1) \), \( (M=3, \sigma=0.5) \), and \( (M=1, \sigma=0.1) \). Notably, the optimized nested \texttt{po} algorithm, denoted as \(\alg^*\), exhibits superior performance in comparison to the BQ nested \texttt{po} algorithm, \(\alg^{BQ}\), with respect to both TR-CR and FA-CR measures.}

\begin{figure}
    \centering
    \begin{subfigure}[c]{0.4\textwidth}
        \centering 
        \includegraphics[width=\textwidth]{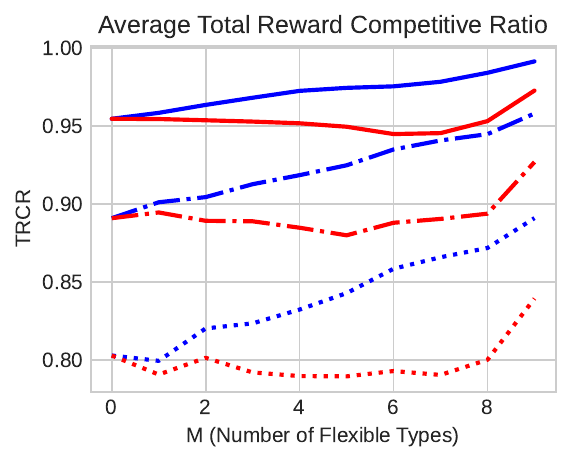}
        \subcaption{Average \textsc{tr-cr} by $M$}
        \label{sim:by_M_left_2}
    \end{subfigure}
    \begin{subfigure}[c]{0.4\textwidth}
        \centering 
        \includegraphics[width=\textwidth]{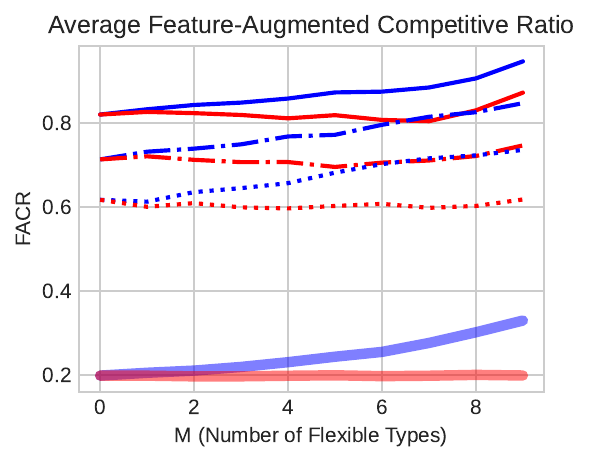}
        \subcaption{Average \textsc{fa-cr} by $M$}
        \label{sim:by_M_right_2}
    \end{subfigure}
    \begin{subfigure}[c]{0.15\textwidth}
        \centering 
        \includegraphics[width=\textwidth]{legend.png}
    \end{subfigure}
    \vspace{0.5em}
    \caption{Average TR-CR and FA-CR across 250 problem instances while varying $M$ for $\sigma \in \{0.1, 0.5, 1.0\}$ with $K = 10$. For FA-CR, the theoretical lower bounds $\Gamma^*$ and $\Gamma^{BQ}$ are also shown.}
    \label{sim:by_M_2}
\end{figure} 

\begin{figure}
    \centering
    \begin{subfigure}[c]{0.42\textwidth}
        \centering 
        \includegraphics[width=\textwidth]{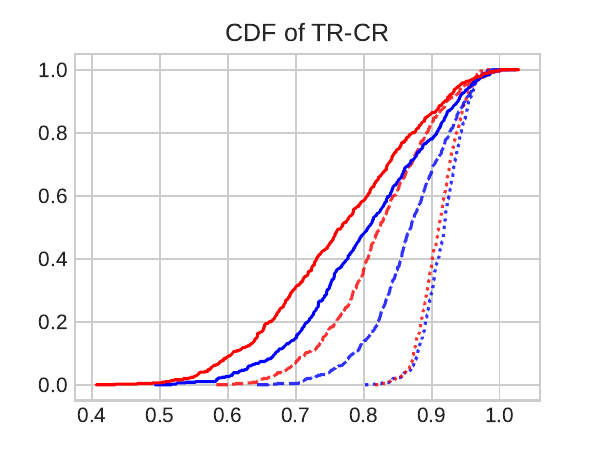}
        \subcaption{Empirical CDF of \textsc{tr-cr}}
    \end{subfigure}
    \begin{subfigure}[c]{0.42\textwidth}
        \centering 
        \includegraphics[width=\textwidth]{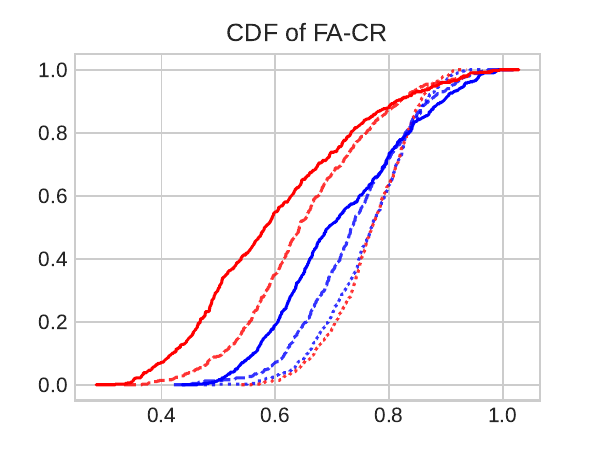}
        \subcaption{Empirical CDF of \textsc{fa-cr}}
    \end{subfigure}
    \begin{subfigure}[c]{0.11\textwidth}
        \centering 
        \includegraphics[width=\textwidth]{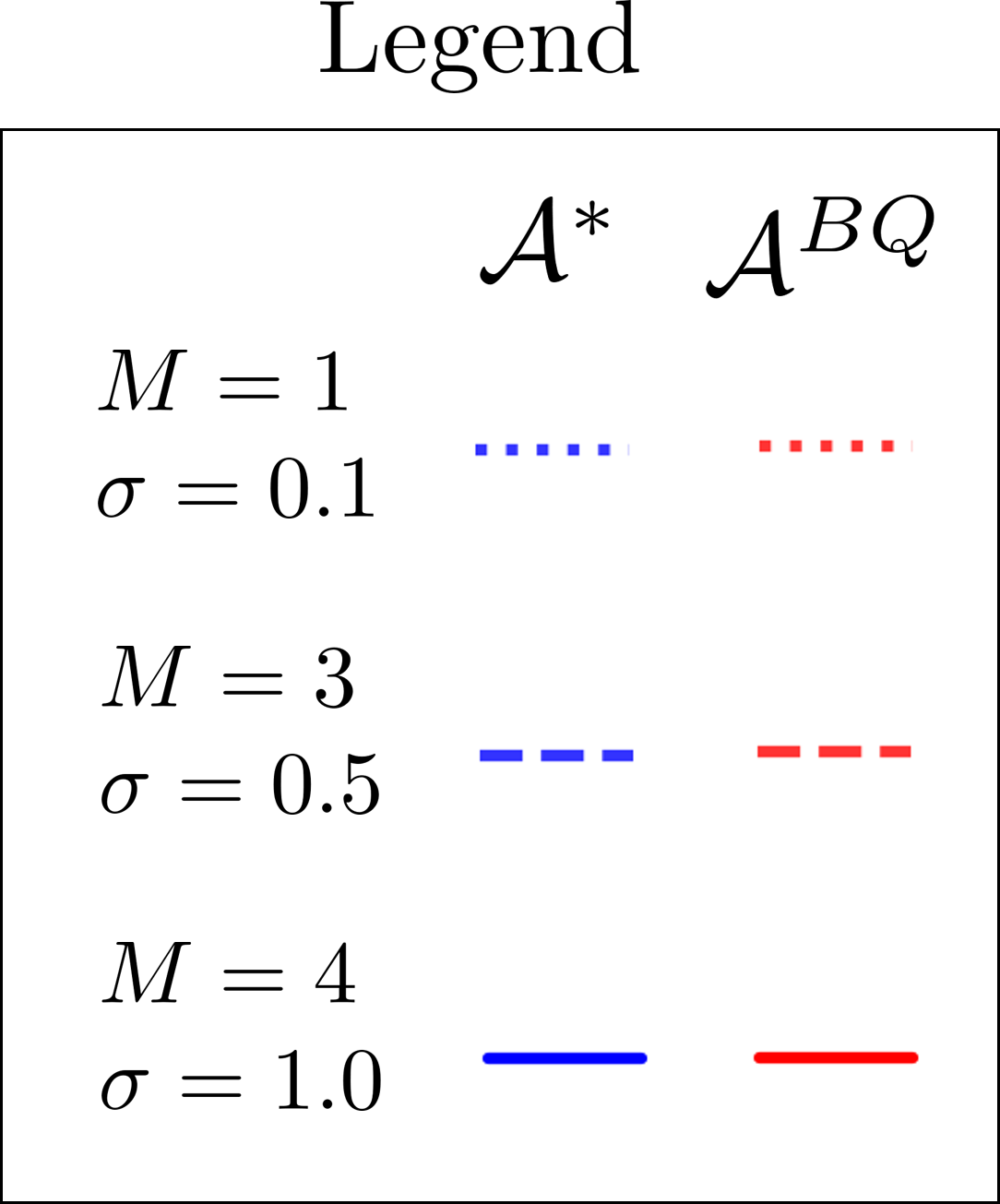}
    \end{subfigure}
    \vspace{0.5em}
    \caption{Empirical cumulative distribution functions of FA-CR and TR-CR across 250 instances for select $(M, \sigma)$ combinations. Here, the number of agent types is $K= 5$. }
    \label{sim:cdf_1}
\end{figure} 
\end{APPENDICES}

\end{document}